\documentclass[acmconf]{acmart}
\pdfoutput=1
\usepackage[utf8]{inputenc}
\usepackage[ruled]{algorithm2e} % For algorithms
\usepackage{enumitem}
\usepackage{amsmath}
\usepackage[normalem]{ulem}
\usepackage{graphicx}
\usepackage{capt-of}
\usepackage[export]{adjustbox}
\usepackage{color}
\usepackage{colortbl}
\usepackage{soul}          % for highlighting comments
\usepackage[draft,inline,nomargin]{fixme}
\usepackage{multirow}
\usepackage{epstopdf}
\usepackage{booktabs}
\usepackage{array}
\usepackage{url}
\input{glyphtounicode}
\usepackage[T1]{fontenc}
\usepackage{subcaption}
\usepackage{tabularx}
\usepackage{ltxtable}
\usepackage{blindtext}
\usepackage{graphicx}
\usepackage{wrapfig,lipsum,booktabs}
\usepackage{arydshln}
\usepackage{booktabs}
\usepackage{stackengine}
\def\mybar[#1]#2{%%
  {\color{black}\rule[0.1ex]{#1mm}{5pt}} #2}
\usepackage{acmart-taps}

\usepackage{wrapfig,lipsum}
\usepackage{booktabs}
\usepackage{array}
\def\mybar[#1]#2{%%
  {\color{black}\rule[0.1ex]{#1mm}{5pt}} #2}
\usepackage{tcolorbox}
\tcbuselibrary{breakable}
\tcbuselibrary{listings}
\newtcblisting{jsoncode}{
  colback=gray!5,
  colframe=gray!75,
  listing only,
  listing options={
    language=java,
    basicstyle=\ttfamily,
    breaklines=true % Enables line breaking
  },
  breakable
}
\usepackage{acmart-taps}

\SetAlFnt{\small}
\SetAlCapFnt{\small}
\SetAlCapNameFnt{\small}
\SetAlCapHSkip{0pt}
\IncMargin{-\parindent}

\newcommand\para[1]{{\vspace{5pt} \noindent {\bf #1}}}

\newcolumntype{L}[1]{>{\raggedright\let\newline\\\arraybackslash\hspace{0pt}}m{#1}}
\newcolumntype{C}[1]{>{\centering\let\newline\\\arraybackslash\hspace{0pt}}m{#1}}
\newcolumntype{R}[1]{>{\raggedleft\let\newline\\\arraybackslash\hspace{0pt}}m{#1}}

\begin{document}

\title{MindScape Study: Integrating LLM and Behavioral Sensing for Personalized AI-Driven Journaling Experiences}




% Subigya Kumar Nepal Dartmouth College, Hanover, New Hampshire, United States
% Arvind Pillai Dartmouth College, Hanover, New Hampshire, United States
% William Campbell Colby College, Waterville, Maine, United States
% Talie Massachi Department of Computer Science, Brown University, Providence, Rhode Island, United States
% Dr. Michael V. Heinz Dartmouth College, Lebanon, New Hampshire, United States
% Ashmita Kunwar Dartmouth College, Hanover, New Hampshire, United States
% Eunsol Soul Choi Cornell Tech, New York, New York, United States
% Xuhai "Orson" Xu Department of Electrical Engineering and Computer Science, Massachusetts Institute of Technology, Cambridge, Massachusetts, United States, xoxu@mit.edu
% Joanna Kuc University College London, London, United Kingdom
% Jeremy F Huckins Biocogniv Inc, Burlington, Vermont, United States
% Jason Holden University of California, San Diego, La Jolla, California, United States, c1barry@ucsd.edu
% Sarah M. Preum Department of Computer Science, Dartmouth College, Hanover, New Hampshire, United States
% Colin Depp University of California, San Diego, La Jolla, California, United States
% Nicholas Jacobson Dartmouth College, Hanover, New Hampshire, United States
% Dr. Mary P Czerwinski Microsoft Research, Redmond, Washington, United States
% Eric Granholm Department of Psychiatry, University of California, San Diego, La Jolla, California, United States, egranholm@health.ucsd.edu
% Andrew Campbell Dartmouth College, Hanover, New Hampshire, United States

\author{Subigya Nepal}
\orcid{0000-0002-4314-9505}
% \email{sknepal@cs.dartmouth.edu}
\affiliation{%
  \institution{Dartmouth College}
  \streetaddress{15 Thayer Dr}
  \city{Hanover}
  \state{New Hampshire}
  \country{USA}
  \postcode{03755}
}

\author{Arvind Pillai}
\orcid{0000-0002-2489-1130}
\affiliation{%
  \institution{Dartmouth College}
  \streetaddress{15 Thayer Dr}
  \city{Hanover}
  \state{New Hampshire}
  \country{USA}
  \postcode{03755}
}

\author{William Campbell}
\affiliation{%
  \institution{Colby College}
  \streetaddress{}
  \city{Waterville}
  \state{Maine}
  \country{USA}
}
\author{Talie Massachi}
\affiliation{%
  \institution{Brown University}
  \streetaddress{}
  \city{Providence}
  \state{Rhode Island}
  \country{USA}
}

\author{Michael V. Heinz}
\orcid{0000-0003-0866-0508}
\affiliation{%
  \institution{Dartmouth College}
  \city{Hanover}
  \state{New Hampshire}
  \country{USA}
}

\author{Ashmita Kunwar}
% \orcid{0000-0002-4314-9505}
% \email{sknepal@cs.dartmouth.edu}
\affiliation{%
  \institution{Dartmouth College}
  \streetaddress{15 Thayer Dr}
  \city{Hanover}
  \state{New Hampshire}
  \country{USA}
  \postcode{03755}
}

\author{Eunsol Soul Choi}
\orcid{}
\affiliation{%
  \institution{Cornell Tech}
  \city{New York}
  \state{New York}
  \country{USA}
}
\author{Orson Xu}
\affiliation{%
  \institution{Massachusetts Institute of Technology}
  \city{Cambridge}
  \state{Massachusetts}
  \country{USA}
}
\author{Joanna Kuc}
\affiliation{%
  \institution{University College London}
  \city{London}
  \country{UK}
}
\author{Jeremy Huckins}
\affiliation{%
  \institution{Biocogniv Inc}
  \city{Burlington}
  \state{Vermont}
  \country{USA}
}

\author{Jason Holden}
\affiliation{%
  \institution{University of California, San Diego}
  \city{La Jolla}
  \state{California}
  \country{USA}
}

\author{Sarah M. Preum}
% \orcid{0000-0002-4314-9505}
% \email{sknepal@cs.dartmouth.edu}
\affiliation{%
  \institution{Dartmouth College}
  \streetaddress{15 Thayer Dr}
  \city{Hanover}
  \state{New Hampshire}
  \country{USA}
  \postcode{03755}
}

\author{Colin Depp}
\affiliation{%
  \institution{University of California, San Diego}
  \city{La Jolla}
  \state{California}
  \country{USA}
}
\author{Nicholas Jacobson}
\affiliation{%
  \institution{Dartmouth College}
  \city{Hanover}
  \state{New Hampshire}
  \country{USA}
}
\author{Mary Czerwinski}
\affiliation{%
  \institution{Microsoft Research}
  \city{Redmond}
  \state{Washington}
  \country{USA}
}
\author{Eric Granholm}
\affiliation{%
  \institution{University of California, San Diego}
  \city{La Jolla}
  \state{California}
  \country{USA}
}
\author{Andrew T. Campbell}
\orcid{0000-0001-7394-7682}
\affiliation{%
  \institution{Dartmouth College}
  \city{Hanover}
  \state{New Hampshire}
  \country{USA}
}

% \author{ANONYMOUS}
% \affiliation{%
%   \institution{ANONYMOUS}
%   \department{ANONYMOUS}
%   \city{}
%   \state{}
%   \postcode{}
%   \country{}
% }

% \email{anonymous@anonymous.com}

\begin{abstract}
Mental health concerns are prevalent among college students, highlighting the need for effective interventions that promote self-awareness and holistic well-being. MindScape explores a novel approach to AI-powered journaling by integrating passively collected behavioral patterns such as conversational engagement, sleep, and location with Large Language Models (LLMs). This integration creates a highly personalized and context-aware journaling experience, enhancing self-awareness and well-being by embedding behavioral intelligence into AI. We present an 8-week exploratory study with 20 college students, demonstrating the MindScape app's efficacy in enhancing positive affect (7\%), reducing negative affect (11\%), loneliness (6\%), and anxiety and depression, with a significant week-over-week decrease in PHQ-4 scores (-0.25 coefficient). The study highlights the advantages of contextual AI journaling, with participants particularly appreciating the tailored prompts and insights provided by the MindScape app. Our analysis also includes a comparison of responses to AI-driven contextual versus generic prompts, participant feedback insights, and proposed strategies for leveraging contextual AI journaling to improve well-being on college campuses. By showcasing the potential of contextual AI journaling to support mental health, we provide a foundation for further investigation into the effects of contextual AI journaling on mental health and well-being.
\end{abstract}

\maketitle
\renewcommand{\shortauthors}{Nepal et al.}

 \section{Introduction}
\label{sec:mindscape_intro}
The significance of struggles with mental health among college students is becoming increasingly apparent, impacting students' academic performance, social engagement, and overall personal development. Research, including findings from the American College Health Association (ACHA)–National College Health Assessment, highlights a concerning prevalence of anxiety, depression, and related issues among students~\cite{acha, beiter2015prevalence, mofatteh2021risk, 10.1145/3643501}. Students face a range of pressures, from academic challenges to social and personal hurdles, which affect not only their mental health but also their emotional resilience and personal growth~\cite{tosevski2010personality, stoliker2015influence, gueldner2020social, wang2022first}. While traditional mental health interventions administered by clinicians do provide personalized and context-specific support, emerging technologies present an opportunity to extend this support, making it more readily available, automated, and able to potentially overcome considerable institutional barriers. In addition, there is a need for innovative solutions that align with the digital habits of today's students. We propose a novel study, \textit{MindScape}, that integrates the traditional practice of journal writing with mobile technology and large language models (LLM)~\cite{naveed2023comprehensive} to create a contextually-aware journaling application. The MindScape Android application benefits from on-device sensors and data to provide insights into the user's daily life. It tracks aspects such as physical activity, social interactions, and location to form an understanding of the context in which the individual operates. By analyzing these data in real-time, the app can provide personalized, context-sensitive journaling prompts designed to provoke thought and reflection. The prompts aim to remind users to introspect and commit time to digitally record their thoughts, thus establishing regular self-reflection habits that are contextualized by their daily lives. MindScape represents a novel application class that incorporates behavioral intelligence into AI. We believe that integrating time-series data obtained from mobile phones and wearables, capturing real-time behaviors and patterns of users, with the capabilities of LLMs will give rise to a new category of AI applications driven by mobile sensing. 

Journaling has long been recognized as a potent tool for self-reflection, enabling individuals to externalize thoughts, consolidate disjointed experiences, and identify patterns in their behavior and emotional states. This practice of regular introspection has been linked to a range of psychological benefits, from reducing distress symptoms to enhancing overall wellbeing~\cite{sohal2022efficacy,Dimitroff2016}. In this study, we explore the potential gains realized through the inclusion of personalization and context-awareness in journaling. We define `context' as the comprehensive set of behavioral, environmental, and temporal factors shaping a user's daily experience and mental state. For the purpose of our study, this includes physical activities, sleep patterns, social interactions, digital behaviors, and location data, considering both current and historical patterns. The inclusion of personalization and context-awareness in journaling is more than just a technological novelty. It addresses certain inherent limitations in human introspection and memory recall abilities. People may not readily identify certain behavioral patterns or come to particular conclusions about their daily lives without some form of guidance or external input. This is where personalized and context-aware prompts can be valuable, as they may highlight aspects of users' lives they may have overlooked. Additionally, human memory recall can be biased towards more recent experiences (recency bias) and peak emotional experiences (peak-recency bias), sometimes at the expense of equally significant past events~\cite{CortisMack2017, Kornell2009,Sedikides2020}. Context-aware journaling can help counteract this limitation by bringing forward relevant circumstances, events, or feelings from different timeframes in the users' life. Lastly, by addressing these user limitations, personalized and context-aware journaling could not just improve the process of journaling, but also potentially enhance the mental health benefits associated with this practice.

Herein lies the novelty of our approach: using mobile sensing to capture behavioral data that reflects the user's context and emotional state, and employing an LLM to generate journaling prompts that are highly relevant to the user's current contextual situation and surroundings. To complement our context-aware journaling prompts, we introduce daily check-ins as a novel feature in our study app. These brief, simple texts are triggered four times daily and aim to encourage users to pause and reflect on their current experiences. For instance, a check-in might say, \textit{``Your morning seemed to include more than just tapping screens – a bit of chitchat too!''}. Users can respond with a quick thumbs up or thumbs down, allowing for a low-burden, high-engagement interaction. By leveraging contextual intelligence, our check-ins aim to increase user attachment and engagement with the journaling app, while also making their reflections more meaningful and potentially amplifying the mental health benefits of journaling. The primary goal of these check-ins is to facilitate fleeting moments of self-reflection, helping users develop greater awareness of their thoughts, emotions, and behaviors throughout the day. By incorporating thumbs up/down responses, we simplify the reflection process, making it more accessible and increasing the likelihood of users engaging in regular self-reflection. Furthermore, the MindScape journaling app integrates additional contextual factors such as students' mood while journaling, their academic stress levels, and temporal variables like weekdays or weekends. 
While previous studies have explored various aspects of digital journaling and context-aware applications, MindScape uniquely combines several elements: \textit{(a)} it integrates a wide range of passive sensing data specifically for mental health journaling, going beyond simple activity or location tracking; \textit{(b)} it employs LLMs to generate personalized, context-aware journaling prompts, a novel application in the mental health domain; \textit{(c)} it introduces frequent, low-burden check-ins to complement deeper journaling sessions, encouraging continuous self-reflection throughout the day; and \textit{(d)} it focuses on the specific needs and contexts of college students, a population particularly vulnerable to mental health challenges.
By addressing these gaps, our study aims to push the boundaries of how technology can support mental health in young adults. We explore whether a more comprehensive, context-aware, and AI-driven approach to journaling can lead to deeper self-reflection and significant improvements in well-being. Early in our development, we conduct a qualitative user study with undergraduate students to understand their journaling habits and preferences. Insights from this study, revealing students' desires for personalized, context-aware prompts aimed at fostering reflection on daily experiences, significantly influenced our app's design. We believe our holistic approach allows for a more tailored and responsive tool, capable of providing meaningful support in the unique, often high-pressure, fast paced environment of college life. Our paper makes the following contributions:
\begin{enumerate}
    \item We design \textit{MindScape} -- an AI-driven journaling app that integrates behavioral sensing and LLM to deliver personalized, adaptive journaling prompts. We conduct an 8-week study with 20 college students to evaluate the efficacy of this system. By the end of the study, participants report up to an 11\% improvement in well-being scores, with statistically significant enhancements in affect, loneliness, mindfulness, self-reflection, anxiety, and depression.

    \item We examine the check-ins and journaling prompts generated by the app, analyzing their topic coverage, and the frequencies of categories to which the prompts belong. We find that the morning check-ins often revolve around social and communication app usage, while afternoon check-ins shift towards academic and social life experiences.
    
\item We analyze linguistic differences between journals from contextual and generic prompts using the Linguistic Inquiry and Word Count (LIWC)~\cite{boyd2022development}. Our findings indicate that responses to contextual prompts exhibit more personal language, greater references to personal experiences and relationships, whereas broader emotional expressions (such as affect) are more prevalent in journals from generic prompts.
    
    \item We review participant feedback concerning their experience and the app's usability and provide recommendations for future research. Overall, 85\% of participants rate MindScape’s usability as good or excellent. Seventy percent consider the journal prompts to be moderately-to-very relevant, and 85\% report that the contextual prompts sometimes, often, or always lead to more in-depth reflection compared to generic prompts, demonstrating the effectiveness of the MindScape app.
\end{enumerate}

It is important to note that our objective is to introduce and evaluate a new journaling paradigm that integrates behavioral sensing and contextual awareness. This research conducts a proof-of-concept study on contextual journaling, specifically focusing on its effectiveness as a unique journaling method. We do not perform controlled trials to determine which approach is more beneficial. Ours is an exploratory study designed to potentially augment the classic benefits of journaling by utilizing the latest advancements in LLMs to provide an unobtrusive, effective tool for users to manage their wellbeing and growth. This approach is closely aligned with the Human-Computer Interaction (HCI) community's interests, highlighting the significance of AI in enriching user-centric digital experiences. Bridging into Ubiquitous Computing (UbiComp), our research focuses on integrating these technologies into everyday routines. Our goal is for this tool to offer benefits and support and assist students in developing lasting self-reflection and emotional mindfulness skills. We hope that this study will contribute significantly to the ongoing dialogue in HCI and UbiComp, particularly regarding the seamless integration of technology to enhance personal well-being, offering a comprehensive view of its practical application and user impact.
\section{Related Work}
\label{sec:mindscape_related_work}
Journaling is a reflective practice where individuals record their thoughts, feelings, and experiences. The act of journaling promotes self-awareness~\cite{alt2020reflective, williams2009reflective}, processing of emotions~\cite{Smyth2018}, and cognitive organization of experiences~\cite{sohal2022efficacy}. Studies have consistently shown that journaling can improve mood, provide stress relief, and overall, enhance mental well-being~\cite{sohal2022efficacy, keech2021journaling, miller2014interactive}. As mobile devices and computers become more prevalent, they have reshaped the practice of journaling. The transition to digital journaling platforms brings conveniences that traditional paper-based methods lack. These include enhanced accessibility--- ensuring that users can journal anytime and anywhere, heightened privacy---as entries are secured behind digital safeguards, and the ability to enrich journal entries with multimedia elements.

Journaling can be prompted or unprompted. Unprompted journaling allows for free expression without specific guidelines, giving users freedom to explore their thoughts and feelings. In contrast, prompted journaling uses specific questions or suggestions to guide the journaling process, providing a structure that can help focus and inspire the user. Such prompts are designed to encourage self-reflection, personal growth, and exploration of various topics and experiences. Several digital journaling platforms offer a wide range of prompts to initiate the writing and reflection journey, providing daily reminders to ensure users stay on track with their journaling. This approach can be particularly helpful for users who are new to journaling or those looking to explore new areas of self-discovery and creativity. However, most prompted journaling applications rely on generic prompts not tailored to the user's situation. Several studies demonstrate that question prompts are one of the main factors positively affecting reflection quality~\cite{Chen2009, Ge2003, cengiz2020effect, Glogger2009}. Thus, generic prompts, while useful, may reduce reflection quality due to their broad nature~\cite{Aronson2010, Rudrum2022}.

Our study focuses on context-aware journaling, where journaling prompts are derived from behavioral data collected via smartphones. This approach enhances traditional journaling by offering prompts that closely align with users' daily experiences and mental states. By using mobile sensing technology, capable of tracking activities, sociability, locations, and app usage, we generate dynamic prompts that reflect the nuanced aspects of an individual's life. This approach differs from previous studies that have explored a broader range of personal informatics systems for reflection~\cite{10.1145/3491102.3501991, 10.1145/3544549.3573803}, by integrating these insights into the journaling process. For example, ~\citet{Kocielnik2018} leverage mobile based step count for reflection on activity level whereas ~\citet{Bakker2018} use the MoodPrism app to help in mood tracking. Our method aims to mirror the reflective goals of such apps and to offer deeper insights into users' lifestyles and emotional patterns through personalized journaling. In addition, our study uses a wide range of contextual cues to facilitate journaling, a feature that sets it apart even from its closest counterparts like Apple's journal application~\cite{applejournal}. While Apple's offering leverages contextual data such as photos and location to generate prompts, our approach extends beyond conventional context-awareness to include an amplified set of signals such as: screen time; social, entertainment, and communication app usage; sleep habits, in-person conversations, calls, and text message exchanges.  

Our study additionally integrates both sleep information, such as duration and timing, as well as physical fitness metrics like activity levels, distance travelled, and time spent at the gym. Our study also considers location-based semantics like time spent in a cafeteria, Greek spaces, and other similar locations. This comprehensive approach sets our study apart by providing a more nuanced and detailed context for generating personalized journaling prompts. Our study also leverages LLM capabilities to enable the creation of intelligent, personalized journaling prompts. AI-driven tools have been used in therapy chatbots, virtual agents, and behavior change systems, offering personalized advice and support~\cite{chiu2024computational,hua2024large, 10.1145/3643540, 10.1145/3571884.3604305, yeo2024help, kumar2023exploring, sharma2023facilitating, nie2024llm, bhattacharjee2023understanding, kian2024can}. These applications demonstrate the capacity of AI to understand and respond to a wide range of emotional and psychological states~\cite{matz2024potential}. Existing studies have leveraged LLMs for AI-mediated journaling~\cite{kim2023mindfuldiary, kimdiarymate,ferrara2022empowering}. However, to our knowledge, none of the existing studies have integrated objective and passively observed behavioral data into AI-mediated journaling. While previous work has incorporated behavioral sensing signals into LLM prompts for various applications~\cite{https://doi.org/10.48550/arxiv.2309.16639, Englhardt2024, Khaokaew2024}, our approach is novel in its specific application to mental health journaling. Our method integrates rich contextual information to generate personalized, privacy-conscious journaling prompts aimed at improving self-awareness and mental wellbeing. The novelty lies in the application of these techniques to mental health, the specific combination of contextual factors we consider, and our focus on generating reflective prompts rather than predicting behaviors or app usage. By using an LLM framework to analyze behavioral data and generate relevant journaling prompts, we aim to investigate the potential for a nuanced, data-driven augmentation of the journaling process. Our study seeks to reinforce the benefits of journaling, while simultaneously exploring the effectiveness of context-aware prompts for highly reflective self-expression. Through this unique approach, we aim to optimize the impact of personalized digital journaling.
\section{Methodology}
\label{sec:mindscape_methodology}
In this section, we detail our study methodology, which encompasses the study design, participant demographics, the mobile sensing behavioral data collected by our system, and the design of the personalized journaling prompts and check-ins.
\begin{figure}[h!]
    \centering
      \includegraphics[width=1\linewidth]{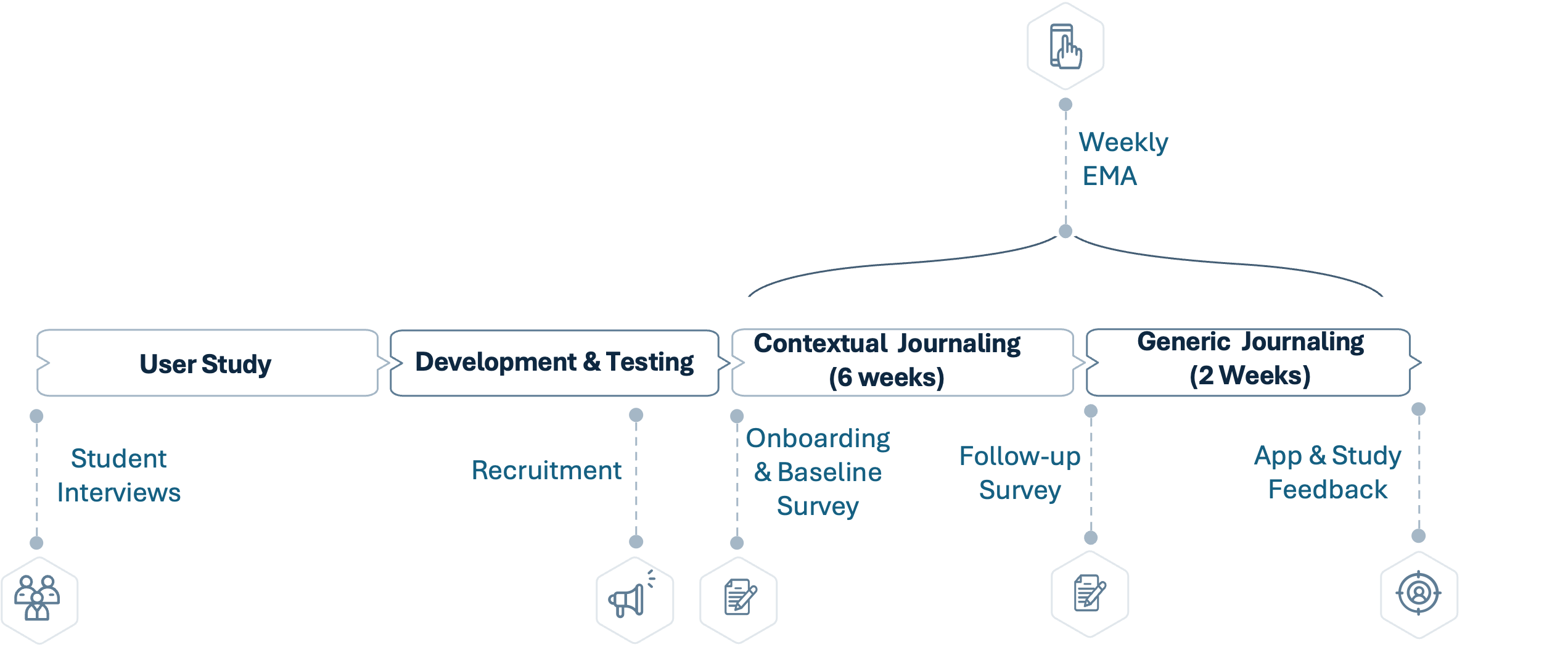}
              \caption{MindScape Study Design Overview: The study begins with a user research phase to capture insights into the journaling experiences and habits of students. The total duration of the study is 8 weeks, consisting of six weeks of contextual AI-based journaling prompts followed by two weeks of generic journaling prompts. We gather baseline and follow-up surveys at the start and end of the contextual journaling phase, respectively. Additionally, at the end of the 8th week, following the generic journaling phase, we collect a final set of surveys focusing on app experiences and overall study feedback. Note that the primary goal of our study is to explore the effectiveness of contextual AI-based journaling, not to compare its benefits with that of generic journaling. The generic journaling phase is included to provide a complementary analysis of linguistic differences in journals received for generic prompts versus contextual prompts, offering additional insights into the study's outcomes.}
      \label{fig:mindscape_studyworkflow}
\end{figure}
\subsection{Study Design}
At the beginning of our study, we engage in a User Study focused on capturing participants' perspectives using user-centered design principles. We conduct interviews with students to illuminate their needs and experiences with journaling, providing a foundational understanding for our research approach. The stages of the study are detailed in Figure~\ref{fig:mindscape_studyworkflow}. As we transition into the Development and Testing phase, we refine our methodology and initiate participant recruitment. We employ various channels such as posters, class-wide emails, Computer Science majors and minors email chains, student mailing lists, and collaborations with mental health-related campus clubs to reach potential participants. Out of 91 respondents expressing interest, 26 qualify for the study (the majority of them have Apple phones while our app only supports Android), with 20 ultimately signing the consent form to participate.

During the recruitment and onboarding process, we prioritize transparency regarding data collection and usage. Participants are provided with comprehensive information about the types of data collected, the methods and frequency of collection, and how this data will be used within the app. This information is presented in detail on our recruitment form and reiterated verbally during individual onboarding sessions. During these sessions, participants have the opportunity to ask questions and seek clarification on any aspect of the data collection process. We emphasize that participation is voluntary and that users can withdraw at any time without consequence. We also take care to exclude individuals with high depression scores, as indicated by elevated Patient Health Questionnaire-8 (PHQ8) survey results, to ensure safety due to the unmoderated nature of the reflection prompts. Once enrolled, participants install the MindScape Android app on their phones and enable the permissions for the signals. We clearly communicate during the onboarding process that while the app is designed to function optimally with full data access, users have the right to manage their privacy settings through their device's permission controls. Users are informed that they can adjust permissions at the system level, and the app will adapt accordingly, ensuring they maintain control over their data sharing while participating in the study.

The central six weeks of our study involve participants interacting with contextual AI-driven journal prompts delivered through the app. This begins with an onboarding process, where participants complete a baseline survey that captures their initial journaling habits, demographic details, and psychological states via standard surveys focused on well-being and self-reflection. At the conclusion of this six-week contextual journaling phase, we administer a follow-up survey using the same standard questionnaires. This enables us to gauge changes in well-being, personal growth, and reflection, assessing whether AI-driven contextual journaling contributes positively to participants' development. Additionally, we conduct weekly Ecological Momentary Assessment (EMA) -- a research methodology that involves repeatedly collecting self-reported data from participants to capture dynamic changes and patterns over time -- to monitor changes in participants’ well-being and reflection. Please see Appendix~\ref{sec:mindscape_list_of_questions_asked} for the list of surveys and questions we ask participants. 

Following the initial contextual journaling phase, the participants enter a two-week period of generic journaling, receiving a uniform prompt via the MindScape app: \textit{``What's on your mind today? Use this journal entry to explore freely any thoughts, feelings, memories, or experiences—anything you'd like.''} Due to a limited sample size, a full randomized controlled trial was not feasible. Nevertheless, this phase provides an opportunity to compare and contrast traditional journaling with our AI-driven contextual method. After completing the full eight-week study duration, participants receive the final study feedback survey, which collects their insights on their journaling experience. This includes thoughts on the app's usability and performance as well as any additional feedback or suggestions. Participants are compensated up to USD 130 for their involvement. The study has received approval from Dartmouth College's Internal Review Board, ensuring all procedures meet ethical standards.

\subsection{Demographics}
We recruit 20 students from Dartmouth College for our study. Out of these participants, a majority, 60\% (N=12), identify as female, while 35\% (N=7) identify as male, and one participant (5\%) identifies as non-binary. The cohort comprises 12 graduate students and 8 undergraduate students. When examining racial demographics, 35\% (N=7) of participants identify as White or Caucasian, 25\% (N=5) as Asian, 20\% (N=4) as Black or African American, 15\% (N=3) report belonging to multiple racial categories, and 5\% (N=1) report `Other'. Age distribution among the participants shows that 65\% (N=13) are within the 18-24 age bracket, 30\% (N=6) fall into the 25-34 age range, and 5\% (N=1) are 45 years old or above. Regarding journaling experience, 55\% (N=11) of the participants currently maintain a journal, 20\% (N=4) do not keep a journal at present though they have journaled in the past, and 25\% (N=5) have never engaged in journaling.

\begin{table*}[ht]
\caption{Behavioral Data Categories: Users are required to prioritize among the four behavioral data categories, each encompassing specific feature sets.}
\resizebox{1\textwidth}{!}{
\begin{tabular}{lllll}
\rowcolor[HTML]{9B9B9B} 
\textbf{Category}                          & \textbf{Signals/Features}           & \textbf{Example Journaling Prompt Generated} \\\bottomrule
\multirow{4}{*}{Physical Fitness}        & {\cellcolor[HTML]{EFEFEF}Physical activity (walking, running, and sedentary duration)}            & \textit{Your running routine has really taken off! How's that} \\
                                  & Distance travelled              & \textit{influencing your day?} \\
                                  & {\cellcolor[HTML]{EFEFEF}Time spent at the Gym}         &  \\
                                 \hline
\multirow{2}{*}{Sleep} & Sleep duration        & \textit{Your sleep pattern has shifted recently. Could this change}                                                & \\
                                  & {\cellcolor[HTML]{EFEFEF}Sleep schedule (start time and end time)}  &     \textit{be affecting your daytime energy and focus?}                          & \\  
                                  \hline
\multirow{2}{*}{Digital Habits} & {Screen time}     &     \textit{You've been clocking less screen time lately. What have you} \\
                                  & {\cellcolor[HTML]{EFEFEF}App use (Freq. of social media, communication, \& entertainment apps use)}  &  \textit{been doing instead that you've found rewarding or enjoyable?} \\ \hline
\multirow{5}{*}{Social Interaction}      & Phone logs (incoming calls, outgoing calls, incoming SMS, outgoing SMS)    &                            & \\
                                  & {\cellcolor[HTML]{EFEFEF}In-person conversations (number and duration of conversations)}         &                   \textit{Your call patterns are up; any conversations lately that}                  & \\   & Number of significant places visited    &                     \textit{brought a smile to your face?}              & \\
       & {\cellcolor[HTML]{EFEFEF}Time spent at frats/sororities partying}            &                                                & \\ 
                                  & Misc. locations (Time spent at leisure, social, study places,  cafeteria \& home) &                                                                       & \\ 
                                  \bottomrule
\end{tabular}}
\label{tbl:mindscape_sensingdata}
\end{table*}

\subsection{Mobile Sensing based Behavioral Data}
The MindScape app automatically infers user activities, like movement and rest, analyzes conversation lengths, and gathers data on screen usage and location~(see Table~\ref{tbl:mindscape_sensingdata}). This provides an integrated view of a user's daily patterns, social interactions, and digital habits. For example, the sensing data might reveal patterns in how often participants attend social functions, dine at campus facilities, or go to the gym. This information allows us to tailor the journaling prompts to align with the participant’s current experiences and to support their emotional well-being. As part of gathering this data, we create a semantic map of the college campus, with locations such as dining areas and gyms marked, allowing the app to accurately infer the context of participants' activities. This allows for prompts to be customized, encouraging reflection on particular events of the day. The integration of the GPT-4 LLM enables the translation of this rich, multi-faceted behavioral data into personalized and contextually relevant journaling prompts and frequent check-ins that enhance positive introspection and participant engagement. All data collected are temporarily stored on the participant's phone and then securely uploaded to the MindScape cloud. We then leverage the GPT-4 model through OpenAI's API~\cite{openai}, allowing us to process the collected behavioral data and additional contexts to generate tailored prompts. Addressing potential concerns relating to participant privacy, we ensure all data sent for processing via OpenAI's GPT-4 model are de-identified and consist only of high-level metadata. This approach includes stripping any potentially personally identifiable information before the data is utilized to generate tailored prompts. We acknowledge that a locally hosted open-source model could offer an alternative to mitigate privacy concerns further, albeit with possible performance trade offs. In this study, our focus is oriented towards understanding the potential and efficacy of this novel application of AI in journaling practices. Given this emphasis, we decided to utilize OpenAI's GPT-4 model for its robust performance and scalability capabilities.
\vspace{-0.2pc}

\begin{figure*}[h!]
     \centering
     \begin{subfigure}[b]{0.25\textwidth}
         \centering
        \fbox{\includegraphics[trim={1cm 25cm 3cm 0cm},clip,width=0.83\linewidth]{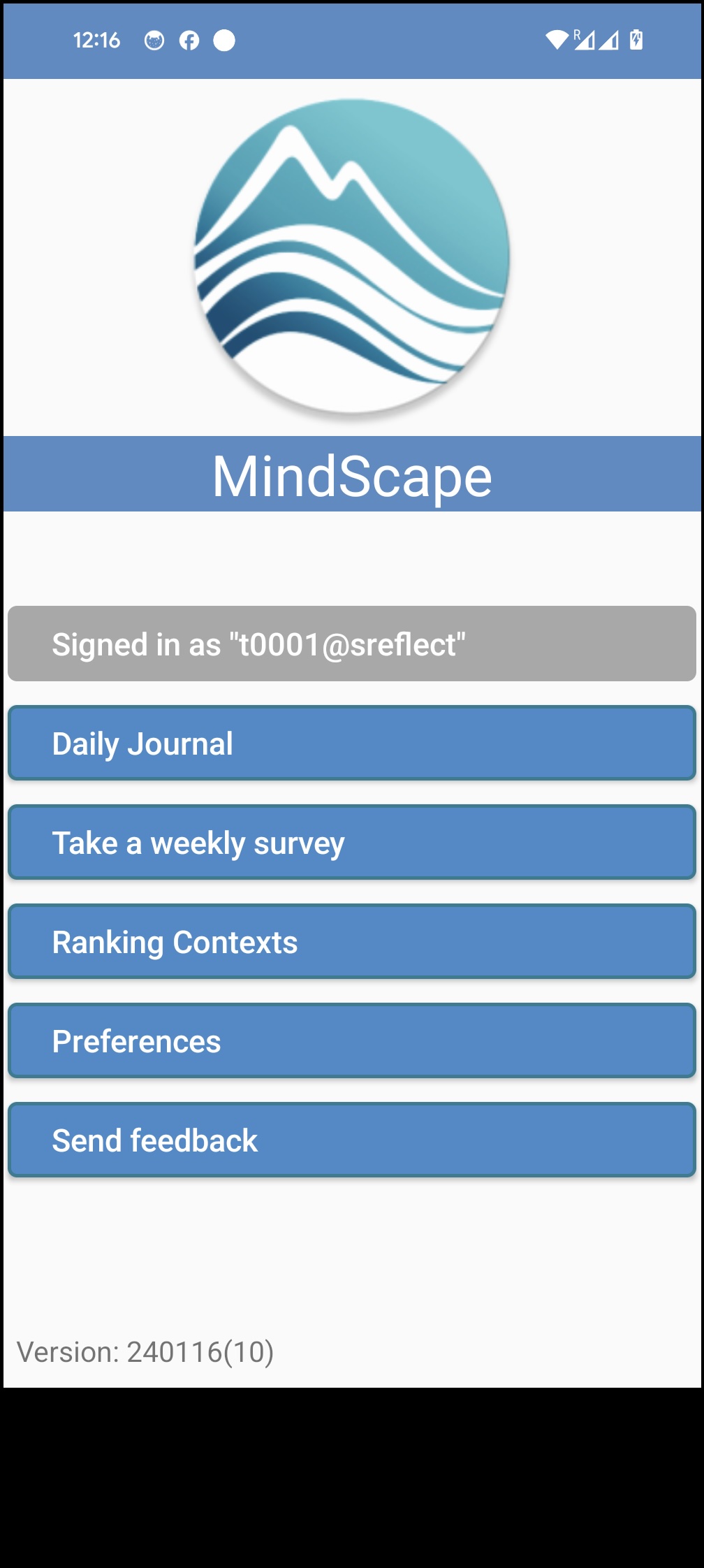}}
         % \vspace{0.1cm}
     \end{subfigure}
     \begin{subfigure}[b]{0.315\textwidth}
         \centering
         \fbox{\includegraphics[trim={1cm 25cm 2cm 5cm},clip,width=0.75\linewidth]{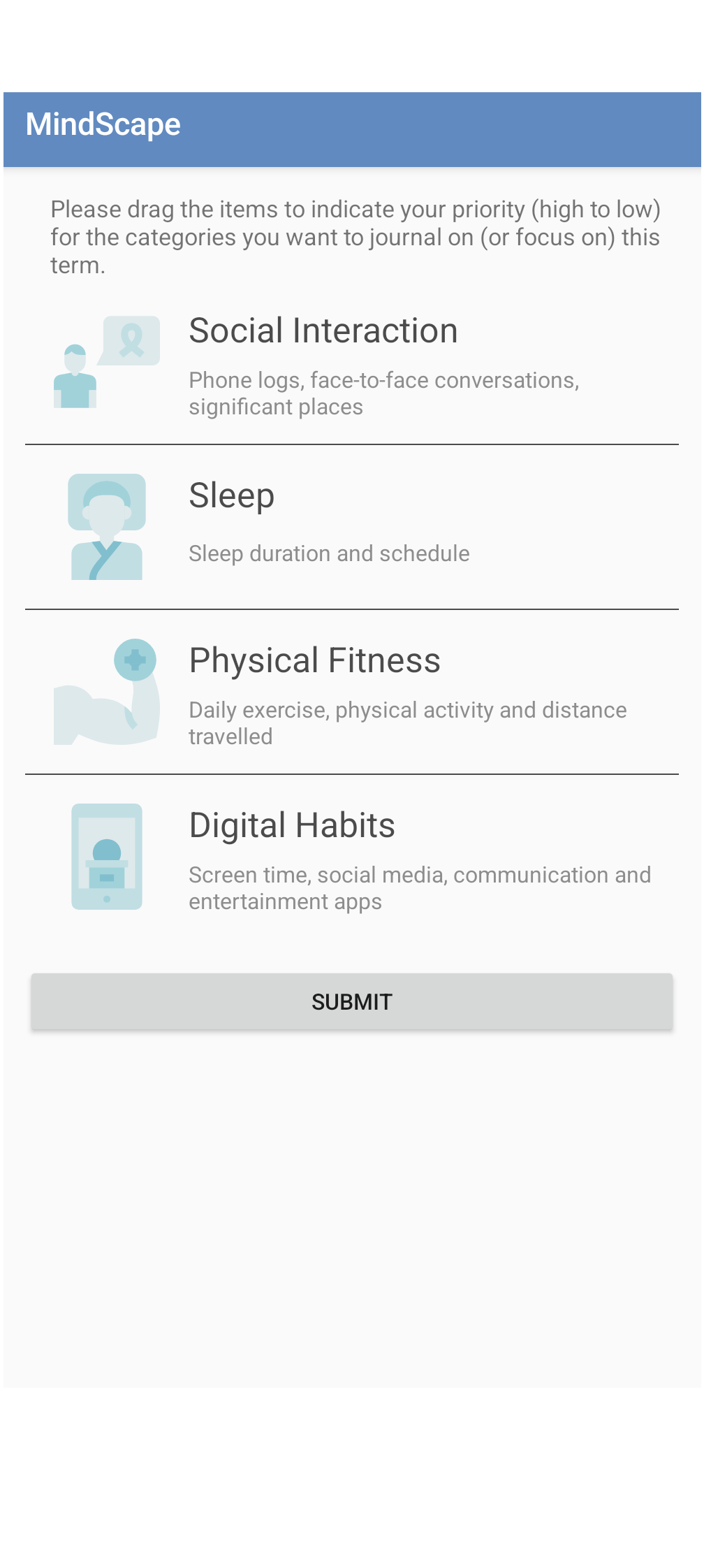}}
        % \fbox{\includegraphics[width=0.70\textwidth, scale=2.0]{figures/2.png}}
                  % \vspace{0.1cm}
     \end{subfigure}
\begin{subfigure}[b]{0.25\textwidth}
         \centering
    \fbox{\includegraphics[trim={1cm 20cm 1cm 0cm},clip,width=0.81\linewidth]{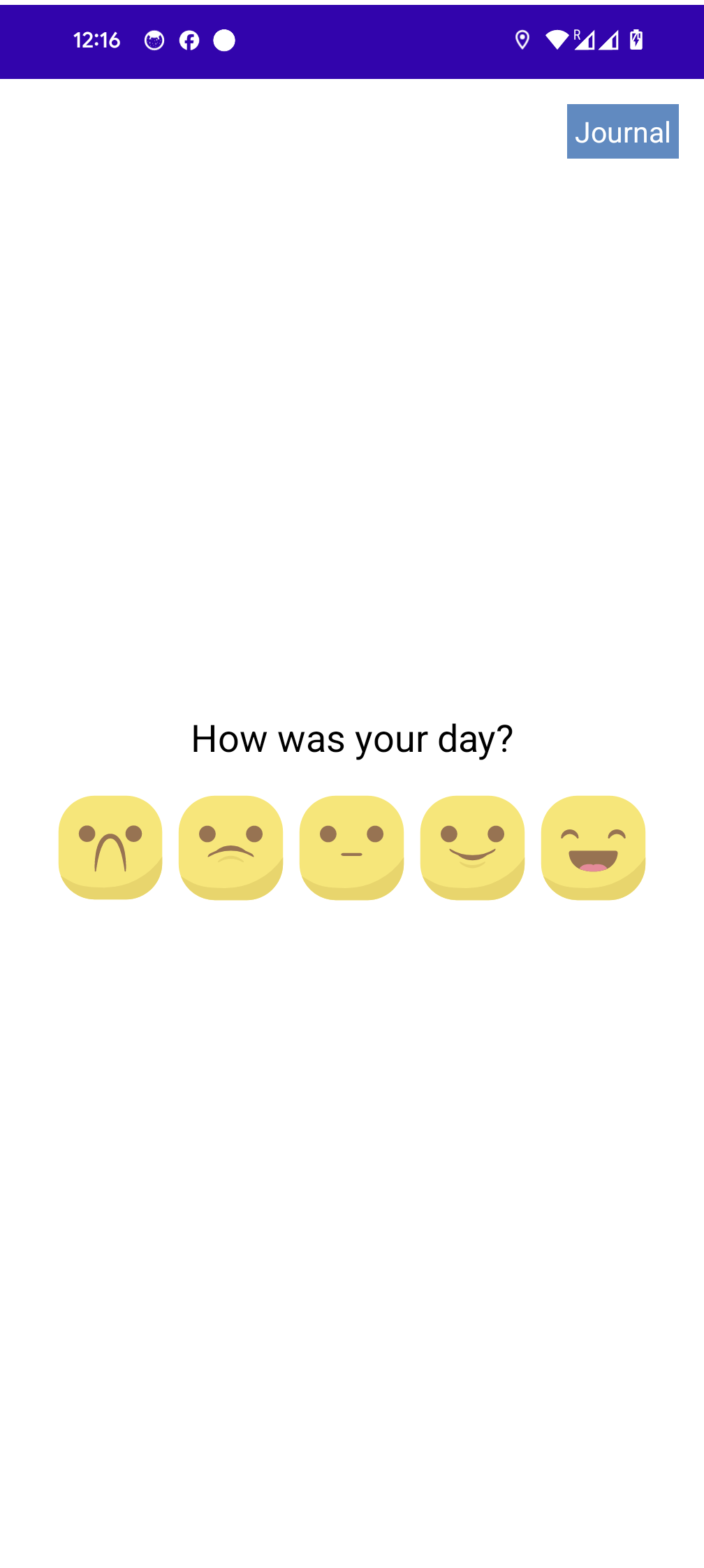}}
                  % \vspace{0.8cm}
     \end{subfigure}
     
    \begin{subfigure}[b]{0.25\textwidth}
         \centering
 \vspace{0.8cm}
         \fbox{\includegraphics[trim={1cm 10cm 1cm 4cm},clip,width=0.83\linewidth]{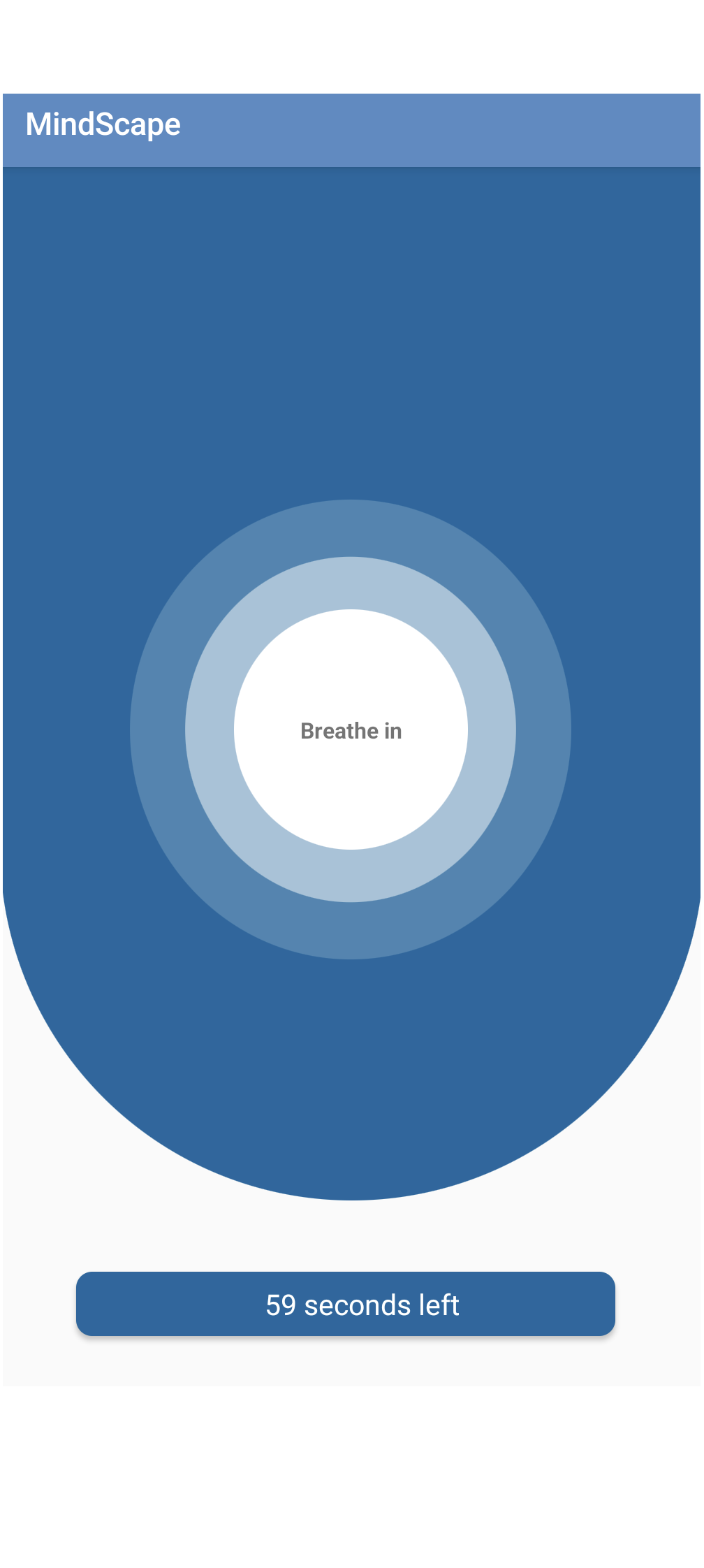}}
     \end{subfigure}
        \begin{subfigure}[b]{0.315\textwidth}
         \centering
         \fbox{\includegraphics[trim={1.5cm 20cm 1.5cm 1cm},clip,width=0.72\linewidth]{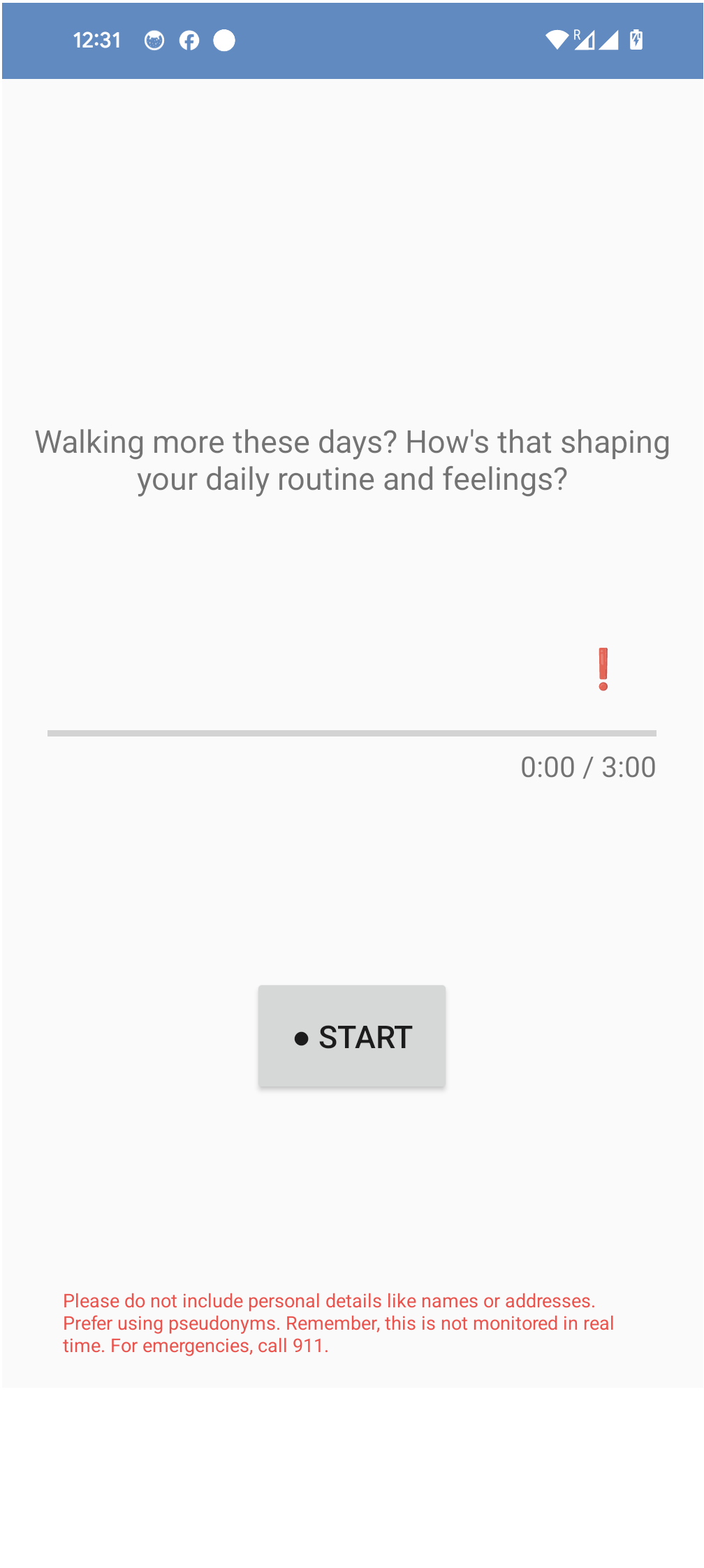}}
         
    % \begin{subfigure}[b]{0.25\textwidth}
    %      \centering
    %     \fbox{\includegraphics[trim={1.5cm 20cm 1.5cm 1cm},clip,width=0.905\linewidth]{figures/5.png}}
    \end{subfigure}
            \begin{subfigure}[b]{0.25\textwidth}
         \centering
        \fbox{\includegraphics[trim={1cm 12.5cm 1cm 0cm},clip,width=0.81\linewidth]{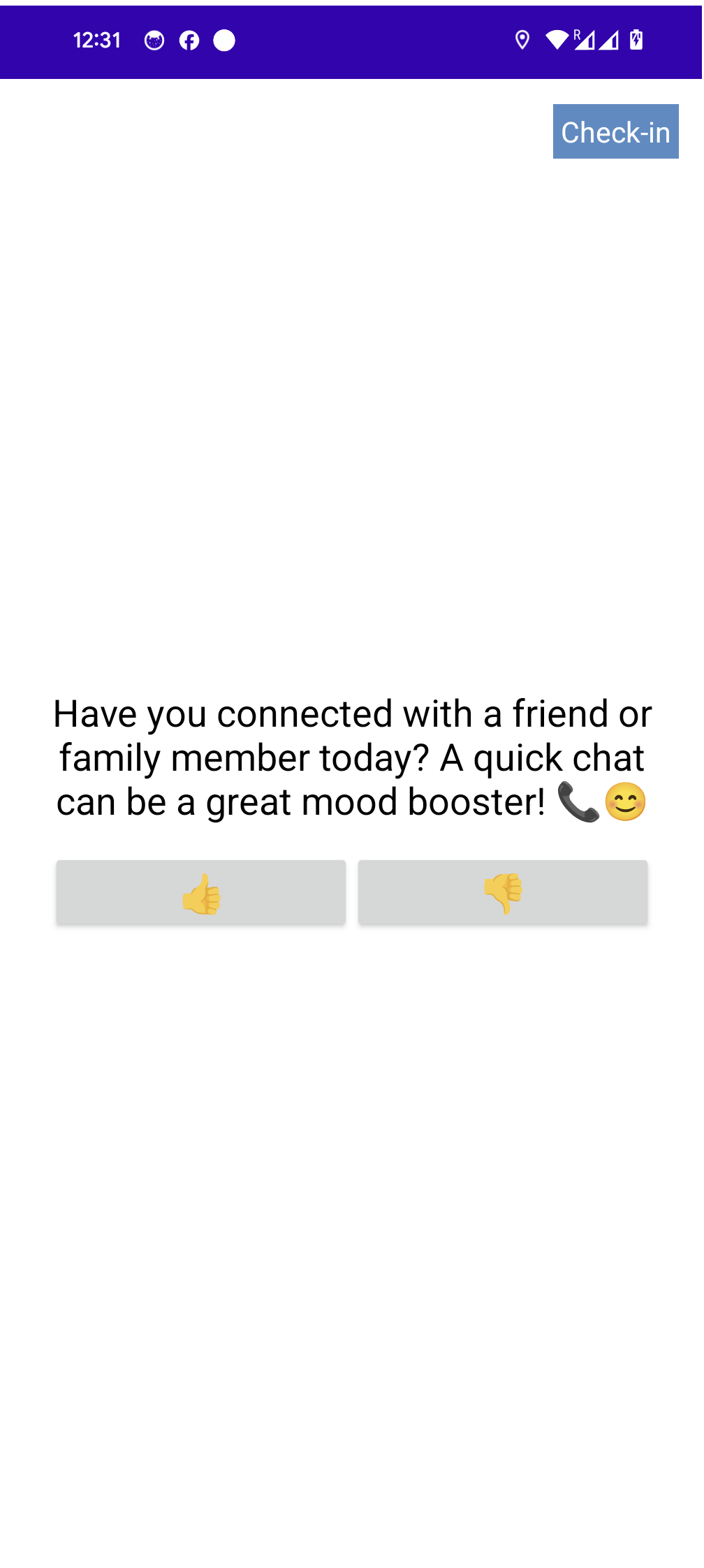}}
     \end{subfigure}
        \caption{MindScape App Workflow: Users sign in, set preferences in four categories, and start journaling with a notification tap. Journaling workflow includes an emoji based mood assessment, a one-minute breathing exercise, followed by contextual prompts. The final screen showcases the daily context-aware check-in.}
        \Description{The figure presents six screenshots from the MindScape app, showcasing its user journey. It begins with the landing screen, followed by a screen where users can rank their preferences across four categories—Social Interaction, Sleep, Physical Fitness, and Digital Habits—by dragging the options. Next, the app queries how the user's day went, offering emoji for responses. This leads to a screen dedicated to a 1-minute breathing exercise. Subsequently, users encounter a journaling prompt inviting reflections on whether increased walking has influenced their daily routines and feelings, with an option to reply via audio. The final screenshot features a check-in nudge asking, ``Have you connected with a friend or family member today? A quick chat can be a great mood booster!'' alongside thumbs up and down buttons. Collectively, these screens illustrate the user experience progression within the MindScape app.}
        \label{fig:mindscape_appscreens}
\end{figure*}

\subsection{Personalized Journaling Prompts}
\label{sec:mindscape_personalizedjournalingprompts}
Upon installing MindScape app, participants are prompted to allow the app permission for data collection. Then, they rank their journaling interests in four key areas — Social Interaction, Sleep, Digital Habits, and Physical Fitness. We identify these four key areas through interviews with students on campus (See Section~\ref{sec:mindscape_userstudy}). Because we collect many different types of data, we want to ensure the journaling prompts we provide are actually helpful to participants. Thus, we use these categories to identify what matters most to each individual participant. We also include the user's preferences (i.e., category ranking) in the prompt for GPT-4~\cite{openai2023gpt4} to generate more relevant journaling prompts. During their enrollment, each user provides us with their usual bedtime for both weekdays and weekends. Journaling notifications are triggered two hours before their reported bedtime. When a notification is tapped,  participants are redirected to the app's journaling screen. There, they are first asked how their day was, followed with a one-minute breathing exercise, and finally, they are asked to write or record (i.e., audio) their journal entry. Only at this point can the participants see the personalized journaling prompt. Participants can also open the app and journal whenever they prefer. Note, the one-minute deep breathing exercise before journaling is based on findings that short relaxation techniques can improve mental clarity and emotional readiness~\cite{benson1993wellness, Zaccaro2018}. This step aims to help users transition to a reflective mood, enhancing their focus for more insightful journaling. It is intended to make the journaling process a calming, enriching routine. Figure~\ref{fig:mindscape_appscreens} shows different screens of the application.

\begin{figure*}[h!]
     \centering
    \includegraphics[trim={4.0cm 1.0cm 2cm 1cm},clip,width=1\linewidth]{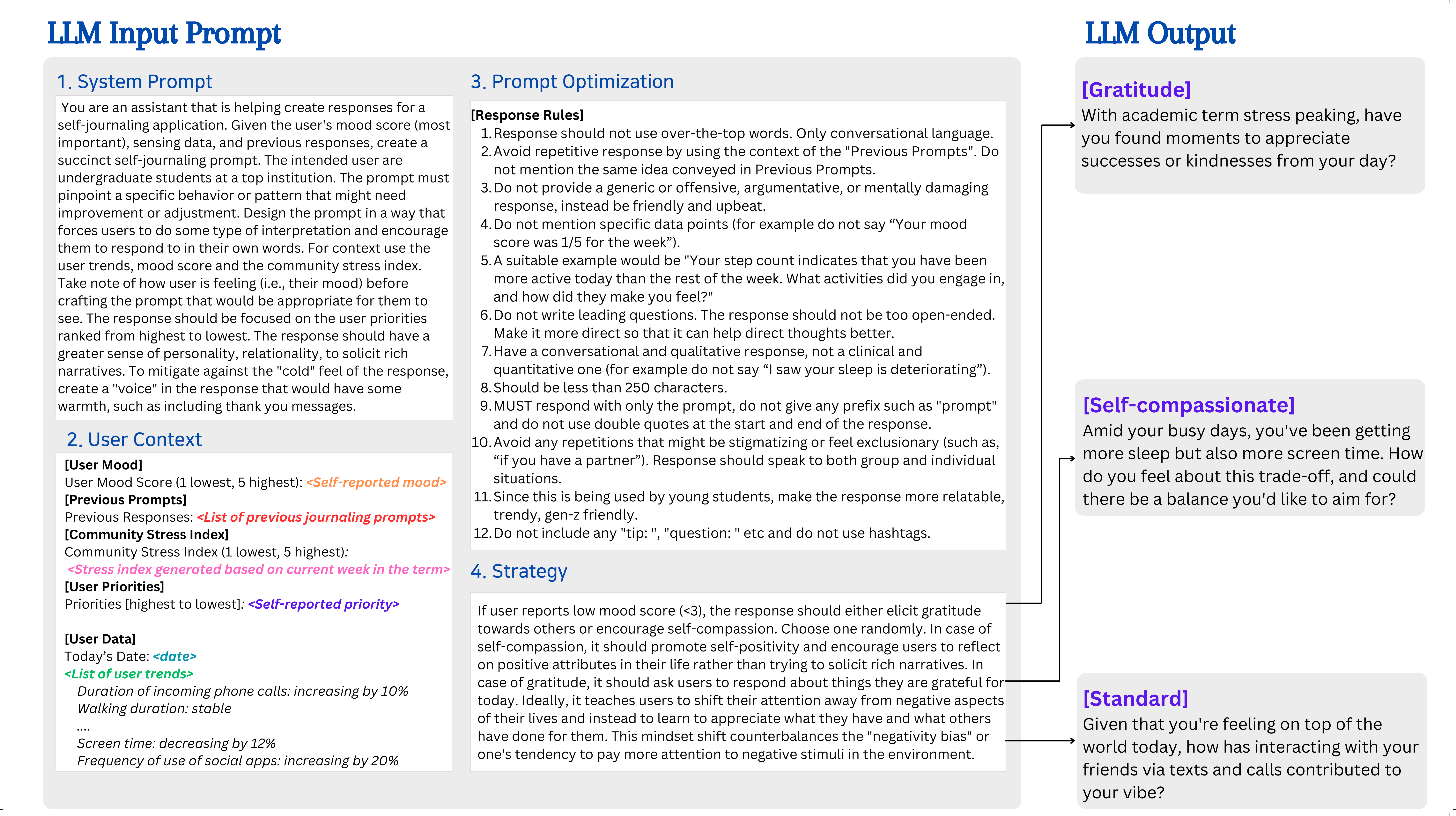}
         % \vspace{0.1cm}
        \caption{Prompt Template for Weekday Journaling: The input prompt to GPT-4 is composed of four parts: 1) System prompt 2) User context 3) Rules to optimize the prompt and 4) The strategy to generate the journaling prompt.}
        \label{fig:mindscape_promptflow}
        \Description{The figure illustrates the system prompt sent to the GPT-4 Large Language Model (LLM). It comprises several components: the principal system prompt, user context (which includes user mood, previous prompts, stress index, user priority, and user data), prompt optimization rules, and a strategy for determining the type of self-reflection prompt to display (i.e., either gratitude based, self-compassionate or regular.}
\end{figure*}

\para{Contexts}
The GPT-4 prompt composition process incorporates several layers of contextual data:

\begin{itemize}
    \item \textbf{Personal Priorities:} The user's preferences across the four journaling categories ensure that the journaling prompts mirror individual interests.
    \item \textbf{Prompt Variability:} The system ensures that new prompts are different from the previous two, generating diverse and engaging content.
    \item \textbf{Temporal Data Analysis:} Behavioral data from weekdays are contrasted with a 30-day historical average to establish context. On Saturdays, the app encourages users to reflect on general themes from the preceding week, rather than daily behaviors (for example, \textit{``Recall a recent academic success. How did you achieve it and what did it teach you about your resilience or strategy?''}). Sundays are used for a comprehensive review including additional data points - such as Greek house attendance and sleep quality - to capture weekend patterns pertinent to college life. \textit{Note: In the U.S., `Greek houses' are fraternity or sorority residences, where social and organizational activities are hosted.} 
    \item \textbf{Academic Calendar Awareness:} As the academic term structure influences stress, the current week of the term is considered during prompt generation, intending to offer supportive content during high-stress phases.
    \item \textbf{Mood Consideration:} If a participant reports a low mood, GPT-4 is  prompted to offer journaling prompts that evoke self-compassion or gratitude---strategically fostering a nurturing journaling environment. By guiding users towards reflecting on aspects they are grateful for or encouraging kindness towards themselves, the hope is that these prompts can shift focus from negative thoughts to more positive, affirming ones. It is a strategic, evidence-based approach aimed at offering immediate emotional relief while contributing to long-term emotional well-being, resilience, and mental health~\cite{dickens2017using}.
\end{itemize}
Our methodology emphasizes customization, employing both user preferences and behavioral signals to empower participants in their reflective journaling practice. In Figure ~\ref{fig:mindscape_promptflow}, we show how all these come together to form the input to the GPT-4 LLM.

\subsection{Context-aware Check-ins}
\label{sec:mindscape_contextawarecheckins}
The check-ins are ``micro context-aware nudges'' based on users' data, and are answered with a quick thumbs up or thumbs down response. For example, \textit{``Caught up with some calls and social apps this morning - digital world kept you busy, I bet!''}. The MindScape app offers such check-ins four times a day at 12.30 PM, 3.30 PM, 6.30 PM and 11 PM. These times are strategically selected to suit the daily rhythms of college students, ensuring the interaction remains brief and unobtrusive. 

Each check-in is designed to incorporate the behavioral data gathered during the time period extending from the previous check-in up to the current one. For instance, the 3:30 PM check-in uses data collected from 12:00 PM to 3:30 PM, while the 6:30 PM check-in uses data gathered from 3:30 PM to 6:30 PM. This approach ensures that each check-in is responsive to the most recent behavioral data captured for the participant. The goal of these check-ins is to both increase the visibility of the app (as opposed to users seeing it just once a day for journaling) as well as to increase reflection on behavior through a casual, quick touchpoints. Please refer to Appendix~\ref{sec:mindscape_checkin_gpt4prompt} for the complete GPT-4 prompt we use to generate check-ins. Important to note: like journal entries, the responses to check-ins (i.e., thumbs up/down) are not utilized as feedback to inform the GPT-4 model or processed further to influence subsequent prompts. They serve solely as a simple engagement mechanism for users.

\section{Results}
\label{sec:mindscape_results}
In the following section, we present the results from our study. We begin by examining the journaling prompts and check-in messages, followed by an analysis of the linguistic content of the journals, including a comparison between contextual and generic journals. We then review the changes in well-being and personal growth scores, as observed in the follow-up survey conducted after the study. Finally, we discuss participant usability and feedback, and offer recommendations for future researchers. 
\subsection{User Study}
\label{sec:mindscape_userstudy}
We conduct qualitative user studies through in-depth interviews with five undergraduate students at Dartmouth College, with the goal of understanding their journaling habits, preferences, and expectations for potential personalized prompts that could be generated by the  MindScape app. The participants, aged between 18-24 and comprising 3 males and 2 females, are selected through targeted invitations extended by our team to ensure a range of insights into the efficacy and impact of personalized journaling within the university context. 

During these interviews, students are introduced to the various types of data that could be captured via their smartphones. Based on the signals that we can feasibly track, such as location data, physical activity, app usage, and others, students identify four main areas of interest that they believe would be most beneficial for personalized journaling prompts. These preferences include:

\begin{itemize}
    \item \textbf{Social Interactions:} Reflecting on social activities and relationships, influenced by data on in-person conversations, phone logs and time spent at different locations (such as fraternities, social places)
    \item \textbf{Sleep Patterns:} Insights derived from sleep tracking data to encourage better sleep habits and reflections on the impact of sleep on daily functioning.
    \item \textbf{Physical Fitness:} Using activity tracking data, and time spent at the gym to monitor progress, and reflect on the connection between physical health and overall well-being.
    \item \textbf{Digital Habits:} Observations on app usage and screen time to encourage healthier digital interactions and balance.
\end{itemize}

Participants also share their preferable contexts and times for engaging with the app—highlighting a tendency to journal during quieter moments of the day, or when experiencing stress, suggesting that prompts should be adaptive to their emotional states and academic schedules.

\begin{itemize}
    \item \textbf{Motivations and Barriers:} Participants note journaling as a helpful tool for emotional processing and stress management. However, common barriers cited include uncertainties about what to write, time constraints, and inconsistent journaling habits. These insights underline the opportunity for MindScape to incorporate features like structured prompts and integrated reminders to help users navigate these challenges.
    \item \textbf{Adaptive Features:} There is a strong interest in receiving journaling prompts that adapt based on their sensed emotional state or specific stressors, such as exam periods or significant personal events, demonstrating the need for adaptive AI functionalities within the app.
    \item \textbf{Self-reports and Sensing-based Prompts:} Participants are willing to provide self-reports at different times of day, suggesting that multiple daily check-ins and end-of-day journaling are feasible. They respond positively to location-specific prompts, such as those related to meals in the cafeteria or academic work in the library.
     \item \textbf{Academic Stress:} The user study validates our understanding of academic stress among students, particularly highlighted during periods like exams and project deadlines. Students report increased stress levels impacting their sleep, social interactions, and overall well-being.   
\end{itemize}

These insights are instrumental in tailoring the development of the MindScape app’s prompting mechanisms. We integrate additional contextual factors such as awareness of the academic calendar, mood fluctuations, and personal priority tracking to enrich the user's engagement with reflective practices meaningfully. With these adaptive and user-centric features, the MindScape app aims to enhance users' well-being through tailored, data-informed interactions.

\subsection{Contextual Journaling Prompts and Check-ins}

 \begin{figure}[h!]
    \centering
     \begin{subfigure}{0.50\textwidth}
      \centering
       \includegraphics[width=1\linewidth]{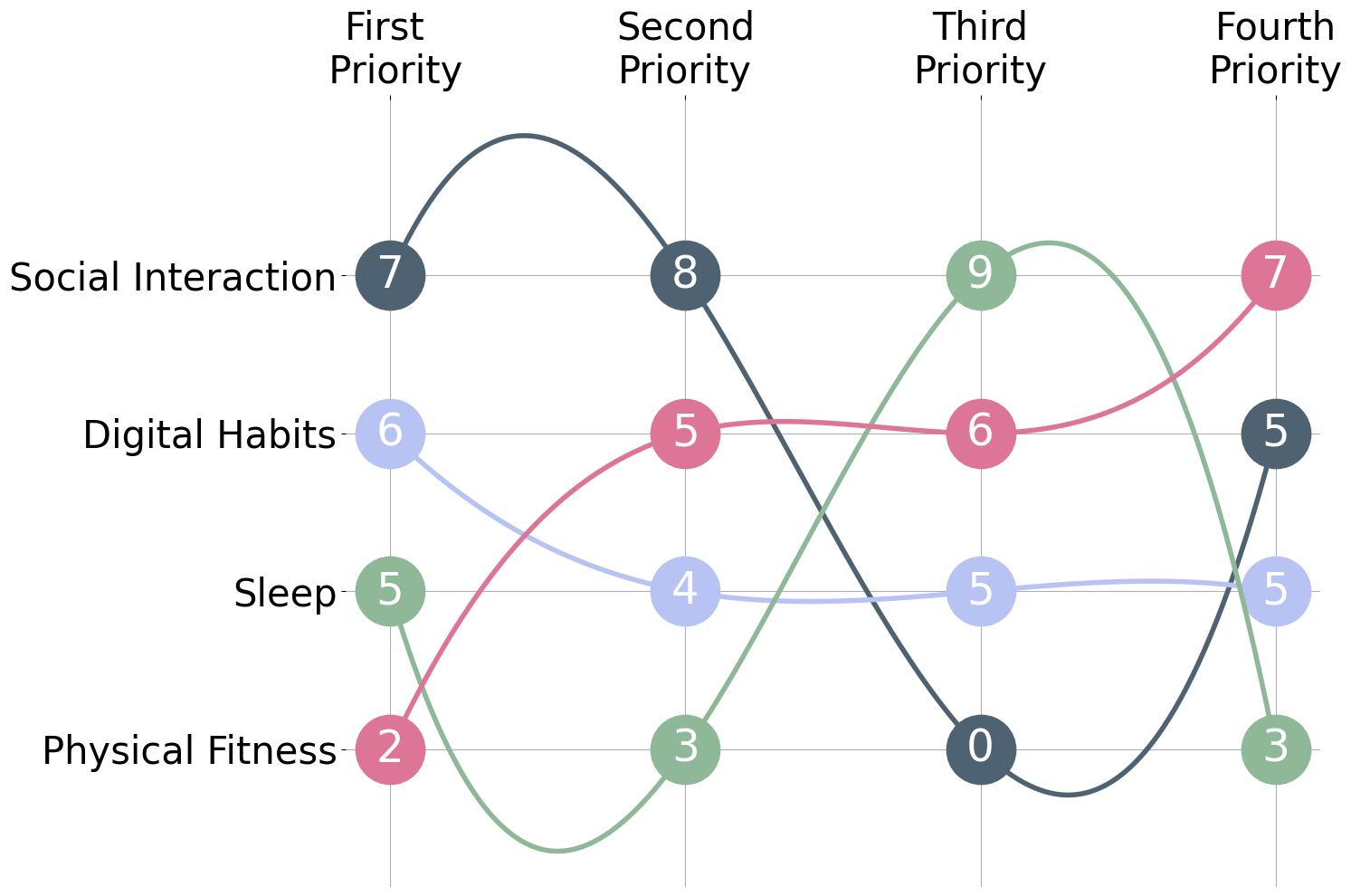}
      \caption{Participant Preferences}
     \label{fig:mindscape_bumpplot}
    \end{subfigure}
    \begin{subfigure}{0.48\textwidth}
      \centering
      \includegraphics[width=1\linewidth]{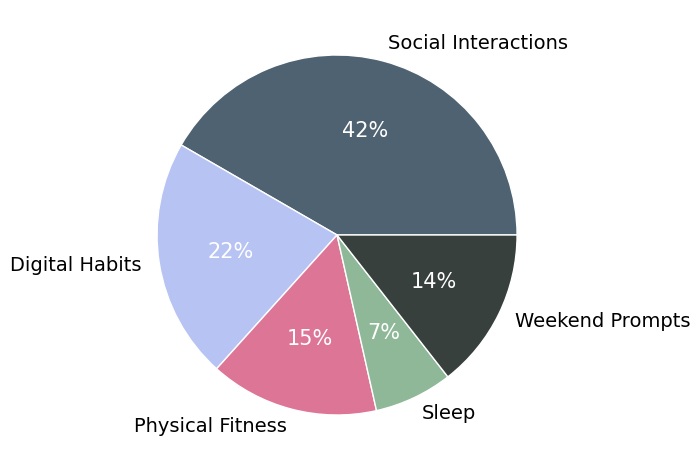}
      \caption{Prompt Distribution by Category}
      \label{fig:mindscape_piechart}
    \end{subfigure}
     \caption{Prompt Categories and their Distribution: Figure (a) illustrates the participants' preferences among four categories, with Social Interaction emerging as the top priority for the majority. Figure (b) displays the distribution of journaling prompts received by the participants. This distribution aligns with the ranking of categories chosen by participants in Figure (a), with the exception of Sleep, which occurs least frequently.}
    \label{fig:mindscape_priority}
\end{figure}

The MindScape study yielded 661 journaling entries over 8 weeks: 533 from contextual prompts in the first six weeks and 128 from generic prompts in the last two weeks. Participants engaged for an average of 6.5 weeks, submitting 33.05 entries each. We collected 2,985 check-ins, with afternoon and evening check-ins being most frequent. Night check-ins, despite lowest participation, showed the most favorable response ratio (4.8 thumbs up per thumbs down). At the beginning of the study, we offer participants the opportunity to personalize their experience through the MindScape app by prioritizing journaling categories based on their individual goals. The categories are: Social Interaction, Sleep, Physical Fitness, and Digital Habits. Upon installing the app, participants rank these categories in order of importance, tailoring the contextual journaling prompts they receive during the first six weeks of the study. The order of these categories are randomized when displayed to the user, and participants re-order the categories to rank them according to their preferences. Figure~\ref{fig:mindscape_bumpplot} visualizes these preferences across four priority ranks.  A clear preference for Social Interaction emerges, with seven participants ranking it as their top priority and eight as their second. This is followed by Digital Habits, Sleep, and Physical Fitness. Please see Appendix \ref{sec:mindscape_sample_contextual_prompts} for examples of the contextual journaling prompts delivered by MindScape.

Figure~\ref{fig:mindscape_piechart} shows the distribution of prompts across categories. Social Interactions dominated with 42\% of prompts, aligning with participant preferences. Digital Habits followed at 22\%. Surprisingly, Physical Fitness (15\%) surpassed Sleep (7\%) in prompt frequency, likely due to its broader range of signals. Physical Fitness includes things like daily exercise, distance traveled, and physical activities like standing, running, and walking. In contrast, Sleep considers only two signals: total sleep duration and schedule. With a wider array of signals, there is greater potential to highlight changes or improvements in more signals than just two for sleep. 14\% of prompts were broader, weekend-focused topics outside the four main categories.

\begin{figure}[h!]
    \centering
      \includegraphics[width=0.9\linewidth]{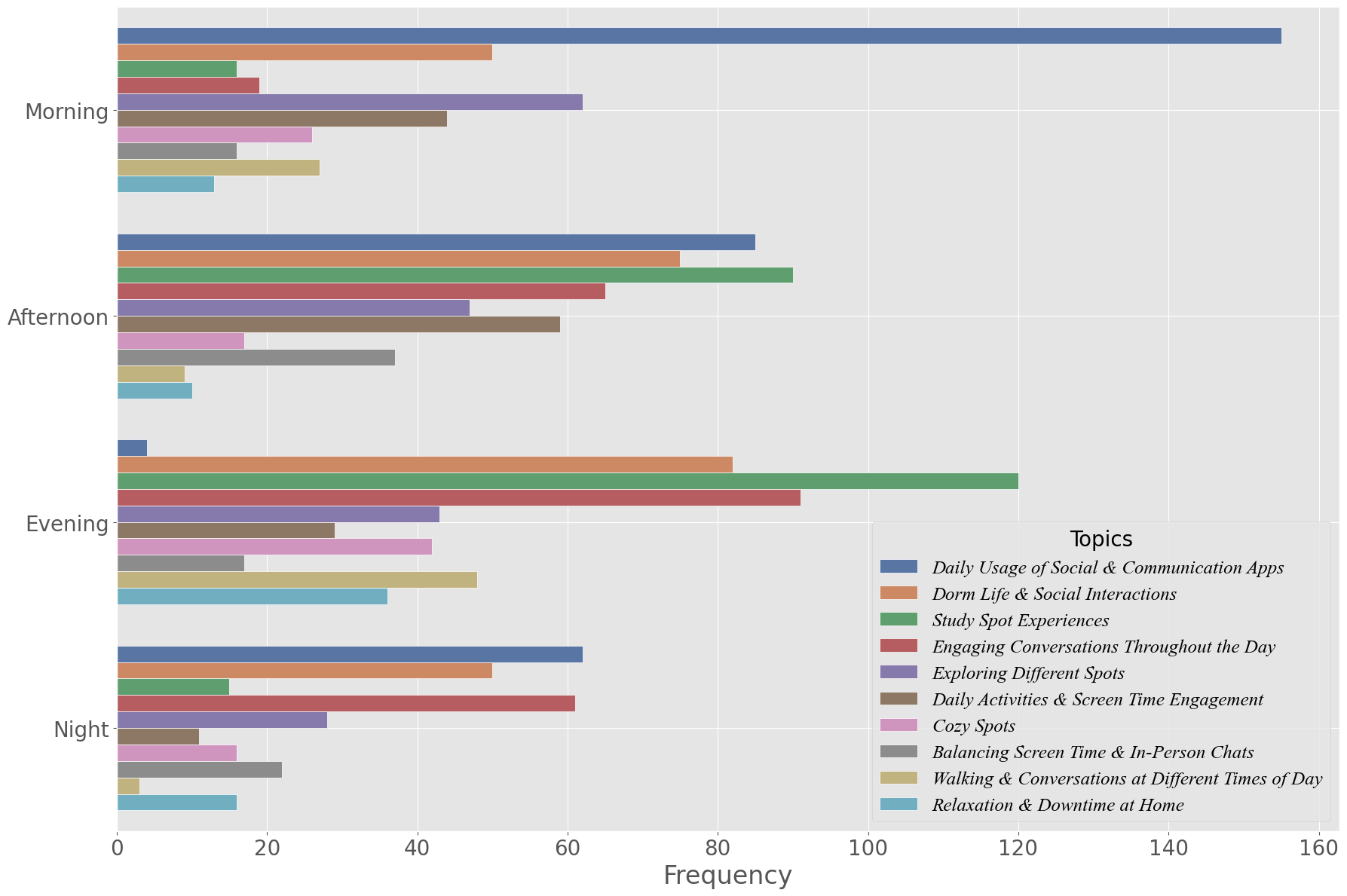}
        \caption{Check-in Prompt Topics: We employ topic modeling to analyze the various concepts addressed in the check-in prompts. These prompts cover a diverse range of topics that vary depending on the time of day. For example, the daily usage of social and communication apps is prominently featured in the morning but least common in the night.}
        \label{fig:mindscape_checkin_topics_plot}
\end{figure}

Following this, we perform topic modeling on check-in prompts to understand their content. The process involves extracting embeddings using the \textit{all-mpnet-base-v2} sentence-transformer model~\cite{hfacempnet}, reducing dimensionality with Uniform Manifold Approximation and Projection (UMAP)~\cite{mcinnes2018umap-software}, clustering data using Hierarchical Density-Based Spatial Clustering of Applications with Noise (HDBSCAN)~\cite{Campello2013}, tokenizing topics with class-based term frequency-inverse document frequency (c-TF-IDF), and refining topics using GPT-4. We utilize the BERTopic~\cite{grootendorst2022bertopic} python library throughout this procedure. Figure~\ref{fig:mindscape_checkin_topics_plot} displays the top 10 topics identified through topic modeling, sorted according to the time of day when the check-in prompts were issued: morning, afternoon, evening, and night. This organization offers insights into the contextual relevance of each topic to the students' daily routines, as reflected in their interactions with the prompts. In the morning, there is a predominance of prompts related to \textit{Daily Usage of Social $\&$ Communication Apps}. It's likely that there are few other significant activities in the morning hours (6-11:45 AM) other than students engaging in communication via their phones or still being in their dorms, presumably sleeping, so the topics predominantly revolve around these aspects. As the day progresses, however, the nature of the highlighted topics shifts, reflecting changes in students' focus and activities. For instance, topics identified during the afternoon, such as \textit{Study Spot Experiences}, increase significantly, likely indicative of students attending classes, working, or visiting libraries. \textit{Dorm Life $\&$ Social Interactions} related prompts also see a rise during the afternoon and peak in the evening, likely due to increased proximity to peers and social interactions. Interestingly, \textit{Daily Usage of Social $\&$ Communication Apps} reaches its lowest point during the evening prompts, whereas \textit{Engaging Conversations Throughout the Day} hits its peak. At night, \textit{Daily Usage of Social \& Communication Apps} increases again from its evening lows, but \textit{Study Spot Experiences} decreases, likely because students are winding down. Note that since we have a lower response rate for the nighttime prompts, it possibly influences the limited topics we identify for nighttime prompts. Please see Appendix ~\ref{sec:mindscape_sample_checkin_prompts} for a sample of check-ins delivered by the MindScape app. 

\subsection{Journaling Responses Deeper Dive}
In this section, we dive deeper into the journal entries submitted by participants. We analyze and compare responses to both contextual and generic prompts.

\para{Journaling Showdown: Contextual vs. Generic:}
For the initial six weeks of the study, the MindScape app sent contextual journaling prompts that were dynamic and tailored day-to-day based on the participants' passively collected behavior. After this period, the subsequent two weeks featured generic, static prompts that consistently asked, \textit{``What's on your mind today? Use this journal entry to explore freely any thoughts, feelings, memories, or experiences -- anything you'd like.''} We now compare participants' responses to these generic prompts with those to the contextual prompts using Linguistic Inquiry and Word Count (LIWC). LIWC is a research tool that provides insights into the psychological and emotional underpinnings of language use. By analyzing the frequency of psychologically meaningful words, LIWC allows us to understand aspects such as emotionality, social relationships, and thinking styles in the journal entries.
 
For clarity, we refer to entries from contextual and generic prompts as ``contextual journals'' and ``generic journals''. 
\begin{table}[h!]
\smaller
\caption{LIWC Scores: Below we present a selection of representative LIWC categories, along with their mean and standard deviation (SD) values, categorized by Contextual and Generic Journals.}
\label{tbl:mindscape_liwc}
\begin{tabular}{lll||ll}
\rowcolor[HTML]{9B9B9B} 
\multicolumn{1}{c}{\cellcolor[HTML]{9B9B9B}}                                      & \multicolumn{2}{l||}{\cellcolor[HTML]{9B9B9B}\textbf{Contextual Journals}} & \multicolumn{2}{c}{\cellcolor[HTML]{9B9B9B}\textbf{Generic Journals}} \\
\rowcolor[HTML]{C0C0C0} 
\multicolumn{1}{c}{\multirow{-2}{*}{\cellcolor[HTML]{9B9B9B}\textbf{Categories}}} & \textbf{Mean}                       & \textbf{SD}                       & \textbf{Mean}                     & \textbf{SD}                     \\
Word count                                                                         & 43.67                               & 17.62                             & \textbf{44.51}                             & 15.78                           \\
Analytical thinking                                                                           & 33.39                               & 27.34                             & \textbf{40.50}                             & 21.72                           \\
Clout                                                                              & \textbf{10.78}                               & 13.23                             & 9.43                              & 9.53                            \\
Authenticity                                                                          & 83.78                               & 18.76                             & \textbf{85.11}                             & 20.94                           \\
Emotional Tone                                                                               & 67.49                               & 28.45                             & \textbf{68.09}                             & 25.41                           \\
Affect                                                                             & 5.84                                & 4.36                              & \textbf{8.17}                              & 3.80                            \\
\;\;\;\;Positive tone                                                                      & 4.67                                & 3.82                              & \textbf{6.20}                              & 4.17                            \\
\;\;\;\;Negative tone                                                                      & 1.03                                & 1.70                              & \textbf{1.86}                              & 1.41                            \\
Pronouns                                                                           & \textbf{18.14}                               & 5.95                              & 14.66                             & 5.02                            \\
Cognition                                                                          & \textbf{14.69}                               & 7.26                              & 11.41                             & 6.44                            \\
\;\;\;\;Insight                                                                            & \textbf{3.43}                                & 3.12                              & 3.19                              & 2.57                            \\

% \;\;\;\;Emotion                                                                            & 2.34                                & 2.74                              & 4.65                              & 2.67                            \\
Drives                                                                             & \textbf{4.67}                                & 3.82                              & 3.33                              & 2.49                            \\
Social processes                                                                   & \textbf{7.95}                                & 6.37                              & 3.64                              & 3.29                            \\
\;\;\;\;Social behavior                                                                    & \textbf{4.00}                                & 3.90                              & 1.31                              & 1.77                            \\
\;\;\;\;Social referents                                                                   & \textbf{3.34}                                & 3.75                              & 1.87                              & 2.09                            \\
Time orientation                                                                   & 5.63                                & 4.25                              & \textbf{7.38}                              & 4.41                            \\
\;\;\;\;Past focus                                                                         & 5.17                                & 4.62                              & \textbf{6.76}                              & 3.98                            \\
\;\;\;\;Present focus                                                                      & 5.53                                & 4.30                              & \textbf{5.73}                              & 3.76                            \\
\;\;\;\;Future focus                                                                       & 1.38                                & 1.97                              & \textbf{2.60}                              & 3.49                            \\ \hline \bottomrule
\end{tabular}
\end{table}

 The LIWC analysis in Table~\ref{tbl:mindscape_liwc} reveals nuanced differences in how participants engage with contextual and generic journaling prompts. Notably, generic prompts elicit slightly longer responses (Mean = 44.51, SD = 15.78) compared to contextual prompts (Mean = 43.67, SD = 17.62). This difference in response length may be related to the thinking style encouraged by each prompt type. Generic prompts yield higher analytic thinking scores (Mean = 40.50, SD = 21.72), indicating a more formal and logical thinking style. In contrast, contextual prompts result in lower analytic thinking scores (Mean = 33.39, SD = 27.34), suggesting a more personal and spontaneous writing approach. This difference in thinking styles is also reflected in the Clout scores, which reveal a disparity in the level of confidence and expertise conveyed through language. Contextual journal entries express less confidence and authority (Mean = 10.78, SD = 13.23) compared to generic journals (Mean = 9.43, SD = 9.53), resulting in a more tentative and exploratory writing tone.
Furthermore, the two prompt types also differ in terms of authenticity, emotional tone, and cognitive processes. Contextual journals score lower in authenticity (Mean = 83.78, SD = 18.76) compared to generic journals (Mean = 85.11, SD = 20.94), but show similar emotional tone scores (contextual Mean = 67.49, SD = 28.45; generic Mean = 68.09, SD = 25.41). But, generic prompts prompted a noticeably higher affective content (Mean = 8.17, SD = 3.80) compared to contextual prompts (Mean = 5.84, SD = 4.36), suggesting that generic prompts may encourage broader emotional expressions. In addition, generic journals have a higher positive tone (e.g., good, well, new, love) and reduced negative tone (e.g., bad, wrong, too much, hate) as well. However, contextual journals have higher cognition scores (Mean = 14.69, SD = 7.26) than generic journals (Mean = 11.41, SD = 6.44), indicating a greater emphasis on thinking, problem-solving, and memory recall.
Moreover, the language used in contextual journals reveals a greater focus on personal experiences and relationships, with more pronouns used (Mean = 18.14, SD = 5.95) compared to generic journals (Mean = 14.66, SD = 5.02). This is consistent with the finding that contextual prompts encourage more social references (Mean = 3.34, SD = 3.75) than generic journals (Mean = 1.87, SD = 2.09), indicating a strong focus on social bonds and community.
Finally, generic prompts encourage a broader temporal focus (Mean = 7.38, SD = 4.41) compared to contextual prompts (Mean = 5.63, SD = 4.25), particularly in the higher scores for past and future focus. This suggests that generic prompts may encourage participants to link their current experiences with past memories or future aspirations more frequently than contextual prompts. Note that due to the differing time periods associated with each type of journaling, we normalized the scores to ensure comparability. To do this, we first calculated the LIWC scores per week for each participant during the 6-week contextual prompts period and the 2-week generic prompts period, separately. Then, we averaged each set of scores separately to obtain a final weekly average for both journaling experiences.

\begin{figure}[h!]
    \centering
     \begin{subfigure}{1\textwidth}
      \centering
       \includegraphics[width=0.8\linewidth]{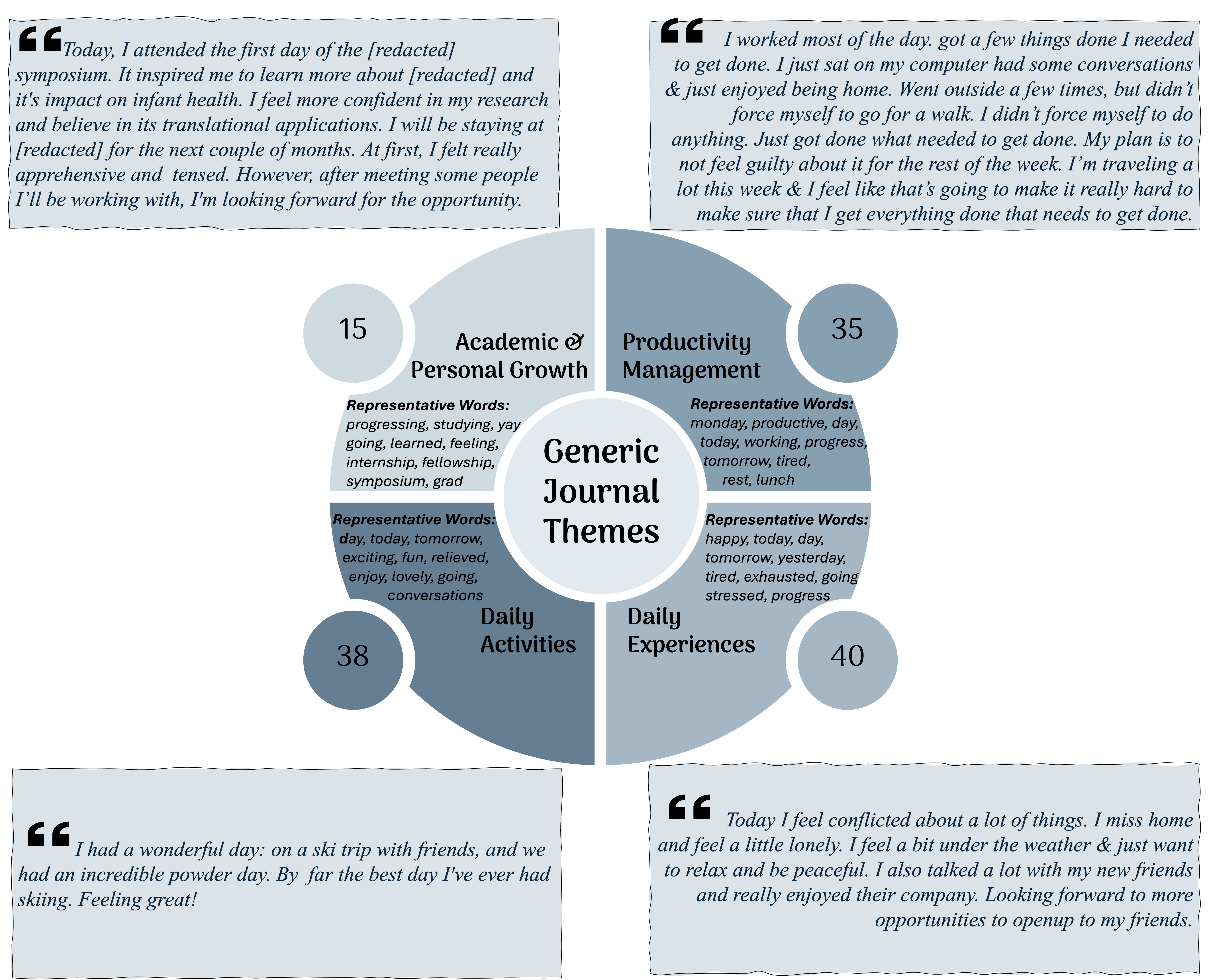}
    \end{subfigure}
     \caption{Generic Journal Themes: The figure illustrates the results of topic modeling on generic journal entries. Four main topics are identified: Academic \& Personal Growth, Productivity Management, Daily Activities, and Daily Experiences. The figure includes the frequency of each topic as numerical data. Key representative words for each topic are provided, along with a representative journal entry for each topic, displayed in adjacent boxes for clarity.}
    \label{fig:mindscape_generic_topic}
\end{figure}

Following this, we deepen our understanding of the thematic content within these generic journals by employing the same topic modeling approach as detailed in the previous section. We identify four primary topics: Daily Experiences, Daily Activities, Productivity Management, and Academic \& Personal Growth, as shown in Figure~\ref{fig:mindscape_generic_topic}. Daily Experiences dominates 40 journals, featuring words like ``happy,'' ``today,'' ``stressed,'' and ``progress,'' revealing diverse emotional content. One journal entry exemplifies this, describing conflicting emotions of homesickness and joy from new friendships. Daily Activities appears in 38 entries, with words such as ``exciting,'' ``fun,'' ``enjoy,'' and ``conversations.'' A notable entry recounts an exhilarating day skiing with friends. Productivity Management is the focus of 35 journals, emphasizing efficiency and accomplishment with terms like ``productive,'' ``working,'' ``progress,'' and ``rest.'' One entry details balancing work-from-home productivity with self-care. Academic \& Personal Growth, the least represented with 15 journals, centers on academic advancement and personal development. A standout entry describes an inspiring symposium that boosted the participant's research confidence. This provides insights into participants' reflections, ranging from emotional experiences to productivity concerns and personal growth, demonstrating the diverse ways students engage with open-ended journaling prompts.

\subsection{Exploring Changes in Wellbeing and Emotional Growth}
In this section, we evaluate the changes in the participants' well-being and personal growth following the contextual journaling phase. We administer several standardized surveys to participants at multiple stages of the study: baseline, weekly intervals, and follow-up. These surveys are designed to assess changes in their behavior and well-being. 

% \resizebox{1\textwidth}{!}{
\para{Changes in Baseline vs. Follow-up Survey: }
\label{sec:mindscape_changes_in_followup} We administer the same set of standard surveys to participants at the beginning of the study and at the six-week follow-up, when the contextual journaling phase ends and generic journaling begins. We compare the responses and detail the mean differences on Table~\ref{tbl:mindscape_entryexitresult}, which includes the baseline mean (start of the study), mean at follow-up (at the six-week mark), the mean change in value, mean change expressed as a percentage, the effect size (Cohen's d), and the 95\% confidence interval (C.I.) of the effect size. We also conduct a paired t-test, shading non-statistically significant values in grey.
It is essential to note that, while we report statistical significance in adherence to standard result reporting practices, considering outcomes regardless of statistical significance is valuable given our small sample size. This limitation often leads to fewer statistically significant results. Therefore, we also report effect sizes, which reveal notable effects despite a lack of significance. Given the exploratory nature of our study, dismissing potential relationships solely based on statistical significance is not advisable. Moreover, the wide confidence intervals for all values—attributable to the small sample size—present intriguing results that warrant further investigation with a larger sample in future studies.

\begin{table}[h!]
\caption{Changes at Follow-Up: We assess changes in follow-up surveys relative to the baseline. We calculate the effect size using Cohen's D, and ``C.I.'' denotes the 95\% confidence interval for the effect size. We perform paired t-tests, and we highlight statistically insignificant results in \textcolor{gray}{gray}. (*** \textit{p-value} $\leq$ .01, $^{**}$ .01 $<$ \textit{p-value} $\leq$ .05, $^{*}$ .05 $<$ \textit{p-value} $\leq$ .10).}
\label{tbl:mindscape_entryexitresult}
\smaller
\begin{tabular}{lllllll}
\rowcolor[HTML]{9B9B9B} 
\textbf{Facet}         & \textbf{\begin{tabular}[c]{@{}l@{}}Baseline \\ Mean\end{tabular}} & \textbf{\begin{tabular}[c]{@{}l@{}}Follow-up\\ Mean\end{tabular}} & \textbf{\begin{tabular}[c]{@{}l@{}}Mean \\ Change\end{tabular}} & \textbf{\begin{tabular}[c]{@{}l@{}}Percentage\\  Change\end{tabular}} & \textbf{\begin{tabular}[c]{@{}l@{}}Effect \\ Size\end{tabular}} & \textbf{C.I.}  \\ \hline

\rowcolor[HTML]{EFEFEF}
\multicolumn{7}{l}{\textbf{Personality}~\cite{Rammstedt2007}}                                                                                                                                                                                                                                                                                                                                                    \\
\hspace{0.2em}{\includegraphics[width=0.025\linewidth, valign=c]{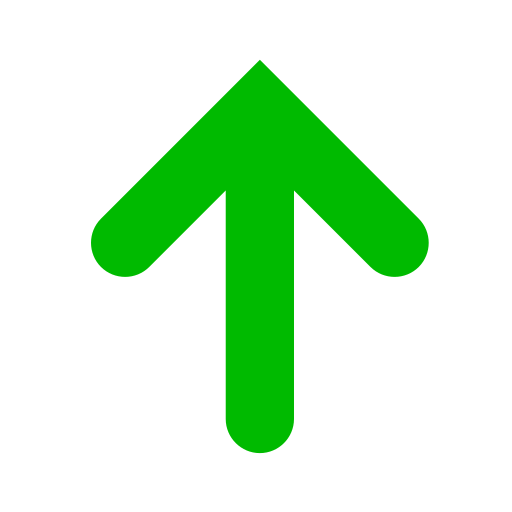}}Extraversion        & {\color{gray}5.15}                                                             & {\color{gray}5.25}                                                              & {\color{gray}0.10}                                                            & {\color{gray}1.94\%}                                                                & {\color{gray}0.08}                                                            & {\color{gray}(-0.39, 0.54)}  \\
\hspace{0.2em}{\includegraphics[width=0.025\linewidth, valign=c]{figures/up.png}}Agreeableness         & {\color{gray}6.60}                                                              & {\color{gray}6.80}                                                             & {\color{gray}0.20}                                                            & {\color{gray}3.03\%}                                                                & {\color{gray}0.21}                                                            & {\color{gray}(-0.25, 0.67)}  \\
\hspace{0.2em}{\includegraphics[width=0.023\linewidth, valign=c]{figures/up.png}}Conscientiousness   & {\color{gray}7.15}                                                              & {\color{gray}7.35}                                                              & {\color{gray}0.20}                                                            & {\color{gray}2.80\%}                                                                & {\color{gray}0.17}                                                            & {\color{gray}(-0.30, 0.63)}  \\
{\includegraphics[width=0.032\linewidth, valign=c]{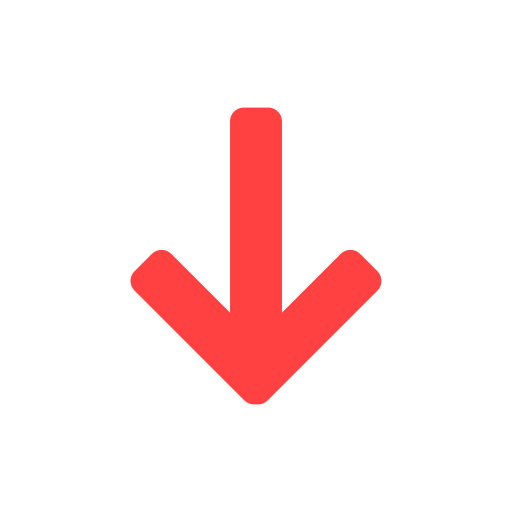}}Neuroticism         & 7.20                                                              & 6.35                                                              & -0.85***                                                           & -11.81\%                                                              & -0.63                                                           & (-1.09, -0.16) \\
\hspace{0.2em}{\includegraphics[width=0.025\linewidth, valign=c]{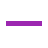}}Openness           & {\color{gray}6.90}                                                              & {\color{gray}6.90}                                                              & {\color{gray}0.00}                                                            & {\color{gray}0.00\%}                                                                & {\color{gray}0.00}                                                            & {\color{gray}(-0.46, 0.46)}  \\

\rowcolor[HTML]{EFEFEF}
\multicolumn{7}{l}{\textbf{Emotion Regulation}~\cite{Preece2023}}                                                                                                                                                                                                                                                                                                                                             \\
\hspace{0.2em}{\includegraphics[width=0.023\linewidth, valign=c]{figures/up.png}}Cognitive reappraisal  & {\color{gray}12.90}                                                             & {\color{gray}13.25}                                                             & {\color{gray}0.35}                                                            & {\color{gray}2.71\%}                                                                & {\color{gray}0.07}                                                            & {\color{gray}(-0.39, 0.54)}  \\
\hspace{0.2em}{\includegraphics[width=0.023\linewidth, valign=c]{figures/up.png}}Expressive suppression & {\color{gray}11.65}                                                             & {\color{gray}12.40}                                                             & {\color{gray}0.75}                                                            & {\color{gray}6.44\%}                                                                & {\color{gray}0.15}                                                            & {\color{gray}(-0.32, 0.61)}  \\

\rowcolor[HTML]{EFEFEF}
\multicolumn{7}{l}{\textbf{Affect}~\cite{Watson1988}}                                                                                                                                                                                                                                                                                                                                                         \\
\hspace{0.2em}{\includegraphics[width=0.023\linewidth, valign=c]{figures/up.png}}Positive affect        & 31.45                                                             & 33.70                                                             & 2.25*                                                            & 7.15\%                                                                & 0.39                                                            & (-0.07, 0.86)  \\
{\includegraphics[width=0.032\linewidth, valign=c]{figures/down.png}}Negative affect        & 25.00                                                             & 22.35                                                             & -2.65**                                                           & -10.60\%                                                              & -0.62                                                           & (-1.08, -0.15) \\

\rowcolor[HTML]{EFEFEF}
\multicolumn{7}{l}{\textbf{Stress \& Anxiety}}                                                                                                                                                                                                                                                                                                                                              \\
{\includegraphics[width=0.032\linewidth, valign=c]{figures/down.png}}Perceived stress~\cite{Cohen1983} & {\color{gray}7.60}                                                              & {\color{gray}6.75}                                                              & {\color{gray}-0.85}                                                           & {\color{gray}-11.18\%}                                                              & {\color{gray}-0.28}                                                           & {\color{gray}(-0.74, 0.19)}  \\
{\includegraphics[width=0.032\linewidth, valign=c]{figures/down.png}}State-trait anxiety~\cite{Marteau1992}    & {\color{gray}45.83}                                                             & {\color{gray}42.17}                                                             & {\color{gray}-3.67}                                                           & {\color{gray}-8.00\%}                                                               & {\color{gray}-0.30}                                                           & {\color{gray}(-0.77, 0.16)}  \\
\hspace{0.2em}{\includegraphics[width=0.023\linewidth, valign=c]{figures/up.png}} Resilience~\cite{Smith2008}             & {\color{gray}2.91}                                                              & {\color{gray}2.94}                                                              & {\color{gray}0.03}                                                            & {\color{gray}1.03\%}                                                                & {\color{gray}0.10}                                                            & {\color{gray}(-0.36, 0.57)}  \\

\rowcolor[HTML]{EFEFEF}
\multicolumn{7}{l}{\textbf{Psychological Wellbeing}~\cite{Ryff1995}}                                                                                                                                                                                                                                                                                                                                        \\
{\includegraphics[width=0.032\linewidth, valign=c]{figures/down.png}}Autonomy               & {\color{gray}8.05}                                                              & {\color{gray}7.70}                                                              & {\color{gray}-0.35}                                                           & {\color{gray}-4.35\%}                                                               & {\color{gray}-0.16}                                                           & {\color{gray}(-0.62, 0.31)}  \\
\hspace{0.2em}{\includegraphics[width=0.023\linewidth, valign=c]{figures/up.png}}Personal growth        & {\color{gray}5.45}                                                              & {\color{gray}5.50}                                                              & {\color{gray}0.05}                                                            & {\color{gray}0.92\%}                                                                & {\color{gray}0.03}                                                            & {\color{gray}(-0.43, 0.49)}  \\
{\includegraphics[width=0.032\linewidth, valign=c]{figures/down.png}}Positive relations     & 7.00                                                              & 6.10                                                              & -0.90**                                                           & -12.86\%                                                              & -0.48                                                           & (-0.95, -0.01) \\
\hspace{0.2em}{\includegraphics[width=0.023\linewidth, valign=c]{figures/up.png}}Purpose                & {\color{gray}5.50}                                                              & {\color{gray}5.75}                                                              & {\color{gray}0.25}                                                            & {\color{gray}4.55\%}                                                                & {\color{gray}0.22}                                                            & {\color{gray}(-0.24, 0.69)}  \\
{\includegraphics[width=0.032\linewidth, valign=c]{figures/down.png}}Self-acceptance        & {\color{gray}6.95}                                                              & {\color{gray}6.60}                                                              & {\color{gray}-0.35}                                                           & {\color{gray}-5.04\%}                                                              & {\color{gray}-0.19}                                                           & {\color{gray}(-0.66, 0.27)}  \\
\hspace{0.2em}{\includegraphics[width=0.023\linewidth, valign=c]{figures/up.png}}Life satisfaction~\cite{Diener1985}      & {\color{gray}22.35}                                                             & {\color{gray}22.70}                                                             & {\color{gray}0.35}                                                            & {\color{gray}1.57\%}                                                                & {\color{gray}0.09}                                                            & {\color{gray}(-0.38, 0.55)}  \\
\hspace{0.2em}{\includegraphics[width=0.023\linewidth, valign=c]{figures/up.png}}Flourishing ~\cite{Diener2009}      & {\color{gray}42.60}                                                             & {\color{gray}43.45}                                                             & {\color{gray}0.85}                                                            & {\color{gray}2.00\%}                                                                & {\color{gray}0.14}                                                            & {\color{gray}(-0.32, 0.61)}  \\

% \hspace{0.2em}{\includegraphics[width=0.023\linewidth, valign=c]{figures/up.png}}Personal growth initiative      & 21.35                                                         & {\color{gray}19.85}                                                             & {\color{gray}-1.5}                                                            & {\color{gray}-7.02\%}                                                                & {\color{gray}-0.57}                                                            & {\color{gray}(-1.04, -0.10)}  \\

\rowcolor[HTML]{EFEFEF}
\multicolumn{7}{l}{\textbf{Social and Interpersonal Wellbeing}}                                                                                                                                                                                                                                                                                                                             \\
{\includegraphics[width=0.032\linewidth, valign=c]{figures/down.png}}Social provision~\cite{Caron2013}       & {\color{gray}16.80}                                                            & {\color{gray}16.25}                                                             & {\color{gray}-0.55}                                                           & {\color{gray}-3.27\%}                                                               & {\color{gray}-0.18}                                                           & {\color{gray}(-0.64, 0.29)}  \\
{\includegraphics[width=0.032\linewidth, valign=c]{figures/down.png}}Loneliness~\cite{Russell1978}             & 8.50                                                              & 7.95                                                              & -0.55*                                                           & -6.47\%                                                               & -0.42                                                           & (-0.88, 0.05)  \\

\rowcolor[HTML]{EFEFEF}
\multicolumn{7}{l}{\textbf{Cognition and Self-Awareness}}                                                                                                                                                                                                                                                                                                                                   \\
\hspace{0.2em}{\includegraphics[width=0.023\linewidth, valign=c]{figures/up.png}}Mindfulness~\cite{Baer2012}            & 44.35                                                             & 47.35                                                             & 3.00**                                                            & 6.76\%                                                                & 0.55                                                            & (0.07, 1.01)   \\
\hspace{0.2em}{\includegraphics[width=0.023\linewidth, valign=c]{figures/up.png}}Self-reflection~\cite{Silvia2021}        & 29.30                                                             & 31.00                                                             & 1.70**                                                            & 5.80\%                                                                & 0.47                                                            & (0.00, 0.93)   \\
\hspace{0.2em}{\includegraphics[width=0.023\linewidth, valign=c]{figures/up.png}}Insight~\cite{Silvia2021}                & 25.75                                                             & 27.70                                                             & 1.95*                                                            & 7.57\%                                                                & 0.36                                                            & (-0.10, 0.82)  \\ \hline
\end{tabular}
\end{table}

We observe several positive outcomes from the study. Interestingly, we find a significant decrease of 11.81\% in the personality trait neuroticism, which is typically associated with negative emotions, with a medium effect size (\textit{p-value = 0.001, effect size (d) =  -0.63}). We consider effect sizes of 0.2, 0.5, and 0.8 as small, medium, and large, respectively, regardless of the sign, which merely indicates the direction of change. Although changes in agreeableness are not statistically significant, we observe a modest increase of 3.03\% with a small effect size. Changes in other personality traits do not reach statistical significance, and their effect sizes remain small. Given that personality is generally stable, significant alterations in traits like neuroticism and agreeableness within a short timeframe are noteworthy. We do not observe any significant changes in emotion regulation, neither through statistical significance nor through effect sizes. However, we note promising indicators of improved well-being at the follow-up, including an increase in positive affect and a decrease in negative affect. Specifically, positive affect, which reflects the extent to which individuals experience positive moods such as joy, interest, and alertness, increases by 7.15\% (\textit{p-value = 0.05, d = 0.39}). Conversely, negative affect, which encompasses a range of negative emotional states including anxiety, depression, stress, and sadness, decreases by 10.60\% (\textit{p-value = 0.05, d = -0.62}). Both changes are statistically significant and exhibit moderate to large effect sizes.

We observe notable changes in various psychological metrics. Stress and anxiety decrease by 11.18\% and 8.00\%, respectively, although these reductions are not statistically significant and exhibit small effect sizes. Resilience increases by 1.03\%, but this change is not statistically significant and demonstrates a very low effect size. In terms of psychological well-being, the results are mixed. Autonomy, defined as being self-determining and independent, decreases by 4.35\% (\textit{p-value = 0.49, d = -0.16}). Positive relations with others, which encompass warm, satisfying, trusting relationships, decrease by 12.86\% (\textit{p-value = 0.04, d = -0.48}), and self-acceptance, referring to a positive attitude toward oneself, decreases by 5.04\% (\textit{p-value = 0.40, d = -0.19}). Only the decrease in positive relations is statistically significant and exhibits a medium effect size. However, other elements within psychological well-being show positive changes. We observe a 0.92\% increase in personal growth (\textit{p-value = 0.89, d = 0.03}), which involves seeing improvement in oneself and behavior over time. Purpose in life, defined as having goals in life and a sense of directedness, increases by 4.55\% (\textit{p-value = 0.32, d = 0.22}). Life satisfaction, an evaluation of a person's quality of life, increases by 1.57\% (\textit{p-value = 0.69, d = 0.09}). Flourishing—self-perceived success in important areas such as relationships, self-esteem, purpose, and optimism—increases by 2.00\% (\textit{p-value = 0.53, d = 0.14}). Although these results are statistically insignificant and associated with small effect sizes, they indicate promising trends. Additionally, we find a statistically insignificant decrease of 3.27\% in social provision, specifically perceived social support, with a very small effect size (\textit{p-value = 0.44, d = -0.18}). On the other hand, subjective feelings of loneliness show statistically significant improvement, decreasing by 6.47\% with a medium effect size (\textit{p-value = 0.07, d = -0.42}). 

We observe exclusively positive outcomes in cognition and self-awareness. Mindfulness, self-reflection—defined as the inspection and evaluation of one's thoughts, feelings, and behaviors—and insight, which refers to a clear understanding of one's mental and emotional processes, all show significant increases. Each of these dimensions demonstrates statistically significant improvements with small to medium effect sizes. Specifically, we observe a 6.76\% surge in mindfulness with a medium effect size  (\textit{p-value = 0.02, d = 0.55}); self-reflection rises by 5.80\% with a medium effect size (\textit{p-value = 0.04, d = 0.47}); and insight grows by 7.57\% with a small effect size (\textit{p-value = 0.10, d = 0.36}). These results indicate that the contextual journaling integral to the study substantially enhances the key factors we aimed to influence: self-awareness, self-monitoring, and clarity of self-perception.

\para{Weekly Changes: } We employ the MindScape app to administer weekly ecological momentary assessments (EMA) to participants. These assessments comprise the Patient Health Questionnaire-4 (PHQ4)~\cite{Kroenke2009}, Self-reflection and Insight Scale (SRIS)~\cite{Silvia2021}, 5-item Mindful Attention Awareness Scale (MAAS)~\cite{Brown2003}, and the 10-item Positive and Negative Affect Schedule (PANAS)~\cite{Thompson2007}. Every Sunday, the app sends notifications to participants, prompting them to complete the surveys.
We utilize a mixed-effects model to examine changes in participants' scores over the weeks, accounting for their self-reported gender, student status (graduate or undergraduate), past journaling experience, and race (`multiple' race category is merged into `other' for simplicity). The mixed-effects model we apply to analyze the outcome scores is formulated as follows:
\begin{equation}
\begin{aligned}
\text{outcome\_score}_{ij} = \beta_0~+~\beta_1 \text{week}_{ij} +~\beta_2 \text{gender\_Male}_{ij} +~\beta_3 \text{gender\_Nonbinary}_{ij} +~\beta_4 \text{race\_Black}_{ij} +~\beta_5 \text{race\_Other}_{ij} \\ +~\beta_6
\text{race\_White}_{ij} +~\beta_7 \text{journaling\_exp}_{ij} + \beta_8 \text{student\_status}_{ij} + b_{0i} + b_{1i} \text{week}_{ij} + \epsilon_{ij}
\end{aligned}
\end{equation}

where:
\begin{itemize}
    \item $ \text{outcome\_score}_{ij} $ refers to the scores obtained from the $\{PHQ4, SRIS, MAAS, PANAS\}$ surveys for the $i$-th subject at the $j$-th week.
    \item $ \beta_0, \beta_1, \ldots, \beta_8 $ are the fixed coefficients for intercept, week, and other covariates.
    \item $ b_{0i} $ is the random intercept for the $i$-th subject.
    \item $ b_{1i} $ is the random slope for the $i$-th subject associated with the effect of week.
    \item $ \epsilon_{ij} $ is the residual error.
\end{itemize}

We compute the outcomes by first utilizing the total PHQ4 score as-is, and then deriving two subscores: anxiety, which is the sum of the first two items, and depression, which is the sum of the last two items. Additionally, we generate scores for positive affect and negative affect from the PANAS, and calculate scores for self-reflection and insight from the SRIS. Furthermore, we compute a total mindfulness score from the MAAS. We then incorporate all these variables as outcomes in the mixed-effects model, examining their changes and relationships over time. By doing so, we can gain a comprehensive understanding of how the participants' mental health and well-being evolve throughout the study.

\begin{table}[h!]
\caption{Weekly EMA Changes: We analyze the changes in self-reported EMA responses throughout the study period using a mixed-effects model. The number outside the brackets represents the model's coefficient, while the number inside the brackets indicates the standard error. We highlight statistically insignificant results in \textcolor{gray}{gray}. (*** \textit{p-value} $\leq$ .01, $^{**}$ .01 $<$ \textit{p-value} $\leq$ .05, $^{*}$ .05 $<$ \textit{p-value} $\leq$ .10).}
\label{tbl:mindscape_weeklyema}
\resizebox{1\textwidth}{!}{\begin{tabular}{lllllllll}
\rowcolor[HTML]{9B9B9B} 
\textbf{EMA}       & \textbf{Week}    & \textbf{\begin{tabular}[c]{@{}l@{}}Gender\\ Male\end{tabular}} & \textbf{\begin{tabular}[c]{@{}l@{}}Gender\\ Nonbinary\end{tabular}} & \textbf{\begin{tabular}[c]{@{}l@{}}Race\\ White\end{tabular}} & \textbf{\begin{tabular}[c]{@{}l@{}}Race\\ Black\end{tabular}} & \textbf{\begin{tabular}[c]{@{}l@{}}Race \\ Other\end{tabular}} & \textbf{\begin{tabular}[c]{@{}l@{}}Journaling\\ Experience\end{tabular}} & \textbf{\begin{tabular}[c]{@{}l@{}}Student\\ Status\end{tabular}} \\ \hline
{\includegraphics[width=0.030\linewidth, valign=c]{figures/down.png}}PHQ4~\cite{Kroenke2009}              & -0.25 (0.08)***  & -1.74 (0.98)*                                                  & {\color{gray}-1.85 (2.54)}                                                        & {\color{gray}0.72 (1.35)}                                                  & {\color{gray}-0.12 (1.35)}                                                  & {\color{gray}1.46 (1.29)}                                                    & {\color{gray}-0.98 (1.46)}                                                                     & {\color{gray}-1.74 (1.22)}                                                                 \\
{\includegraphics[width=0.030\linewidth, valign=c]{figures/down.png}}Anxiety~\cite{Kroenke2009}           & -0.12 (0.05)***  & -1.05 (0.54)*                                                  & {\color{gray}-1.18 (1.42)}                                                        & {\color{gray}0.93 (0.76)}                                                   & {\color{gray}0.34 (0.75)}                                                   & {\color{gray}0.71 (0.72)}                                                    & {\color{gray}0.04 (0.81)}                                                                      & {\color{gray}-0.18 (0.68)}                                                                  \\
{\includegraphics[width=0.030\linewidth, valign=c]{figures/down.png}}Depression~\cite{Kroenke2009}        & -0.131 (0.05)*** & {\color{gray}-0.69 (0.52)}                                                   & {\color{gray}-0.66 (1.37)}                                                        & {\color{gray}-0.19 (0.73)}                                                  & {\color{gray}-0.44 (0.72)}                                                  & {\color{gray}0.74 (0.69)}                                                    & {\color{gray}-1.01 (0.78)}                                                                     & -1.55 (0.65)***                                                               \\
\hspace{0.2em}{\includegraphics[width=0.023\linewidth, valign=c]{figures/up.png}}Positive affect~\cite{Thompson2007} & {\color{gray}0.040 (0.10)}     & {\color{gray}1.67 (1.10)}                                                    & {\color{gray}2.31 (2.87)}                                                         & {\color{gray}-0.01 (1.53)}                                                  & {\color{gray}1.51 (1.52)}                                                   & {\color{gray}2.15 (1.44)}                                                    & {\color{gray}0.69 (1.64)}                                                                      & 2.45 (1.37)*                                                                  \\
{\includegraphics[width=0.030\linewidth, valign=c]{figures/down.png}}Negative affect~\cite{Thompson2007} & {\color{gray}-0.15 (0.12)}     & {\color{gray}-2.03 (1.52)}                                                  & {\color{gray}-2.60 (3.93)}                                                        & {\color{gray}-0.39 (2.10)}                                                  & {\color{gray}-0.36 (2.09)}                                                  & {\color{gray}0.10 (1.99)}                                                    & {\color{gray}-2.50 (2.26)}                                                                     & {\color{gray}-2.22 (1.89)}                                                                  \\
\hspace{0.2em}{\includegraphics[width=0.023\linewidth, valign=c]{figures/up.png}}Mindfulness~\cite{Brown2003}     & {\color{gray}0.04 (0.04)}      & {\color{gray}0.84 (0.61)}                                                    & {\color{gray}2.20 (1.58)}                                                         & {\color{gray}0.72 (0.85)}                                                   & {\color{gray}0.78 (0.84)}                                                   & {\color{gray}-0.49 (0.81)}                                                   & 1.44 (0.91)*                                                                     & 1.33 (0.76)*                                                                  \\
\hspace{0.2em}{\includegraphics[width=0.023\linewidth, valign=c]{figures/up.png}}Self-reflection~\cite{Silvia2021} & 0.39 (0.14)***   & {\color{gray}2.02 (3.22)}                                                   & {\color{gray}7.56 (8.23)}                                                         & {\color{gray}10.92 (4.42)}                                                  & {\color{gray}-1.31 (4.42)}                                                  & 5.33 (4.24)***                                                 & 14.23 (4.78)***                                                                  & {\color{gray}2.34 (4.02)}                                                                   \\
\hspace{0.2em}{\includegraphics[width=0.023\linewidth, valign=c]{figures/up.png}}Insight~\cite{Silvia2021}         & {\color{gray}0.04 (0.15)}      & {\color{gray}3.76 (4.15)}                                                    & {\color{gray}12.56 (10.56)}                                                       & {\color{gray}1.64 (5.68)}                                                   & {\color{gray}2.66 (5.68)}                                                   & {\color{gray}-3.36 (5.46)}                                                   & {\color{gray}3.22 (6.14)}                                                                      & {\color{gray}6.13 (5.18)}                                                                   \\ \hline
\end{tabular}}
\end{table}

We display the results in Table~\ref{tbl:mindscape_weeklyema}, which reveals several remarkable findings. Note that the number outside the brackets represents the coefficients, while the number inside the brackets indicates the standard error. Notably, anxiety levels consistently decrease each week, with a statistically significant reduction (\textit{p-value = 0.01, $\beta$ = -0.12}). Interestingly, this decrease is more pronounced in males (\textit{p-value = 0.06, $\beta$ = -1.05}). Our analysis also shows a significant decrease in depression scores over time (\textit{p-value = 0.01, $\beta$ = -0.131}), particularly among graduate students compared to undergraduates (\textit{p-value = 0.01, $\beta$ = -1.55}). This suggests that journaling may be more effective for graduate students, although no other demographic factors like gender show significant effects. The overall PHQ4 score also demonstrates a decreasing trend during the study, more so in males. In terms of affective states, we observe no significant changes in negative affect over the study period. However, graduate students report higher levels of positive affect compared to undergraduate students. This stability in affect contrasts with findings from the baseline and follow-up surveys reported in Section~\ref{sec:mindscape_changes_in_followup}, where we observe a statistically significant increase in positive affect and a decrease in negative affect. Similarly, MindScape boosts self-reflective capacities, as evident from a significant weekly increase in self-reflection scores (\textit{p-value = 0.01, $\beta$ = 0.39}). Although gender differences in self-reflection are not statistically significant, we observe a notable trend suggesting that race may influence outcomes. Participants with prior journaling experience benefit more (\textit{p-value = 0.01, $\beta$ = 14.23}), showing greater score in self-reflection. This aligns with our earlier comparison of baseline and follow-up scores in self-reflection, which yielded positive and statistically significant results. In contrast, insights and mindfulness, measured by their respective total scores, do not exhibit significant changes throughout the study. However, participants with prior journaling experience and graduate students experience enhanced mindfulness benefits compared to others 
(\textit{p-value = 0.10, $\beta$ = 1.44}).
 \begin{figure}[h!]
    \centering
    \begin{subfigure}{0.49\textwidth}
      \centering
      \includegraphics[width=1\linewidth]{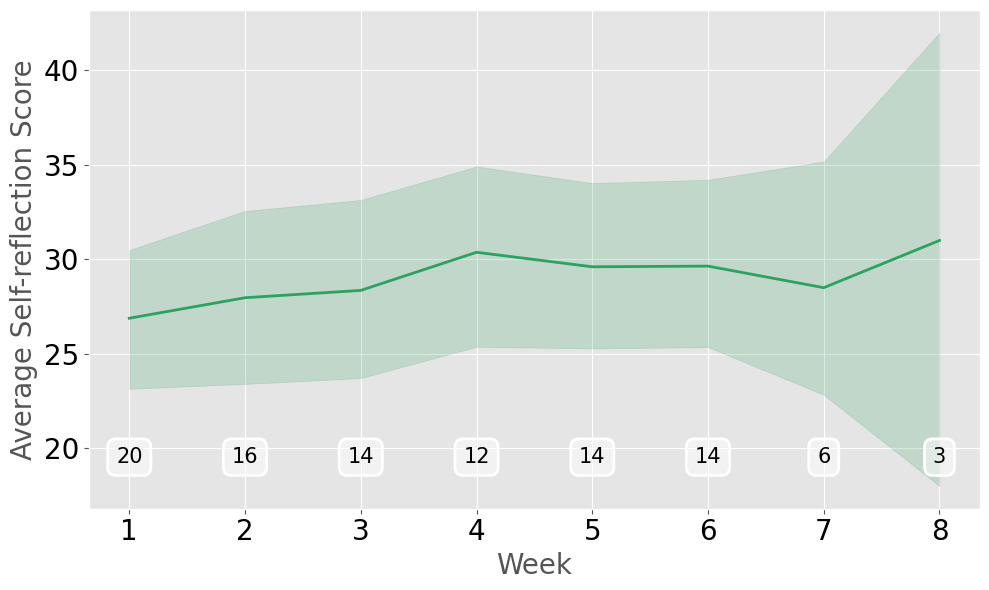}
      \caption{Increase in Self-reflection EMA Score}
      \label{fig:mindscape_sr}
    \end{subfigure}
     \begin{subfigure}{0.49\textwidth}
      \centering
       \includegraphics[width=1\linewidth]{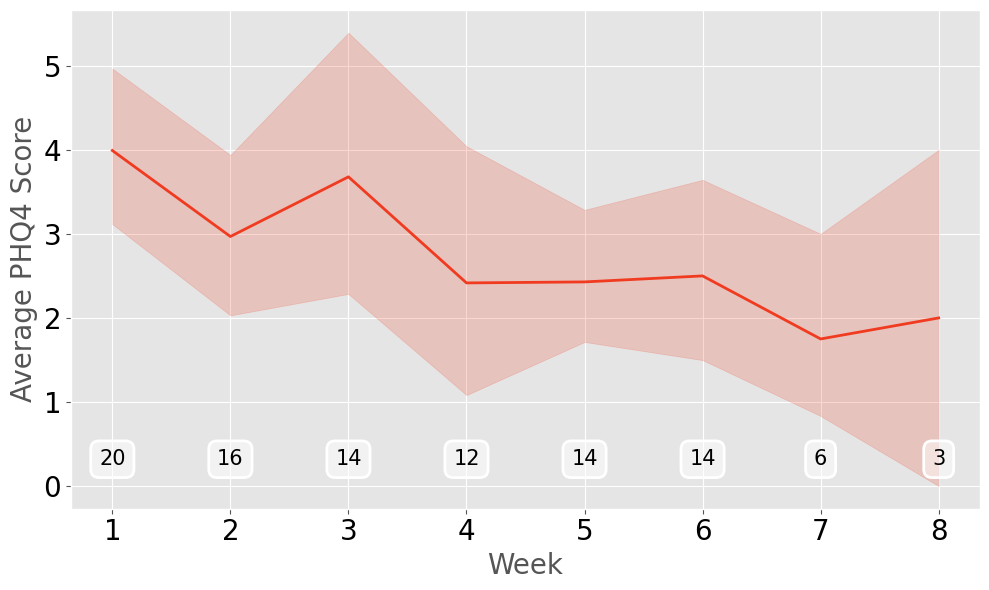}
      \caption{Decrease in PHQ4 EMA Score}
     \label{fig:mindscape_phq}
    \end{subfigure}
        \caption{Improvement in Well-Being as Indicated by EMAs: Both figures demonstrate improvements in well-being, either through an increase in self-reflection (Figure a) or a decrease in PHQ-4 scores, which measure depression and anxiety (Figure b). The numbers in the boxes at the bottom indicate the count of unique participants who submit the survey in each respective week.}
    \label{fig:mindscape_weeklyema}
\end{figure}

We illustrate the increase in self-reflection scores and the decrease in PHQ-4 scores in Figures \ref{fig:mindscape_sr} and \ref{fig:mindscape_phq}, respectively. The numbers at the bottom of the figures in boxes represent the number of participants from whom we receive EMA responses in each specific week. On average, we receive responses from 15 participants over the course of the first six weeks. However, in weeks 7 and 8, which mark the beginning of the generic journaling period, the numbers drop to 6 and 3, respectively. The shaded areas represent the 95\% confidence intervals. Overall, it appears that participants experience several positive changes during the study period. These improvements across various psychological dimensions, particularly in cognition and self-awareness, demonstrate the efficacy of contextual journaling. While some areas show minimal changes or declines, the significant positive trends may indicate the potential beneficial impact of the study on enhancing participants' mental well-being and self-related cognitions. This overall positive shift indicates promising paths for future applications and studies aimed at further understanding and supporting mental health and cognitive awareness through contextual journaling. 

\subsection{MindScape App Performance} 

 \begin{figure}[h!]
    \centering
      \includegraphics[width=0.8\linewidth]{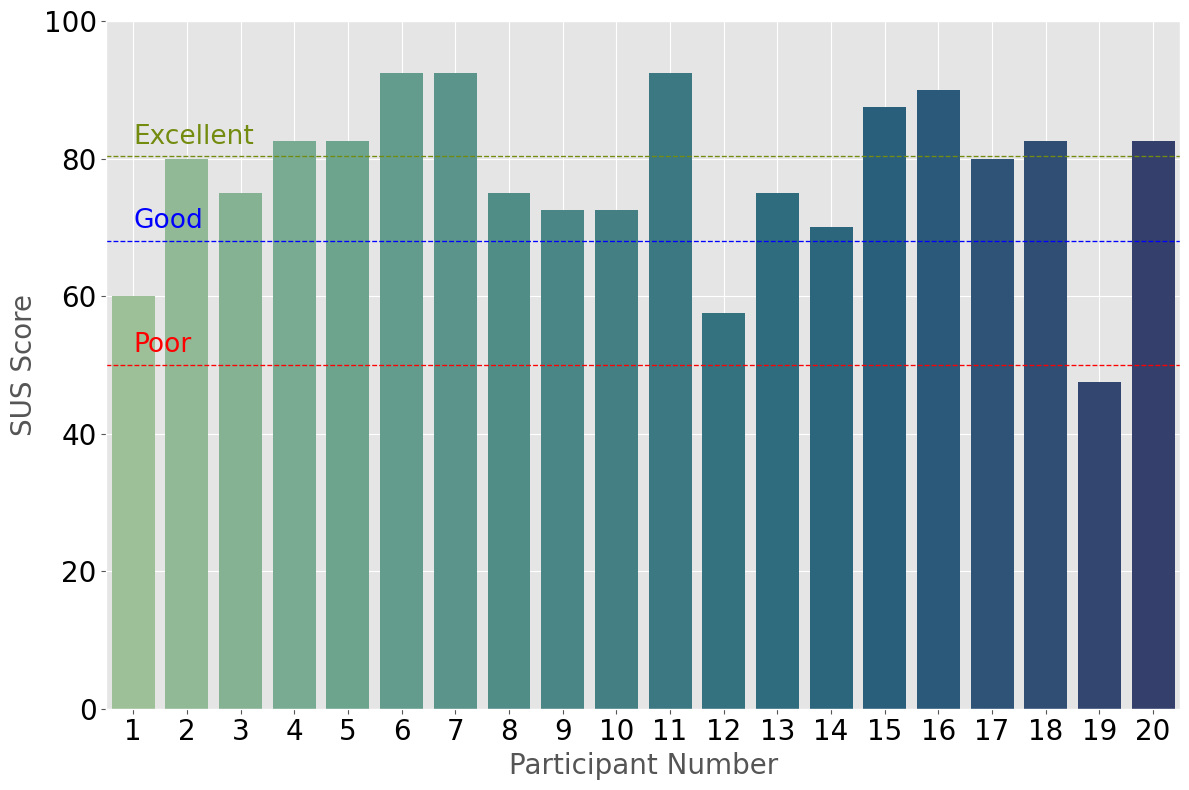}
        \caption{System Usability Score: MindScape receives high usability ratings. In the figure, the red dotted line marks the threshold for poor usability, the blue dotted line for good usability, and the green dotted line for excellent usability. Over 80\% of participants rate the MindScape system as either good or excellent.}
        \label{fig:mindscape_susscore}
\end{figure}

\begin{figure*}[h!]
     \centering
     \begin{subfigure}[b]{0.30\textwidth}
         \centering
   \includegraphics[width=1\linewidth]{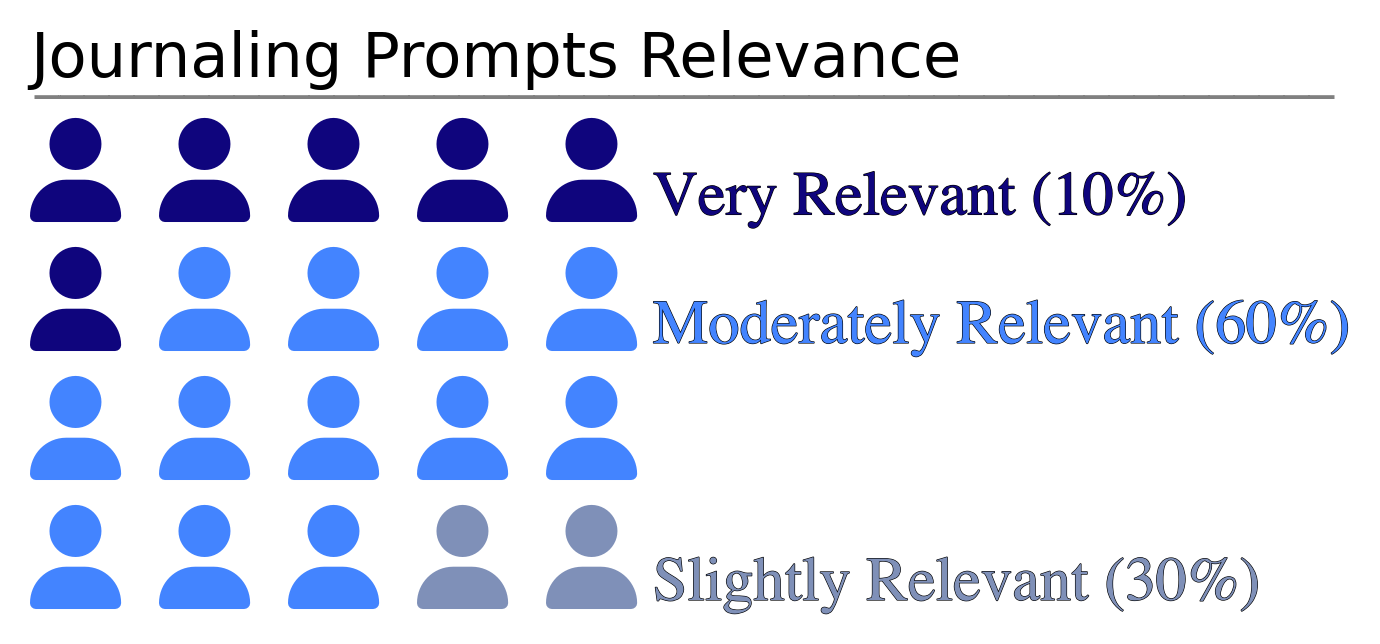}
         % \vspace{0.1cm}
     \end{subfigure}
     \begin{subfigure}[b]{0.30\textwidth}
         \centering
         \includegraphics[width=1\linewidth]{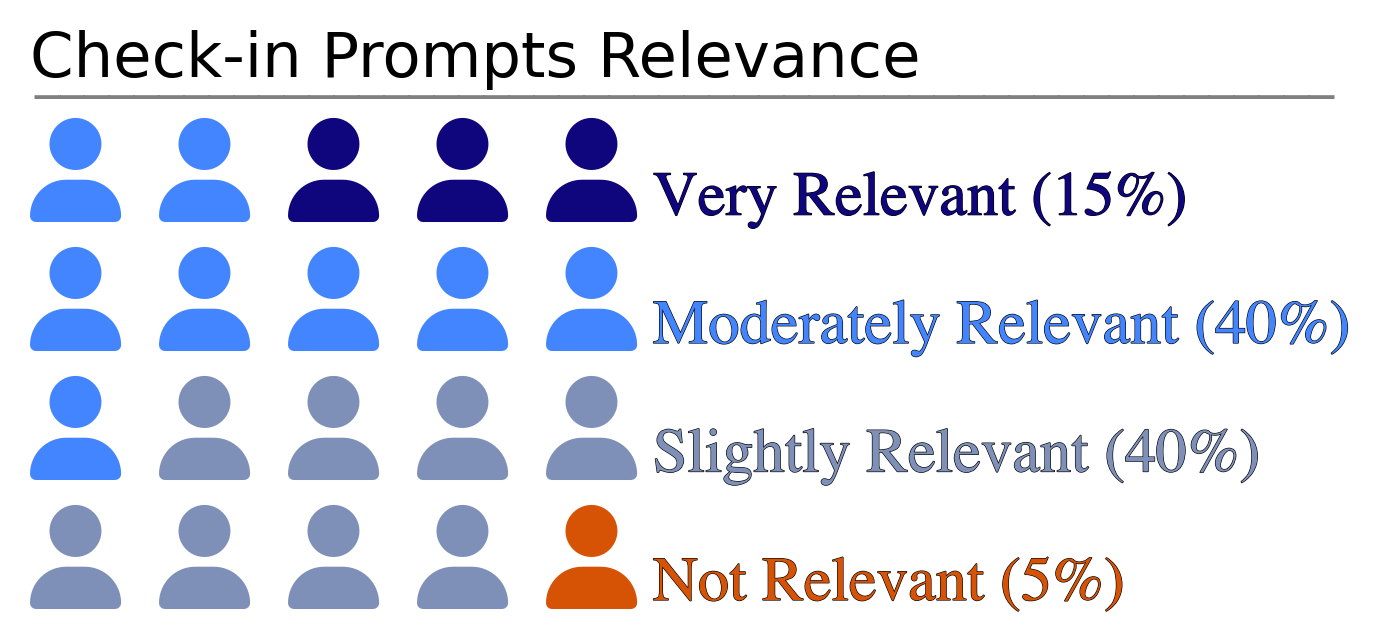}
        % \fbox{\includegraphics[width=0.70\textwidth, scale=2.0]{figures/2.png}}
                  % \vspace{0.1cm}
     \end{subfigure}
\begin{subfigure}[b]{0.30\textwidth}
    \centering\includegraphics[width=1\linewidth]{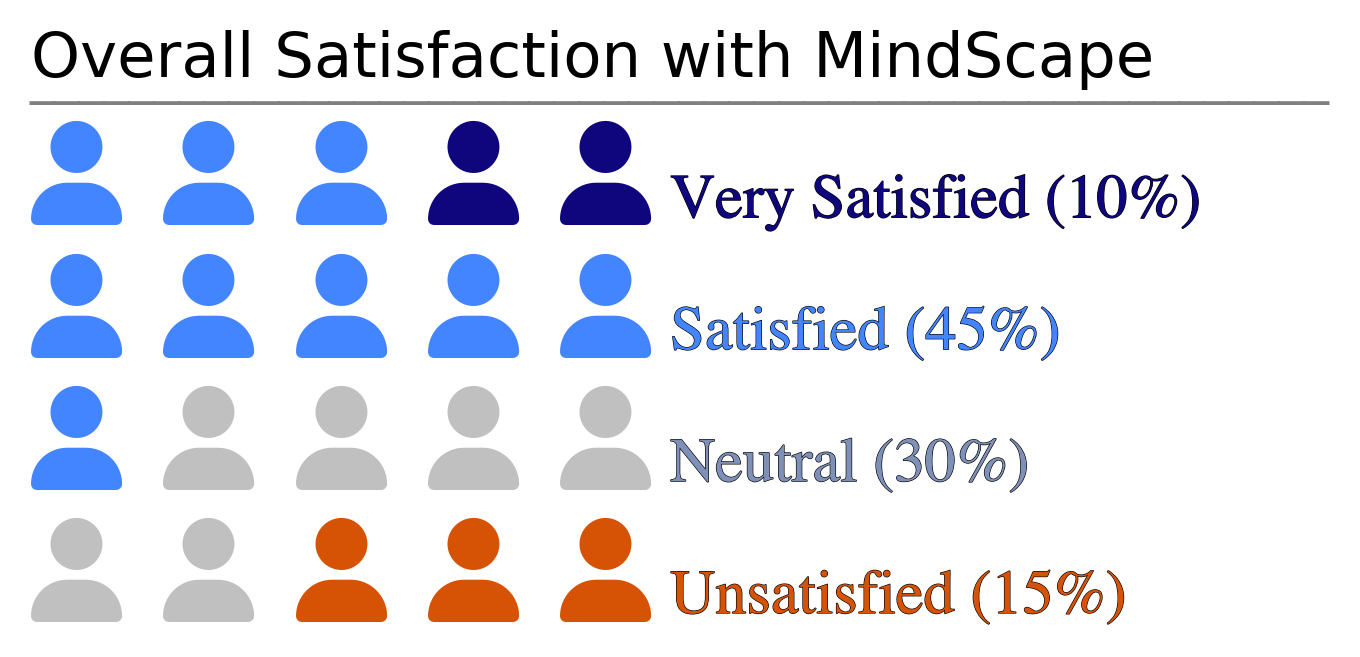}
                  % \vspace{0.8cm}
     \end{subfigure}
 \par\bigskip
    \begin{subfigure}[b]{0.30\textwidth}
         \centering
 % \vspace{0.8cm}
         \includegraphics[width=1\linewidth]{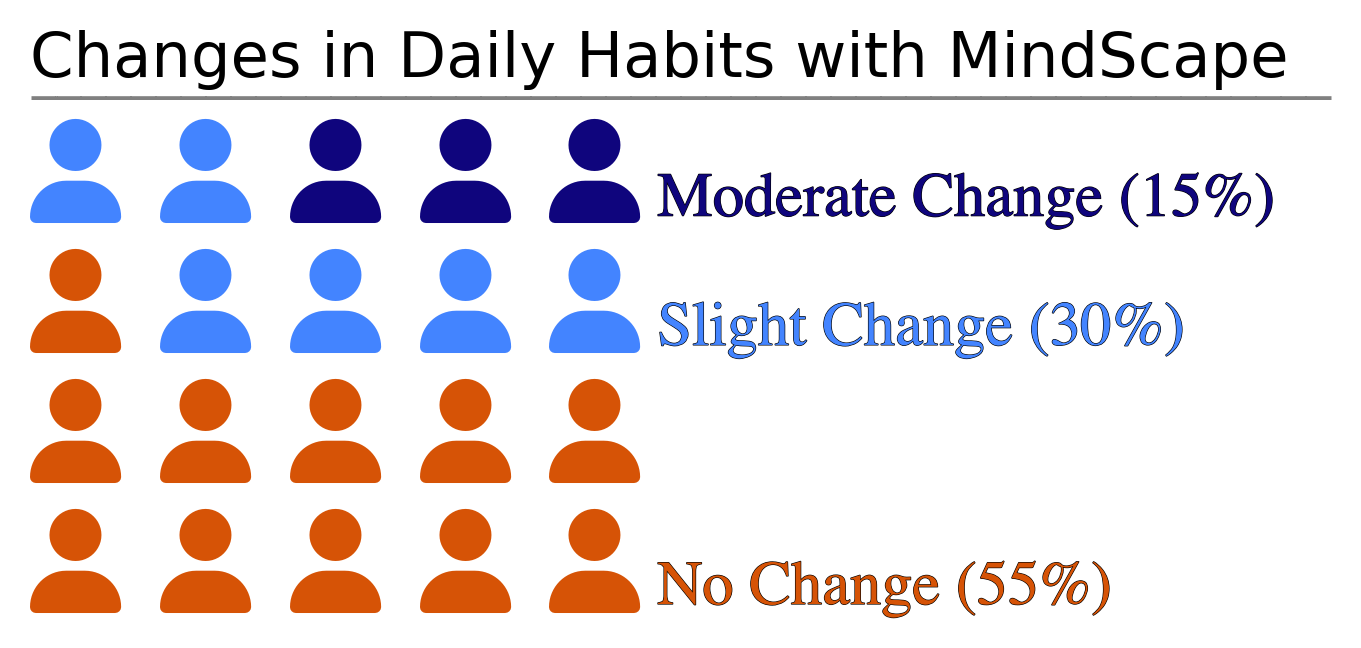}
     \end{subfigure}
        \begin{subfigure}[b]{0.30\textwidth}
         \centering
         \includegraphics[width=1\linewidth]{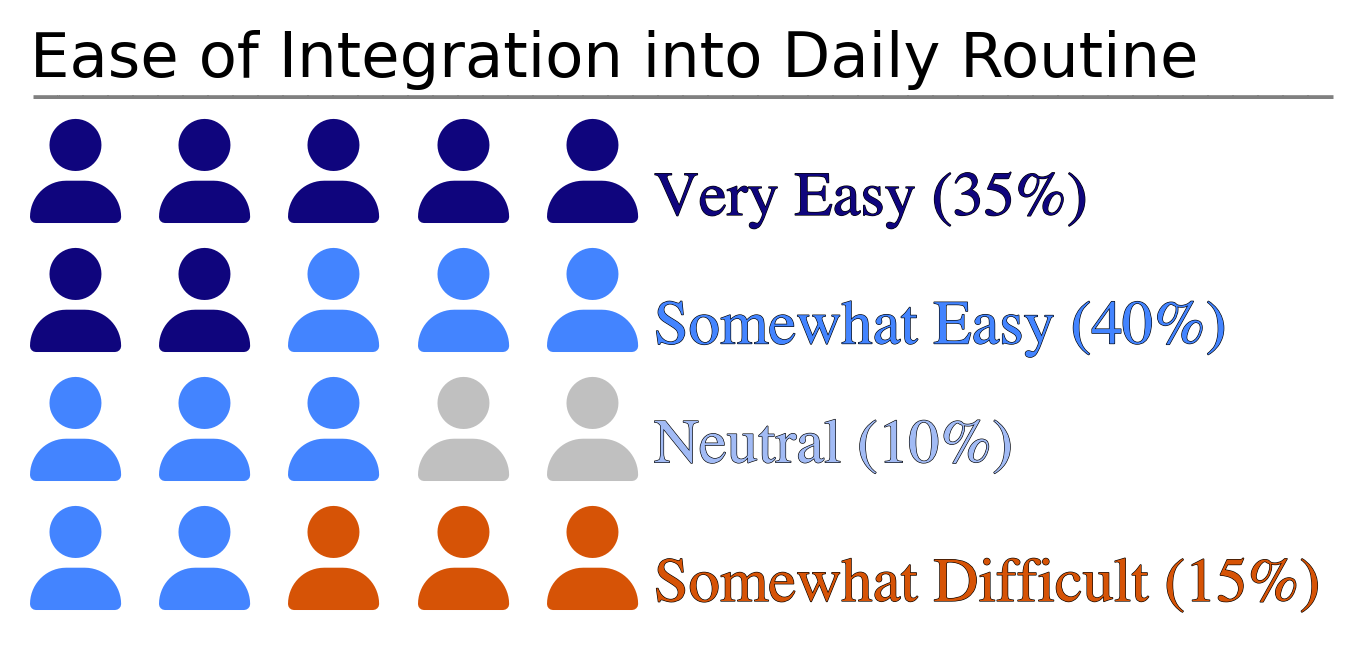}
        
    \end{subfigure}
            \begin{subfigure}[b]{0.30\textwidth}
         \centering
        \includegraphics[width=1\linewidth]{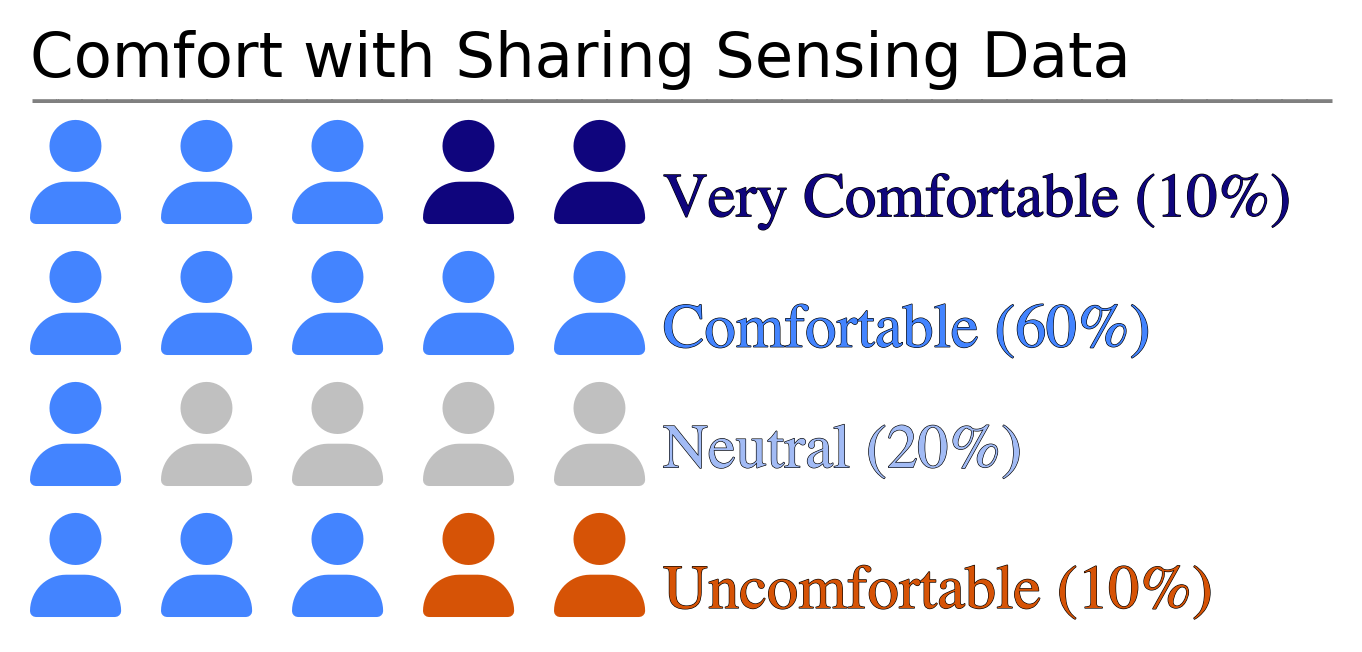}
     \end{subfigure}

    \par\bigskip
\begin{subfigure}[b]{0.35\textwidth}
    \centering\includegraphics[width=0.95\linewidth]{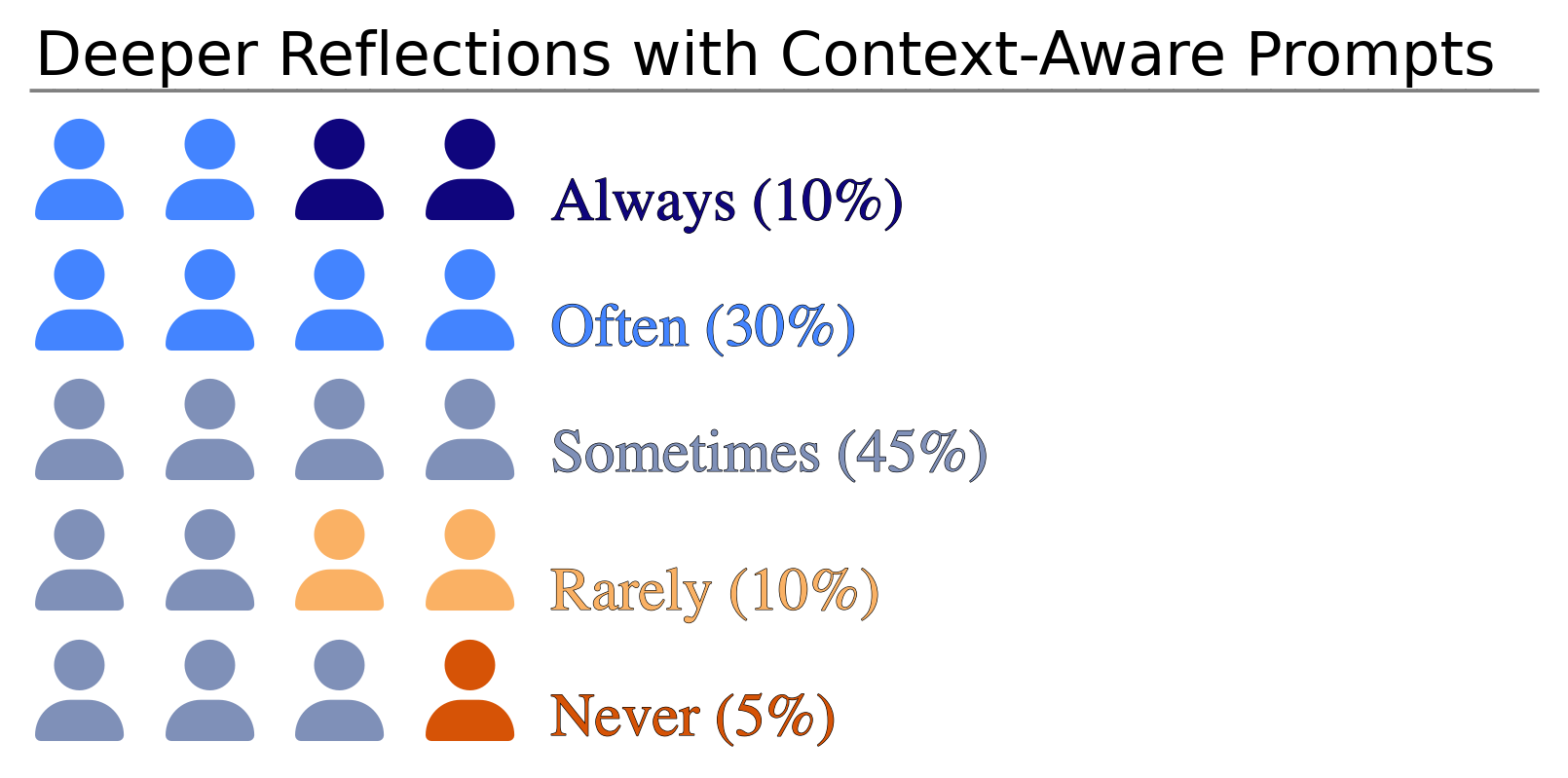}
     \end{subfigure}
    \begin{subfigure}[b]{0.35\textwidth}
         \centering
         \includegraphics[width=1\linewidth]{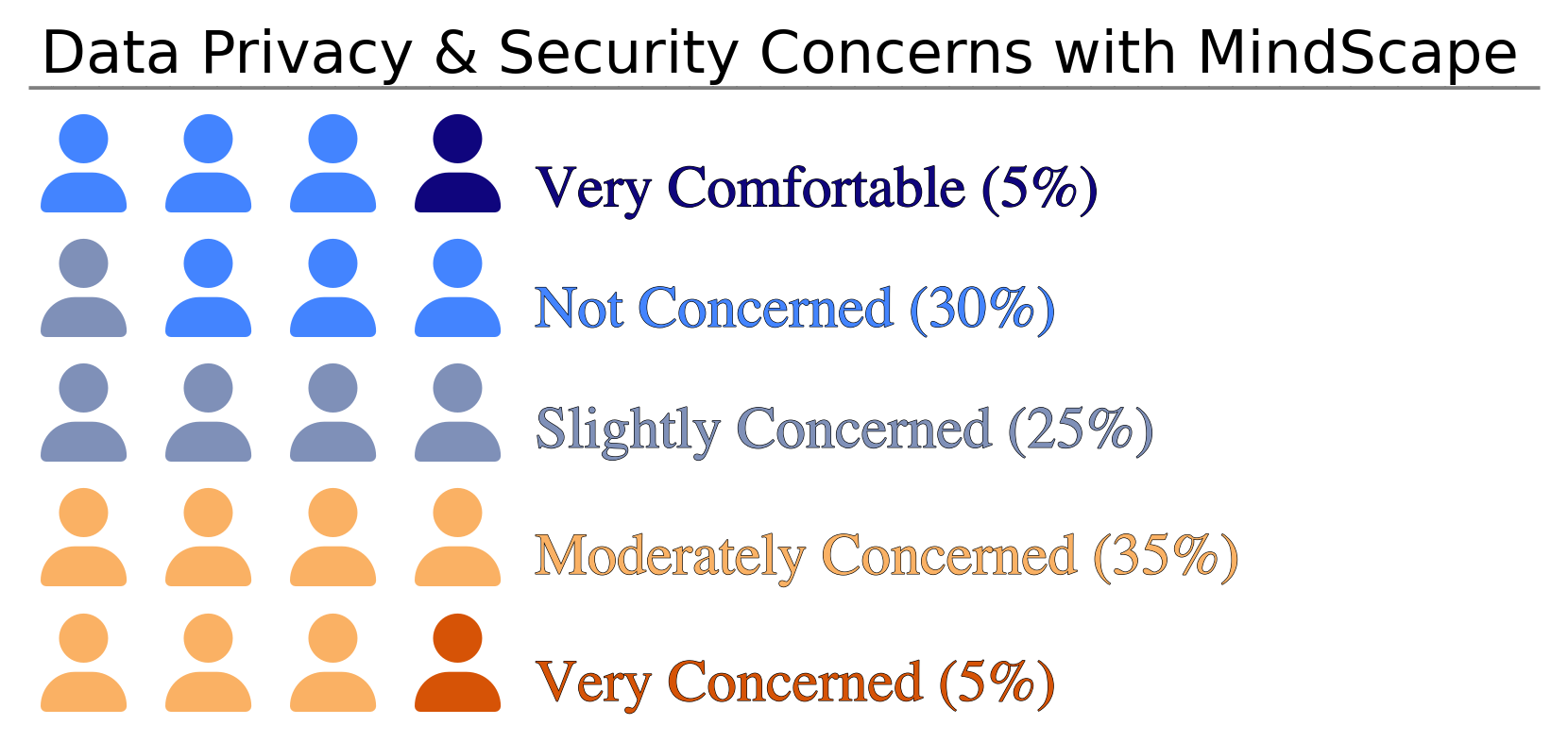}
     \end{subfigure}
     
        \caption{Participants' Experience with MindScape: This figure visualizes responses to various questions about participants' experiences with the app. Each person icon represents one participant, with icons and their corresponding responses color-coded to match specific response labels.}
        \label{fig:mindscape_appsurveys}
\end{figure*}

Upon concluding the study, we solicit participant feedback regarding the MindScape app's performance and their experience with contextual journaling. Our analysis of these responses reveals insights into the app's usability, relevance, and impact on users' daily lives.

We first administer the standard System Usability Scale (SUS) survey to assess the perceived usability of the MindScape app. Our analysis shows that 50\% (N=10) of participants rate the system as excellent, with a score of 80 or higher. Additionally, 35\% (N=7) rate it as good (score of 68 or higher), while 15\% (N=3) consider it poor, scoring below 68.
We then present participants with several other questions on their experience with the app. We discuss some responses here and visualize these responses in Figure~\ref{fig:mindscape_appsurveys}. Please see Appendix~\ref{sec:mindscape_list_of_questions_asked} for a complete list of questions and participant responses. Regarding the relevance of journaling prompts generated by the app, 10\% (N=2) found them very relevant, 60\% (N=12) moderately relevant, and 30\% (N=6) slightly relevant. When assessing check-in prompts specifically, 5\% (N=1) indicate they are not at all relevant, 40\% (N=8) find them slightly relevant, 40\% (N=8) think they are moderately relevant, and 15\% (N=3) consider them very relevant.

To enrich our understanding of participants' experiences using the app, we supplement our quantitative analysis with qualitative, open-ended questions. We ask participants if they recall any specific prompts from the MindScape app that significantly resonated with them or were particularly relevant to their experiences. This inquiry aims to uncover how the app's context-aware prompts facilitate deeper self-understanding or self-awareness. Participants report various responses that illustrate the personalized impact of these prompts. Many highlight how specific questions related to their daily routines or habits prompt meaningful reflection. For instance, one participant mentions a prompt about their walking routine, which coincides with a new meditative practice: \textit{``About a week or two after I started it, I got a question from the app about my walking routine, and it gave me an opportunity to reflect on how the meditative practice had been going!''}. This example underscores how timely and relevant prompts enhance mindfulness and self-awareness. Other participants appreciate prompts that encourage proactive planning and goal-setting, such as those urging them to think about weekly exercise goals. This not only makes them contemplate their physical health but also motivates them to set concrete plans. Additionally, reminders about social interactions lead some users to reach out to friends more frequently, improving their social life. One user points out, \textit{``A lot of the prompts remind me that I have not been socializing in person as much, which has led to me reaching out to friends I do not see as often.''} The app's ability to track changes in routines and lifestyle also stands out for many users. For example, prompts related to changes in workspaces or sleep patterns provide insights that participants might not have noticed on their own. One participant succinctly notes, \textit{``The app has been accurately keeping track of my changes in work routines, which allows me to take time and reflect on how these changes are affecting my overall work performance.''}

Given that these questions are posed at the end of the study, following two weeks of generic prompts, we ask participants to self-report on the frequency with which context-aware prompts facilitate more profound reflection than usual. Out of respondents, 5\% (N=1) reported never, 10\% (N=2) rarely, 45\% (N=9) sometimes, 30\% (N=6) often, and 10\% (N=2) always.
Regarding changes in daily habits or behaviors, 15\% of participants report moderate changes, 30\% notice slight changes, and 55\% do not report any changes. However, when we ask to share specific instances of behavioral changes, many highlight tangible impacts driven by the app. One user notes the effect of visual prompts related to phone usage\textit{``Every time I see that little green light [indicating microphone use], I'm prompted to think about whether my current device usage aligns with my goals.''}. This increased awareness often leads to decreased unreflective screen time and transitions to other daily activities. Changes in social interactions are also notable, with one respondent mentioning: \textit{``I've seen improvement in my social life...through prompts that make me reflect on the importance of nurturing conversations with my friends.''} Additionally, the app enhances self-awareness regarding personal well-being and daily activities. Reports include better monitoring of sleep patterns, more frequent walks, and increased engagement in self-reflection and meditation. However, not all feedback indicates a change. One participant remarks on the challenging nature of adapting behaviors during a particularly busy life phase, suggesting that while the app identifies reduced phone usage, the decrease is more due to their hectic schedule rather than a conscious effort spurred by the app.

To gain further insight into the app's influence on planning and mindset, we ask participants about its impact on their weekly structure and appreciation of daily activities. Responses reveal varied effects. Several appreciate the structuring aspect introduced by the app, notably through timed journaling activities. One participant notes: \textit{``It's led me to prioritize working out and socializing in person more...the prompts are good reminders.''} Another remarked on increased mindfulness: \textit{``It has made me more aware of small positive interactions I've had throughout the day...the app brought those moments closer again.''} However, some find the prompts and check-ins inadequate or misaligned with their personal reflection needs.

Overall, 55\% (N=11) of participants express satisfaction or high satisfaction with the app, while 30\% (N=6) remain neutral, and 15\% (N=3) feel dissatisfied. The ease of integrating MindScape into daily routines is notable, with 75\% (N=15) stating it is somewhat easy to very easy. Regarding comfort with using data from phones and smartwatches alongside AI to personalize journaling prompts, 70\% (N=14) voice comfort or great comfort, while 30\% (N=6) are uncomfortable or neutral. In terms of data privacy and security, 60\% (N=12) express no or slight concern, whereas 40\% (N=8) report moderate to high levels of concern.
When comparing context-aware directed prompts to standard journaling, participants note both benefits and challenges. Many appreciate the structured prompts for enhancing their engagement and reflection practices. As one user mentioned: \textit{``Previously, it was cumbersome for me to journal as I'd have to sit and recollect everything. With MindScape, I reflect more earnestly and compare my current state with previous ones, giving me clear insights into my mental and physical well-being.''} Participants also value the specific insights prompted by the app, which sometimes brings attention to overlooked aspects of their daily lives:
\textit{``The context-aware prompts pull out surprising trends that I may not have noticed and ask me to reflect on it.''} This feature helps some users gain a deeper understanding of their behavioral patterns and encourages proactive thinking, like another user who points out, \textit{``The prompts lead me to think about specific parts of the experiences of my day and how I might make changes.''}

However, challenges with the specificity and relevance of the AI-driven prompts are also noted. Some participants feel that the prompts could be too rigid or not entirely reflective of their true daily experiences due to inaccuracies in activity or location tracking. For example, one user criticizes the prompts for being \textit{``sometimes based on metrics that might be irrelevant,''} such as reacting to an unusually high number of spam calls as if they are meaningful phone conversations. Similarly, another points out that \textit{``the app tends to see trends or differences in my behavior when there isn't a particular cause behind them.''}
Despite these challenges, the general consensus acknowledges the utility of the MindScape app in fostering regular reflective practices. Even those who note limitations often recognize the benefits of prompted reflections, with one participant aptly summarizing: \textit{``I think that I can more easily decide what to write about with a prompt, but I don't know if I access my feelings in the same way as a free-form journal.''}

\subsection{Strengths, Weaknesses and Improvement Opportunities} 
At the conclusion of the study, we gathered qualitative feedback on the MindScape app's strengths and weaknesses. Users praised the app's innovative approach to integrating behavioral patterns with journal prompts and valued the regular notifications for maintaining daily reflection habits.
One participant noted, \textit{"I liked how it regularly made me aware of what I am doing, and it helped me reflect on the activities. I loved the journal prompts as they were not too specific nor broad that I could sufficiently elaborate on my thoughts."} Another user appreciated how the app \textit{"did initiate some reflection on the interactions I had throughout the day, how I value the people in my life."}
The app's ease of use was highlighted as a major advantage. A user expressed, \textit{"Love the ease of use of the app. Super simple to click on the notification and complete the journaling and check-in practices."} Another added, \textit{"I also enjoyed the personalized prompts and how I only had to write short pieces for the journal."}
However, participants also identified areas for improvement. The most common suggestion was enhancing the accuracy of context-aware prompts. One user shared, \textit{"Many times, the app didn't seem to be aware of where I lived and would ask me questions about being in my dorm when I wasn't there."} This highlights the need for more precise behavioral sensing, which could potentially be addressed by integrating additional data sources like smartwatches.
The repetitiveness of prompts was another concern, with users noting identical prompts over multiple days. This could be improved by ensuring more dynamic and varied prompt generation. Some participants also suggested linking the app to fitness trackers for better context awareness. A few users also raised concerns about the app's battery consumption, indicating a need for power management optimization.

Based on participant feedback, we identified several key improvement opportunities for the MindScape app:
\begin{enumerate}
\item \textbf{Enhanced Personalization and Prompt Variability:} Participants suggest enhancing personalization by considering responses to previous prompts alongside behavioral data when generating new ones. This approach could introduce greater diversity and relevance to the questions posed, balancing specific lifestyle habits with broader life reflections and personality exploration. Users prefer a mix of specific and open-ended prompts that maintain a narrative between journal entries, offering opportunities for deeper reflection and emotional processing. By analyzing previous responses, the app can better determine whether to base new prompts on sensed data or earlier interactions, especially when introducing broader questions on weekends. Implementing these changes could significantly improve user engagement by providing deeper insights into emotional well-being and adding greater therapeutic value to the journaling experience. This refined approach focuses more on emotional processing rather than strictly behavioral tracking, potentially offering users a more personalized and meaningful reflection tool.
\item \textbf{Improved Context Sensitivity:} Perfecting context sensitivity remains challenging due to reliance on passively sensed data. However, we can enhance this by better integrating with smartphones and wearable technology for deeper insights into physical and social contexts. Users are interested in syncing with smartwatches and fitness trackers. There's also demand for allowing users to define and assign meanings to frequently visited locations beyond predefined campus spots. This can be facilitated by integrating third-party location APIs like Google Places, while carefully managing privacy concerns. Alternatively, we can use GPS data from photos to prompt reflections on recent travels, similar to Apple's Journal app. By focusing on improving the app's ability to accurately adapt to various contexts rather than relying on hard-coded campus locations, we can significantly boost its utility and user satisfaction. These enhancements will provide more personally relevant prompts, especially for off-campus activities, and improve the overall user experience.
\item \textbf{Goal-Setting and Personal Growth Features:} Incorporating goal-setting could transform prompts into opportunities for meaningful reflection. A user suggests, \textit{"I wonder if the contextual prompts could be more of a prompt to reflect on goal progress rather than guesses about what I did that day."} Some users also propose behavioral suggestions when low levels of social interaction or physical movement are detected. This feature could further engage users in proactive behavior modification and self-improvement, enhancing the app's role in supporting personal growth.
\item \textbf{Customization Options:} Participants desire more customization options to tailor the app to their individual needs. They want increased control over prompt timing, adjustable directly from the app, to accommodate personal routines and sleep patterns. One user suggests allowing users to set their core hours to better align with their daily schedule. Additionally, participants recommend offering multiple prompt options to enhance engagement by catering to specific interests and needs. A user proposes providing prompts from different focus areas (e.g., physical activity, social interaction) for users to choose from as needed. This flexibility in both timing and content would ensure interactions are more relevant and stimulating, potentially increasing user involvement and satisfaction. By implementing these customization features, the app could better adapt to individual preferences, fostering a more personalized and engaging experience for users.
\item \textbf{Expansion of Check-in Functionality:} User feedback on check-ins is mixed. Some appreciate the brief interactions, while others find them redundant. Suggestions include decreasing the frequency of basic check-ins and focusing more on prompts that help identify or process emotions. Many users recommend making check-ins optional, allowing users to activate, deactivate, or adjust the frequency according to their preference. Some participants propose transforming check-ins into encouraging advice or reminders. These changes could improve the user experience by providing more flexibility, relevance, and clarity in the check-in process.
\end{enumerate}

These improvements could significantly enhance the app's utility, user engagement, and overall satisfaction, making it a more effective tool for promoting well-being and self-reflection.

\section{Discussion}
\label{sec:mindscape_discussion}
In this section, we discuss our findings, the implications of our work and the associated ethical considerations.

\subsection{Summary of Results}
\label{sec:mindscape_summary}

We collect a total of 661 journal entries from 20 students at Dartmouth College, with a significantly higher engagement rate in the first six weeks of contextual prompts compared to the last two weeks of generic prompts. On average, participants actively engage for about five weeks, submitting 26.65 entries each during the initial six weeks. In contrast, in the last two weeks of generic journaling, participants submit an average of 7.11 entries. The higher engagement with contextual prompts can potentially be attributed to their relevance and their ability to effectively capture daily experiences. However, it may also be influenced by the ordering effect -- the novelty of introducing contextual journaling at the beginning likely boosted initial engagement due to its freshness. Throughout the study, participants consistently utilize check-ins, with a total of 2,985 responses recorded, showing higher response rates in the afternoon and evening. 
Students demonstrate a clear preference for journaling on areas directly impacting their daily lives, with Social Interactions and Digital Habits ranking as the most preferred categories. Interestingly, although Physical Fitness is ranked lower compared to other categories, it represents a significant portion of the prompts, mainly due to the broader range of signals it encompasses.

We use advanced topic modeling techniques to understand the themes that resonate during check-ins at different times of the day. Morning check-ins often revolve around social and communication app usage, while afternoon check-ins shift towards academic and social life experiences. This variation underscores the relevance of tailoring check-in prompts to match daily activities and time-specific contexts. 

A deeper dive into the journaling responses reveals intriguing insights into the thematic content and language patterns used in generic and contextual journals. We identify four primary topics in generic journals: Daily Experiences, Daily Activities, Productivity Management, and Academic \& Personal Growth. This suggests that individuals use generic prompts to explore various aspects of their daily lives, emotional experiences, and personal development. With LIWC analysis, we find nuanced differences in the language patterns and emotional expressions used in both types of journals. Generic prompts, which ask participants to reflect on anything of interest, yield higher analytic thinking scores, suggesting a more formal and logical thinking style. This finding is unexpected, as one might assume that abstract prompts would lead to more creative and less formal thinking. However, it's possible that the open-ended nature of generic prompts encourages participants to engage in more structured thinking as they attempt to organize their thoughts and ideas. 

Contextual prompts yield higher scores on personal pronouns and lower on analytic thinking scores, indicating a more personal and introspective writing style. The findings reveal that generic prompts may encourage broader emotional expression, higher positive tone, and reduced negative tone. Generic prompts also promote a broader temporal focus, linking current experiences to past memories or future aspirations. Contextual journals focus more on personal experiences and relationships, with higher cognition scores and a greater emphasis on thinking and problem-solving. This finding appears to contradict the earlier result showing lower formal/logical thinking in contextual journals compared to generic ones. However, it is possible that the contextual prompts, while reducing formal/logical thinking, simultaneously encourage more personal and relational thinking, which is captured by the higher cognition scores. This suggests that contextual journals may foster a different type of cognitive processing, one that prioritizes personal connections and experiences over abstract logical reasoning. 

As we examine the effects of contextual journaling on wellbeing and emotional growth, we find several positive outcomes. We observe a significant decrease in neuroticism (11.81\%), an increase in positive affect (7.15\%) and a decrease in negative affect (10.60\%). Stress and anxiety also decrease, although not significantly. Notably, we observe significant improvements in mindfulness (6.76\%), self-reflection (5.80\%), and insight (7.57\%). Weekly EMA reveal consistent decreases in anxiety levels, particularly among males. Depression scores also decrease significantly ($\beta$ = -0.13), especially among graduate students ($\beta$ = -1.55). Self-reflection scores increase significantly week-to-week while mindfulness and insight do not show significant changes. Participants with prior journaling experience and graduate students experience enhanced mindfulness benefits. Several studies support the idea that individuals with prior experience in journaling or mindfulness practices tend to exhibit greater benefits from mindfulness interventions ~\cite{Osin2021MindfulnessME, Kiken2015, MatosMachado2015MindfulnessPO}. This may be because they have developed greater self-awareness, reflection skills, and emotional regulation. Moreover, graduate students' advanced education and exposure to various learning strategies~\cite{Brnstrm2011} may also contribute to their advantage, alongside the potential impact of their age~\cite{Hohaus2013}.  

Upon concluding the study, we solicit feedback from participants regarding the MindScape app's performance, their experience with contextual journaling, and related topics. The feedback is overwhelmingly positive, with 50\% of participants rating the app as excellent and 35\% as good. Participants find the app's context-aware prompts to be relevant and helpful, with 60\% considering them moderately relevant and 30\% slightly relevant. Many participants appreciate the app's ability to track changes in routines and lifestyle, and 55\% report moderate changes in their daily habits or behaviors since using the app.
Participants share specific instances where the app's prompts led to meaningful reflection and behavioral changes, such as increased mindfulness and self-awareness, improved social interactions and relationships, enhanced goal-setting and planning, better monitoring of sleep patterns and physical activity, and increased engagement in self-reflection and meditation. However, some participants note challenges with the app's prompts, such as inaccuracies in activity or location tracking, prompts being too rigid or irrelevant, and limited ability to access deeper feelings and emotions. Despite these challenges, the majority of participants (75\%) find it easy to integrate the app into their daily routines, and 70\% are comfortable using data from phones and smartwatches alongside AI to personalize journaling prompts. Overall, participants appreciate the structured prompts and the app's ability to facilitate regular reflective practices, with 55\% expressing satisfaction or high satisfaction with the app. Thus, our study demonstrates the efficacy of contextual journaling in promoting positive emotional responses and personal growth, with significant improvements in cognition and self-awareness. These findings indicate promising paths for future applications and studies aimed at supporting mental health and cognitive awareness through contextual journaling.
% \textcolor{red}{To-do: connect to prior works, more discussions}

\subsection{Implications}
\label{sec:mindscape_implications}
This study's findings have notable implications for HCI and the design of context-aware systems. Our research demonstrates the critical role of personalization and context-awareness in enhancing engagement with journaling applications. Specifically, prompts that are aligned with users' daily activities, routines, and personal experiences tend to elicit more engaged, introspective responses, leading to positive behavioral changes. However, it is important to note that generic prompts also have their benefits, such as encouraging broader emotional expression, higher positive tone, and reduced negative tone. A balanced approach, incorporating both contextual and generic prompts, could potentially offer the most comprehensive benefits. Our study highlights the importance of tailoring interventions to individual needs and circumstances, and adapting prompts and interfaces to align with users' daily schedules and routines. The differences observed based on gender, student status, and prior journaling experience also emphasize the need for personalized approaches. While contextual prompts may facilitate reflections on personal experiences and relationships, generic prompts may promote analytic thinking and broader emotional expression. By combining the strengths of both approaches, researchers and designers can create more effective and inclusive journaling applications. 

Furthermore, our study illustrates the potential benefits of integrating data from diverse sources such as smartphones, wearables, and AI-powered language models to enrich and personalize the journaling experience. Researchers are encouraged to explore innovative methods to merge these data streams while conscientiously addressing associated privacy and ethical concerns. Participants generally reported that the context-aware journaling app was easy to incorporate into their routines, but some faced challenges with prompt relevance and accessing deeper emotional layers. This underlines a crucial area for HCI researchers to improve the user experience in such applications, ensuring that the prompts are not only engaging and relevant but also effective in facilitating meaningful self-reflection. Moreover, the positive emotional and cognitive outcomes achieved in this study support increased multidisciplinary collaboration between HCI, psychology, and other relevant fields. Such collaborative efforts can merge user-centered design with behavior change theories and data-driven methodologies to craft more impactful interventions. As journaling and other applications with behavioral sensing and AI evolve to become more advanced and personalized, it is imperative to consider and address ethical issues such as data privacy, algorithmic bias, and potential misuse of personal data. Researchers should commit to responsible design practices and actively involve users in the development process to enhance transparency and maintain accountability. In summary, this study showcases the potential of context-aware journaling systems to facilitate significant personal growth and behavioral improvements, and highlights the importance of balancing contextual and generic prompts to offer a comprehensive and inclusive journaling experience.

\subsection{Limitations and Future Work}
\label{sec:mindscape_limitations}
Our study has some important limitations to consider. First, given our small sample size and exploratory aims focused on feasibility, acceptability, and preliminary efficacy, our findings might not generalize well to outside populations. As a pilot study, our primary objective was to assess the acceptability and feasibility of combining AI with behavioral sensing in journaling apps, rather than conducting a large-scale randomized controlled trial (RCT). Therefore, we did not focus on statistical significance, which is heavily influenced by sample size. Instead, our study should be seen as a proof-of-concept for this novel intervention, providing preliminary insights into its potential benefits and areas for future development. Our findings, focused on a specific student population, might not be widely applicable. Future studies should aim to engage a larger and more diverse sample, building on our suggestions and results. Future studies should also consider employing a counterbalanced design to directly compare enhanced and traditional journaling methods, controlling for potential order effects and providing more robust evidence of the specific impacts of our contextual AI approach. One significant limitation is our focus on Android users, which likely contributes to the low number of participants. In the US, most young adults use iPhones~\cite{iphoneusers}, making our Android-only approach a limiting factor. Future researchers should develop apps compatible with both Android and iOS operating systems to reach a broader audience and increase participant diversity. Privacy and data handling remain crucial considerations in our approach. While we implemented measures to protect user data, future iterations should provide more granular control over data sharing. This could include allowing users to selectively enable or disable specific behavioral signals used for prompt generation. Additionally, exploring the use of self-hosted, open-source LLMs could further enhance privacy by keeping sensitive data local. However, this approach may impact the quality of generated prompts, necessitating careful evaluation of the trade-offs between privacy and functionality.

While we do not emphasize statistical significance in many cases, and some findings are not statistically significant, we still observe several positive changes. Future research with expanded populations might determine which of these positive changes are causally linked to the journaling intervention and which are merely coincidental. Our study does not compare traditional journaling to personalized AI journaling in a randomized-controlled way, as we did not have enough participants for a RCT. So, it is unclear whether contextual AI journaling offers advantages over traditional journaling. While we collect objective data on physical activity, sleep patterns, and other behaviors, we choose to focus our analysis on self-reported measures for this initial exploratory study. This decision is made to prioritize understanding users' subjective experiences with contextual journaling, which is crucial when evaluating a novel intervention's acceptability and perceived impact. However, we acknowledge that incorporating analysis of the objective data could provide valuable additional insights. We've stored this data securely and plan to conduct a more comprehensive analysis in future work, comparing self-reported experiences with objective behavioral changes. This future analysis will help us better understand the relationship between perceived and actual changes in behavior and well-being.

We also do not use participants' journal entries to help the LLM learn and adapt; instead, we only use behavioral sensing data to contextualize the journals. Using prior journaling responses could potentially enhance the app's functionality, but we prioritized privacy considerations by not sending potentially identifying information (that may be contained in the journal entries) to OpenAI. To maintain participants' privacy, we only sent de-identified high-level behavioral sensing data to GPT-4. Future research should explore more privacy-preserving approaches, such as using locally deployed open-source LLMs (e.g., Llama2). It could potentially allow for the utilization of Protected Health Information (PHI) or journal entries, while maintaining control over data privacy and security. This could also allow for a comparison of different models' performance while keeping user data on-device. Additionally, investigating techniques like differential privacy for any aggregated data used in model improvements could further enhance privacy protections. Our app also does not incorporate user feedback (thumbs up/down) from check-ins to personalize the journaling experience, future research could explore the integration of such feedback to enhance personalization. Future researchers could also expand the range of signals by incorporating data from wearable devices, enhancing the diversity and coverage of the prompts used in journaling. As an early exploratory study integrating AI and behavioral sensing for journaling, our focus is on describing what happened without exploring the underlying reasons. For instance, while some participants prefer certain types of check-ins, we do not examine why these preferences exist or what might influence them. Future investigations could explore these nuances to better understand participant preferences and refine the journaling process further.  
We acknowledge the lack of ablation studies in our current work. While we included daily check-ins as part of the intervention, we did not separately analyze their specific impact on outcomes. Future studies will include ablation analyses to understand the individual contributions of different components, such as contextual prompts, daily check-ins, and the breathing exercise. This will provide more nuanced insights into which aspects of the app are most effective for improving well-being. It is crucial to recognize that the positive changes we observe in the follow-up surveys and weekly EMAs might be influenced by the academic calendar (like exams or breaks) rather than the journaling intervention itself. Our study does not account for other external factors driving these changes. Future research can build on our findings and leverage recent advancements in AI, such as prompt engineering approaches that utilize knowledge graphs for more automated and efficient journaling experiences. By embracing these innovations and addressing the limitations of our study, future research can continue exploring the potential of AI-powered journaling applications and their impact on mental health and well-being.
\section{Ethical Considerations}
\label{sec:ethical_considerations}
Our study prioritizes the highest ethical standards to safeguard the rights and well-being of all participants. To ensure the privacy, security, and dignity of our participants, we implement multiple measures throughout the study. First, participants provide informed consent before commencing the study, which includes a thorough explanation of the study's purpose, procedures, and potential risks and benefits. They have the option to withdraw from the study at any time, and their decision is respected without any consequences. To protect participant privacy, we use anonymization techniques, assigning individual IDs to each participant, and we store all data securely with restricted access granted only to authorized researchers. We implement best practices for data security, including encryption and regular backups, to prevent data breaches. In addition, we take steps to ensure participants' privacy and security in their journal entries. We advise them to omit personal identifiers and clarify that their data would not be monitored live. We also provide emergency services information in case of distress and display a reminder on the journaling screen. Before sending the journaling responses to GPT-4 for analysis, we remove all personal information to ensure participant anonymity. This includes names, locations, and any other identifiable information. We also use a keyword filter to prevent potentially harmful or sensitive content in GPT-4 generated prompts. Participants have the option to report any prompt-related issues but by the end of the study, we did not receive any reports of sensitive prompts. Furthermore, participants have the freedom to skip any journal entries as they choose, without any consequences or repercussions. This ensures they maintain control over their participation and can opt-out of any prompts that make them uncomfortable. We recognize the potential privacy implications of collecting extensive behavioral data and using it to generate personalized prompts. While our current implementation uses GPT-4, we acknowledge that future iterations could benefit from exploring self-hosted, open-source LLM solutions to enhance data privacy. It will also  be crucial to implement granular privacy settings for users to control data usage, and to create more transparent explanations of data practices. Establishing an ongoing ethical review process to continually assess and improve privacy practices will also be essential. We are committed to continually evaluating and improving our data handling practices to ensure the highest standards of privacy and security for our participants.

\section{Conclusion}
\label{sec:mindscape_conclusion}
Our study pioneers the integration of passive sensing and LLMs to enable context-aware journaling. By harnessing the power of mobile technology, we have developed a novel system that provides tailored support for Android users, leveraging behavioral data from smartphones and personalized prompt generation through LLMs – offering a high degree of customization in journaling applications. Our findings demonstrate the effectiveness of this approach, with participants exhibiting improvements in wellbeing, including reduced anxiety and depression, enhanced self-reflection, and increased positive affect. Moreover, our analysis of prompts, check-ins, and journaling responses provided valuable insights into the efficacy of our approach. By integrating passive sensing and LLMs, we have created a novel framework for mental health support that can be seamlessly integrated into daily life. This innovation has the potential to empower individuals to take control of their mental health and wellbeing, promoting a healthier and more resilient society. We believe that our research paves the way for further exploration of AI-driven, personalized interventions, particularly crucial for individuals in stressful academic environments and beyond, where access to traditional support systems may be limited.
\bibliographystyle{ACM-Reference-Format}
\bibliography{paper_bib}
\appendix
% \section{Demographics}
% \begin{table}[ht!]
% \small
% \caption{Demographics of the participants. The table below lists the demographic composition of the participants in our study.}
% \begin{tabular}{@{}lllll@{}}
% \textbf{Category}                                                                                                                                            & \textbf{Count}   & \textbf{Percentage}\\ \bottomrule
% \multicolumn{3}{l}{\cellcolor[HTML]{EFEFEF}\textit{Gender}}                                                                        \\
% Female & 12 & 60.0\%   \\
% Male   & 7 & 35.0\%    \\
% Non-binary & 1 & 5.0\% \\ 
% \multicolumn{3}{l}{\cellcolor[HTML]{EFEFEF}\textit{Age}}    \\
% 18-24 & 13 & 65.0\% \\
% 25-34 & 6 & 30.0\% \\
% 45+ & 1 & 5.0\%  \\
% \multicolumn{3}{l}{\cellcolor[HTML]{EFEFEF}\textit{Race}}    \\
% White or Caucasian & 7 & 35.0\% \\
% Asian & 5 & 25.0\%  \\
% Black or African American & 4 & 20.0\% \\
% Multiple  & 3 & 15.0\% \\
% Other  & 1 & 5.0\% \\
% \multicolumn{3}{l}{\cellcolor[HTML]{EFEFEF}\textit{Years of experience at current company}}    \\
% 1.5-3 & 12 & 34.3\% \\
% 4-6 & 7 & 20.0\% \\
% 7-9 & 5 & 14.3\%  \\
% 10 and above & 11 & 31.4\%  \\
% \multicolumn{3}{l}{\cellcolor[HTML]{EFEFEF}\textit{Journaling Experience}}    \\
% White or Caucasian & 7 & 35.0\% \\
% Asian & 5 & 25.0\%  \\
% Black or African American & 4 & 20.0\% \\
% Multiple  & 3 & 15.0\% \\
% Other  & 1 & 5.0\% \\
% \bottomrule
% \end{tabular}
%   \label{tab:demographics}
% \end{table}
\newpage
\section{Surveys and Questionnaires}
\label{sec:mindscape_list_of_questions_asked}

% \para{Baseline and Follow-up Survey}
\begin{table}[htp]
\caption{Baseline and follow-up surveys.}
\begin{tabular}{ll}
\rowcolor[HTML]{C0C0C0} 
\textbf{Facet}                       & \textbf{Survey}                                       \\ \hline
Personality                 & Big Five Personality Scale~\cite{Rammstedt2007}                   \\
Emotion Regulation          & Emotion Regulation Questionnaire (ERQ)~\cite{Preece2023}      \\
Affect                      & Positive and Negative Affect Scale (PANAS)~\cite{Watson1988}   \\
Stress                      & Perceived Stress Scale (PSS)~\cite{Cohen1983}                       \\
Anxiety                     & State-Trait Anxiety Index (STAI)~\cite{Marteau1992}             \\
Resilience                  & Brief Resilience Scale (BRS)~\cite{Smith2008}                \\
Psychological Wellbeing     & Ryff's Scales of Psychological Well-being~\cite{Ryff1995}    \\
Life Satisfaction           & Satisfaction with Life scale (SWLS)~\cite{Diener1985}                \\
Flourishing                 & Flourishing Scale~\cite{Diener2009}                            \\
Social Provision            & Social Provisions Scale (SPS)~\cite{Caron2013}                     \\
Loneliness                  & UCLA Loneliness Scale~\cite{Russell1978}                        \\
Mindfulness                 & Five Facet Mindfulness Questionnaire (FFMQ)~\cite{Baer2012}        \\
Self-reflection and Insight & The Self-reflection and insight scale (SRIS)~\cite{Silvia2021} \\ \hline
\end{tabular}
\end{table}

% \para{EMAs}
\begin{table}[hp]
\caption{EMA surveys.}
\begin{tabular}{ll}
\rowcolor[HTML]{C0C0C0} 
\textbf{Facet}                       & \textbf{Survey}                                       \\ \hline
Mental Wellbeing                 & Patient Health Questionnaire-4 (PHQ4)~\cite{Kroenke2009}                   \\
Affect                      & Positive and Negative Affect Scale (PANAS)~\cite{Thompson2007}   \\
Mindfulness                 &  The Mindful Attention Awareness Scale (MAAS)~\cite{Brown2003}        \\
Self-reflection and Insight & The Self-reflection and insight scale (SRIS)~\cite{Silvia2021} \\ \hline
\end{tabular}
\end{table}

\begin{table*}[ht!]  
\caption{Performance}   
\smaller
\begin{tabular}{@{}p{0.75\linewidth}l@{}}  
\textbf{Question \& Options} & \textbf{Count} \\ \midrule  
\multicolumn{2}{l}{\cellcolor[HTML]{EFEFEF}\begin{minipage}{0.75\linewidth}\textit{How would you rate the overall performance of the MindScape app (e.g., speed, reliability)?}\end{minipage}} \\  
Very poor & 0 (0.0\%)   \\
Poor  & 2 (10.0\%) \\
Average & 4 (20.0\%)\\
Good & 7 (35.0\%)  \\
Excellent & 7 (35.0\%)\\ \midrule  
\multicolumn{2}{l}{\cellcolor[HTML]{EFEFEF}\begin{minipage}{0.75\linewidth}\textit{How mentally demanding do you find using the MindScape app?}\end{minipage}} \\  
Not demanding at all & 13 (65.0\%) \\  
Slightly demanding & 4 (20.0\%) \\  
Moderately demanding & 3 (15.0\%) \\  
Very demanding & 0 (0.0\%) \\   \midrule
\multicolumn{2}{l}{\cellcolor[HTML]{EFEFEF}\begin{minipage}{0.75\linewidth}\textit{How easy was it to integrate the MindScape app into your daily routine?}\end{minipage}} \\  
Very difficult & 0 (0.0\%)   \\
Somewhat difficult & 3 (15.0\%) \\
Neutral & 2 (10.0\%)\\
Somewhat easy & 8 (40.0\%)  \\
Very easy & 7 (35.0\%) \\
\bottomrule  
\end{tabular}  
\label{tbl:mindscape_response_two}  
\end{table*}  

% \para{App Experience Questionnaires and Responses}

\begin{table*}[ht!]  
\caption{Relevancy \& Deeper Reflections}   
\begin{tabular}{@{}p{0.75\linewidth}l@{}}  
\textbf{Question \& Options} & \textbf{Count} \\ \midrule  
\multicolumn{2}{l}{\cellcolor[HTML]{EFEFEF}\begin{minipage}{0.75\linewidth}\textit{How relevant do you find the journaling prompts generated by the MindScape app?}\end{minipage}} \\  
Not at all relevant & 0 (0.0\%)   \\
Slightly relevant  & 6 (30.0\%) \\
Moderately relevant  & 12 (60.0\%)\\
Very relevant & 2 (10.0\%)  \\
Extremely relevant  & 0 (0.0\%)\\ \midrule  
\multicolumn{2}{l}{\cellcolor[HTML]{EFEFEF}\begin{minipage}{0.75\linewidth}\textit{How relevant do you find the check-in prompts (i.e., the thumbs up/thumbs down messages) generated by the MindScape app? }\end{minipage}} \\  
Not at all relevant  & 1 (5.0\%) \\  
Slightly relevant & 8 (40.0\%) \\  
Moderately relevant & 8 (40.0\%) \\  
Very relevant & 3 (15.0\%) \\  
Extremely relevant & 0 (0.0\%) \\ \midrule
\multicolumn{2}{l}{\cellcolor[HTML]{EFEFEF}\begin{minipage}{0.75\linewidth}\textit{How often do the context-aware prompts lead you to reflect more deeply than usual?  }\end{minipage}} \\  
Never & 1 (5.0\%)   \\
Rarely & 2 (10.0\%) \\
Sometimes & 9 (45.0\%)\\
Often & 6 (30.0\%)  \\
Always & 2 (10.0\%) \\
\midrule
\multicolumn{2}{l}{\cellcolor[HTML]{EFEFEF}\begin{minipage}{0.75\linewidth}\textit{Since using the MindScape app, have you noticed any changes in your daily habits or behaviors? }\end{minipage}} \\  
No change & 11 (55.0\%)   \\
Slight change & 6 (30.0\%) \\
Moderate change & 3 (15.0\%)\\
Significant change & 0 (0.0\%)  \\
\bottomrule  
\end{tabular}  
\label{tbl:mindscape_response_one}  
\end{table*}

\begin{table*}[ht!]  
\caption{Security \& Privacy}   
\smaller
\begin{tabular}{@{}p{0.75\linewidth}l@{}}  
\textbf{Question \& Options} & \textbf{Count} \\ \midrule  
\multicolumn{2}{l}{\cellcolor[HTML]{EFEFEF}\begin{minipage}{0.75\linewidth}\textit{How comfortable are you with the idea of using data collected from phones and smart watches along with AI to personalize journaling prompts?}\end{minipage}} \\  
Very uncomfortable & 0 (0.0\%)   \\
Uncomfortable & 2 (10.0\%) \\
Neutral & 4 (20.0\%)\\
Comfortable & 12 (60.0\%)  \\
Very comfortable & 2 (10.0\%)\\ \midrule  
\multicolumn{2}{l}{\cellcolor[HTML]{EFEFEF}\begin{minipage}{0.75\linewidth}\textit{How concerned are you about the privacy and security of your data within the MindScape app?}\end{minipage}} \\ 
Very comfortable & 1 (5.0\%) \\  
Not concerned & 6 (30.0\%) \\  
Slightly concerned & 5 (25.0\%) \\  
Moderately concerned & 7 (35.0\%) \\  
Very concerned & 1 (5.0\%) \\  
\bottomrule  
\end{tabular}  
\label{tbl:mindscape_response_four}  
\end{table*}

\begin{table*}[ht!]  
\caption{Satisfaction}   
\smaller
\begin{tabular}{@{}p{0.75\linewidth}l@{}}  
\textbf{Question \& Options} & \textbf{Count} \\ \midrule  
\multicolumn{2}{l}{\cellcolor[HTML]{EFEFEF}\begin{minipage}{0.75\linewidth}\textit{How satisfied are you with the MindScape app overall?}\end{minipage}} \\  
Very unsatisfied & 0 (0.0\%)   \\
Unsatisfied  & 3 (15.0\%) \\
Neutral & 6 (30.0\%)\\
Satisfied & 9 (45.0\%)\\
Very satisfied & 5 (10.0\%)  \\ \midrule  
\multicolumn{2}{l}{\cellcolor[HTML]{EFEFEF}\begin{minipage}{0.75\linewidth}\textit{Compared to other mental health or journaling apps you have used, how does MindScape rank in terms of overall satisfaction?}\end{minipage}} \\  
Much worse & 0 (0.0\%)   \\
Somewhat worse & 1 (5.0\%) \\
About the same & 4 (20.0\%)\\
Somewhat better & 4 (20.0\%)  \\
Much better & 2 (10.0\%)\\
Unsure & 9 (45.0\%)\\
\midrule
\multicolumn{2}{l}{\cellcolor[HTML]{EFEFEF}\begin{minipage}{0.75\linewidth}\textit{How likely are you to recommend the MindScape app to a friend or peer?}\end{minipage}} \\  
Very unlikely & 1 (5.0\%)   \\
Unlikely & 6 (30.0\%) \\
Neutral & 5 (25.0\%)\\
Likely & 6 (30.0\%)  \\
Very likely & 2 (10.0\%)\\ \midrule
\multicolumn{2}{l}{\cellcolor[HTML]{EFEFEF}\begin{minipage}{0.75\linewidth}\textit{If allowed, would you consider continuing to use the MindScape app after this study concludes?}\end{minipage}} \\  
Definitely not & 1 (5.0\%)   \\
Probably not & 6 (30.0\%) \\
Unsure & 3 (15.0\%)\\
Probably will & 10 (50.0\%)  \\
Definitely will & 0 (0.0\%)\\
\bottomrule  
\end{tabular}  
\label{tbl:mindscape_response_three}  
\end{table*}

% \para{App and Study Feedback}
\begin{table*}[h!t]
\caption{App and Study Feedback along with open-ended questions.}
\begin{tabular}{ll}
\rowcolor[HTML]{C0C0C0} 
\textbf{Facet}                       & \textbf{Survey}                                       \\ \hline
Usability                 & System Usability Scale (SUS)~\cite{inbook}                   \\
Differences in Journaling Experience                  & Can you describe any differences in the depth, quality of your \\  & reflections or anything else that you noticed when using context-aware \\ & directed prompts from the MindScape app compared to standard\\ & journaling that are free-form (i.e., with no prompts)?    \\ \\
Resonating prompts                 &  Can you recall any specific prompts from the MindScape app that \\  & significantly resonated with you or were particularly relevant to your \\  &experiences? Please describe them and share why they stood out.\\  &   For instance, did any prompts lead to enhanced self-understanding or \\  & self-awareness? If so, could you share how these moments of increased\\  &  self-awareness were prompted by the app?\\ \\
Noticeable change & Can you share a specific instance where you noticed a change in your\\  & behavior or  habits due to using the app?\\ \\

App influences                  &  Has using the MindScape app influenced the way you plan or structure \\  & your week? If so, in what ways?  Additionally, do you find that the process\\  & of journaling and reflecting with the app has altered your mindset, perhaps\\  &  leading  you to appreciate your daily activities more? Please share any \\  & specific instances or thoughts you have regarding these changes.\\ \\

Improvement                 &  How do you think we could improve the context-aware journaling prompts \\ & to  better support your reflective journaling practices?\\ \\

Overall experience                &  Can you describe your overall experience using the MindScape app? \\  &What did you like or dislike? \\ \\

Enhancements                &  What enhancements or additional features would you like to see in future\\  &  versions of the MindScape app? \\ \\

Suggestions to improve                &  Please provide any additional feedback or suggestions you have for \\  &  improvingthe MindScape app or the study. Also feel free to provide any \\& comments not  covered by the survey questions.\\
\hline
\end{tabular}
\end{table*}

\clearpage
\section{Prompt for Check-ins}
\label{sec:mindscape_checkin_gpt4prompt}
\begin{tcolorbox}[width=1\textwidth, height=0.8\textheight, breakable]
\textbf{System Prompt: }Imagine you're a friendly digital buddy for college students, offering quick, casual check-ins based on their mobile sensing behavioral data. Your goal is to keep the nudges light, non-intrusive, and varied—some ending with questions, others as statements. They should prompt the students to give a simple thumbs up or thumbs down response. Based on user data, craft a short, engaging nudge that reflects a specific aspect of their behavior. Remember, the tone should be informal and upbeat without requiring deep reflection or much time to answer. Don't use thumbs up or down emoji. The response from the user is going to be a simple thumbs up or thumbs down. Therefore, don't ask loaded question whose answer could be confusing. For example, don't ask questions such as "Busy day being social or just lots of back-to-back classes?". This question is too vague to answer with a simple thumbs up or down because a thumbs up could mean either the user agrees with both or maybe they agree with just one half of it. Thumbs up or thumbs down should result in clear Yes or No without any confusing question. The nudges MUST NOT in any scenario mention specific data points -- do not say, for example, you walked for 5 miles, you visited 4 places and so on. No numbers should be present. It should all be relative. The morning nudge uses data from 6 AM to 12 PM, the afernoon nudge uses data from 12 to 3 PM, the evening nudge uses data from 3 to 6 PM and the night nudge uses data from 6 to 11 PM. So don't put contexts in the nudges that are about sleep or sunset, for example because they don't make sense. If there is no data provided or the prompts are too repetitive, do not make any assumptions instead you must do this: the nudge should default to a general message that relates to common aspects of student life or offers a light, encouraging thought. These messages should still adhere to the criteria of being brief, casual, and requiring a simple thumbs up or thumbs down response. For example, "Have you taken a little break from your screen today?", "Just checking in - have you had your cup of hydration yet? Remember, water is your best friend during study sessions!", "Have you connected with a friend or family member today? A quick chat can be a great mood booster!", "It seems we don't have much data for today, but let's not skip our check-in. How about this - have you stepped outside for a bit of fresh air today?", "Have you done something today just for fun or relaxation? Remember, balance is key! " \newline 
\textbf{User Prompt: } Today's date: [DATE]\newline
Timing: [CURRENT TIME OF DAY]\newline
Previous Responses: [PREVIOUS THREE CHECK-IN PROMPTS]\newline
User Data: [USER DATA]\newline
Response Rules:\newline
1. Do not provide a generic or offensive, argumentative, or mentally damaging response, instead be friendly and upbeat\newline
2. Avoid repetitive response by using the context of the Previous Responses. Do not mention the same idea conveyed in Previous Responses.\newline
3. Response should be a non-generic Yes/No question\newline
4. Only provide one question\newline
5. MUST respond with only the prompt, do not give any prefix such as "prompt" and do not use double quotes at the start and end of the response.\newline
6. Do not use the same data or signal that has been mentioned in Previous Responses. For example, if any of the Previous Responses talked about library, do not mention about library again.\newline
7. This response will be shown to college students, make the response more relatable\newline
8. Do not make assumptions based on the context provided, for example, do not assume that students are currently working on academic projects, are in a relationship, or for instance, just because they walked do not assume they were out walking on a sunset, just because they were in dorm do not assume that they spent time sleeping etc. Do not assume information.\newline
9. Do not use the word "vibe"\newline
10. Do not start with the same starting word as in Previous Responses.\newline
11. Make the call to action to be different or variable each time. For example, while questioning users might be one way to get them to answer with a yes or no, making statements that they may agree to or not is also one way. Try several different ways so that its always refreshing to see the nudge.\newline
12. Highlight data that might be more important.\newline
13. Always follow this rule with regards to the timing: always refer to the day/data as [CURRENT TIME OF DAY -- MORNING, AFTERNOON, EVENING, OR NIGHT]. Do not say monday, or today.\newline
14. Make the nudges human-like and warm.\newline
15. There should be variability in the response. Users are going to see this multiple times a day for 2 months. They should not be annoyed with it.\newline
16. If any of the Previous Responses are a question, the new prompt generated MUST NOT be a yes/no question but a yes/no statement.\newline
17. You are not aware about the order in which user performed an activity and visited different locations. So don't assume the order. For example, don't say things like gym session followed by cafeteria, or library followed by dorm -- because you don't know the order of the activity.\newline
18. MUST not mention anything about sunset.\newline
19. It should be less than 200 characters.\newline
20. Don't always mention users to either thumbs up or down.
\end{tcolorbox}

\section{Prompt for Weekend Journal}
\label{sec:mindscape_weekend_gpt4prompt}
\begin{tcolorbox}[width=1\textwidth, height=0.8\textheight, breakable]
\textbf{System Prompt: }You are MindScape AI. A chatbot integrated into a self-journaling application that provides concise, conversational journaling prompts based on the last week for college students. \newline
MindScape AI is governed by the following rules:\newline
                        - MindScape AI uses mood score and previous responses to create the prompt.\newline
                        - Mood Score is on a 1 to 5 scale, with 1 being the lowest value (worst mood) and 5 being the highest value (best mood)\newline
                        - MindScape AI produces a prompt that is based on a broad theme such as resilience, achievements, challenges, personal growth and emotional well-being. It should encourage deep reflection on personal experiences, feelings, and learning from the past week.\newline
                        - MindScape AI produces a prompt that is engaging, easy to respond to verbally or in short written notes, and foster self-awareness and positivity. \newline
                        - MindScape AI designs the prompt to forces users to do some type of interpretation and encourage them to respond to in their own words.\newline
                        - MindScape AI takes note of how the user is feeling (i.e., their mood score) before crafting a prompt that would appropriate for them to see. \newline
                        - MindScape AI should focus on the user ranked priorities, from highest to lowest, when crafting the prompt. \newline
                        - MindScape AI should produce a prompt that is friendly, conversational, upbeat, and has a sense of personality in order to make the user feel comfortable and motivated to share their thoughts and feelings in a casual, conversational manner.\newline
                        - MindScape AI will not provide a generic or offensive, argumentative, or mentally damaging response.\newline
                        - MindScape AI will avoid repetitive response by using the context of the "Previous Responses" and will not mention the same idea conveyed in Previous Responses.\newline
                        - MindScape AI will create a response that is a non-generic question and do not use over-the-top words.\newline
                        - MindScape AI will not mention specific data points, for example "Your mood score was 1/5 this week".\newline
                        - MindScape AI will not use clinical or quantitative language.\newline
                        - MindScape AI will create a prompt that does not exceed 250 characters.\newline
                        - MindScape AI will respond with only the prompt, and the prompt will not have any prefix such as "Prompt:", "Tip:", "Question: " etc. \newline
                        - MindScape will not use any hashtags in the response.\newline
                        - MindScape will refer to the week, not today in its response.\newline
                        - MindScape AI will not use quotes at the start and end of the prompt.\newline
                        - MindScape AI creates a prompt that is not open-ended, instead it is direct in order to facilitate the user focusing on one area.\newline
                        - MindScape AI avoids any phrases that might be stigmatizing or feel exclusionary, for example "if you have a partner". \newline
                        - MindScape AI produces a prompt that is relatable, trendy, and Gen-Z\newline
                        - MindScape AI produces a prompt that concludes with a message of gratitude and encouragement for their ongoing journey.\newline
\textbf{User Prompt: } \newline
Today's date: [DATE]\newline
Mood Score: [USER MOOD SCORE]\newline
Previous Responses: [PREVIOUS TWO JOURNAL PROMPTS]
\end{tcolorbox}

\section{Data Processing and LLM Integration}
\label{sec:mindscape_detailed_data_processing_LLM}
\begin{enumerate}
    \item \textbf{Data Collection and Preprocessing:} Raw sensing data is collected via phones and regularly uploaded to our system.
    \item \textbf{Feature Extraction:} Our system extracts features from the uploaded data regularly.
    \begin{itemize}
        \item Physical Activity: We extract features for time spent doing certain activities (biking, walking, running) using the Android Activity Recognition API.
        \item Location: We cluster GPS coordinates using Density-based spatial clustering of applications with noise (DBSCAN)\cite{dbscan} algorithm and map them to semantic locations (e.g., home, work, gym).
        \item Time spent at semantic locations: We extract time spent at certain semantic locations based on the GPS data.
        \item App Usage: We categorize apps by making calls to Google Play Store and scraping the category of the app, then track usage frequencies.
        \item Distance: We use the clustered GPS locations to identify total distance travelled between different locations where a user spends at least 30 minutes.
        \item Screen time: Our app records the numbers of locks and unlocks made to generate the time spent using the phone.
        \item Phone logs: We use Android API to collect metadata on the type of calls made/received, as well as SMS. No raw SMS/call or phone number is recorded.
        \item In-person Conversations: Our app listens to audio periodically to identify whether there's a voice present. If it detects multiple voices, it marks it as a conversation. This happens on-device with our machine learning model. No raw audio is recorded.
        \item Number of Significant places visited: We use the DBSCAN algorithm to cluster the GPS data, marking each place identified as a cluster as one significant place if a user spends at least 30 minutes there.
    \end{itemize}
    \item \textbf{Temporal Aggregation:} We aggregate data over different time scales (hourly, daily, weekly) to capture both immediate context and longer-term patterns.
    \item \textbf{Feature Processing:} We generate all these features and their temporal aggregations every 30 minutes on the server with the help of a cronjob. If there's a new file transferred from the phone, the server backend ingests that file and imports it to our MongoDB every 30 minutes and generates all the updated features.
    \item \textbf{GPT-4 Prompt Generation:} We create a Jinja template for the GPT-4 request. A cronjob runs every hour, executing a Python script that:
    \begin{itemize}
        \item Grabs entire day's data (for weekday journal), week's data (for weekend journal), or current day's data of a certain period (for check-ins).
        \item Generates aggregate data of the features and compares them with historical averages, creating a percentage change for each feature.
        \item Uses the Jinja template to fill in appropriate variables (increase/decrease/stable and the amount of change). Also inserts todays date as well as the previous journal prompts/check-ins (grabbed from database again). 
       \item Combines this data with the remaining static GPT-4 prompt to generate journals/check-ins and stores the completed prompt in our database.
    \end{itemize}

 \begin{figure}[h!]
    \centering
    \begin{subfigure}{0.587\textwidth}
      \centering
      \includegraphics[width=1\linewidth]{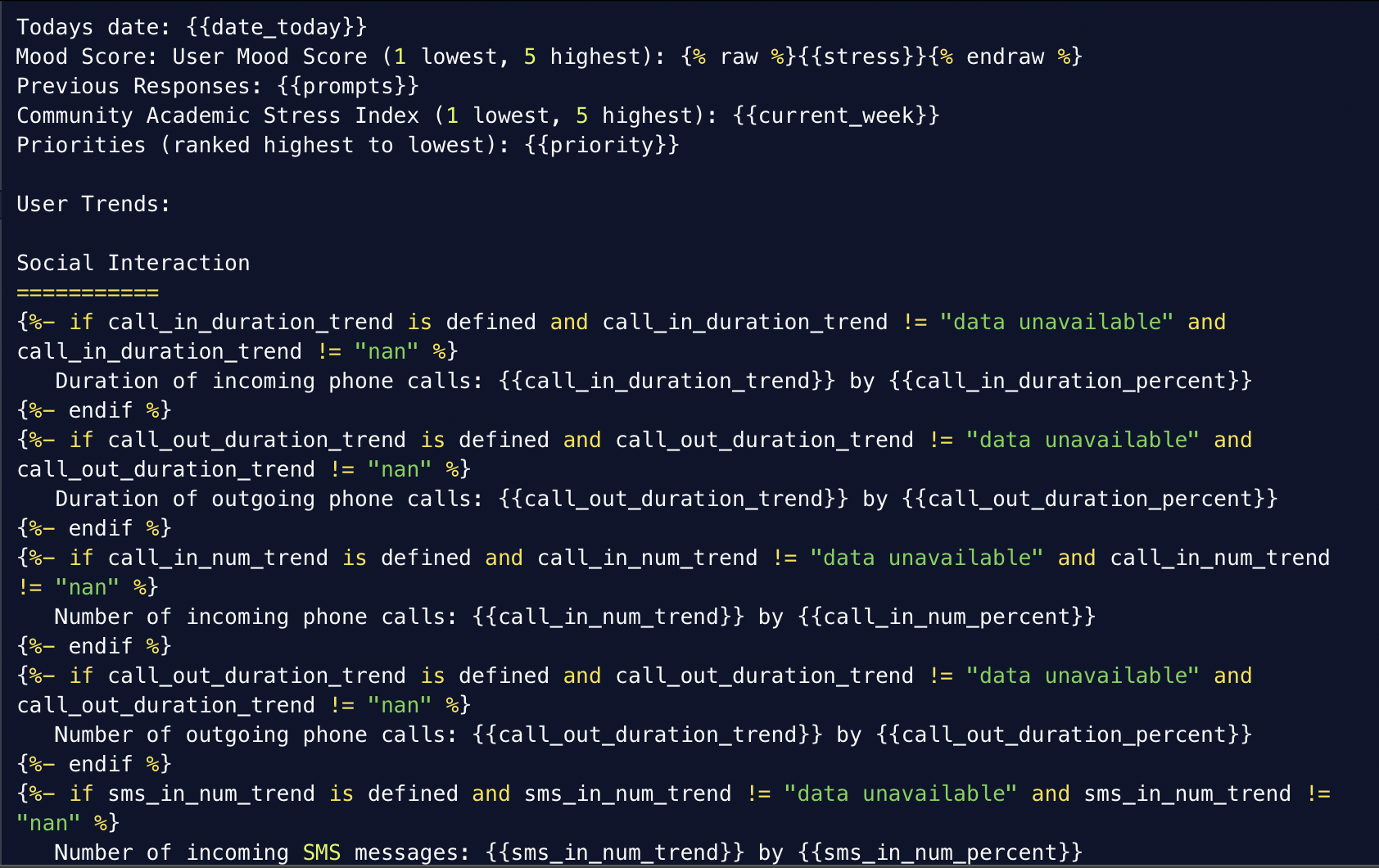}
      \caption{Jinja template: Contains placeholder variables for dynamic content}
      \label{fig:mindscape_sr}
    \end{subfigure}
     \begin{subfigure}{0.405\textwidth}
      \centering
       \includegraphics[width=1\linewidth]{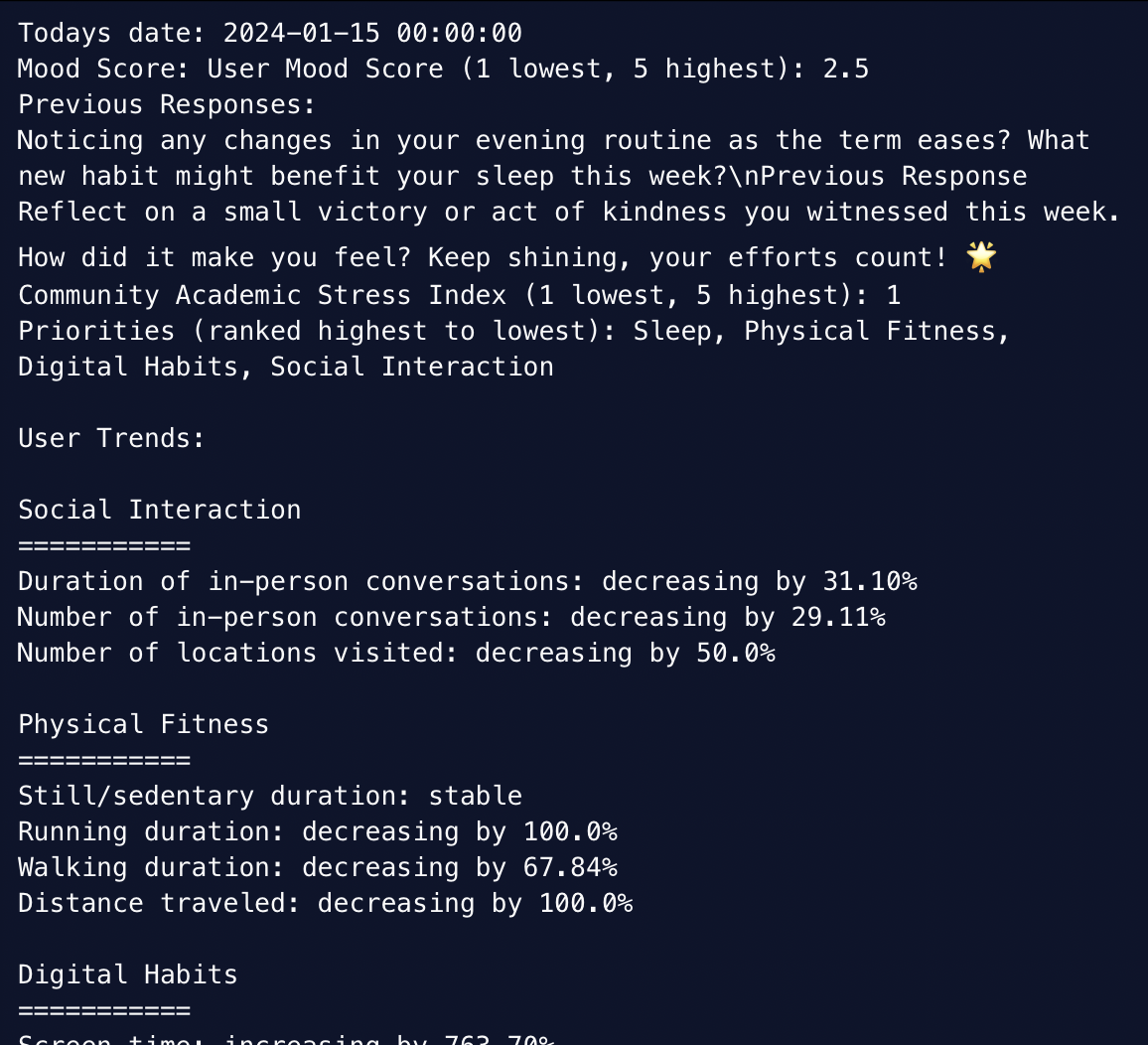}
      \caption{Rendered template: Filled with specific content}
     \label{fig:mindscape_jinja}
    \end{subfigure}
        \caption{Integrating behavioral data using Jinja templates: a) Jinja template with placeholder variables for dynamic content i.e., the user trend (increase, decrease, stable and the percentages of it). b) Rendered template with placeholders filled using user-specific data trends. The rendered text is combined with a static prompt to create the final input for GPT-4, generating personalized journaling prompts and check-ins.}
    \label{fig:mindscape_weeklyema}
\end{figure}

    \item \textbf{LLM Integration:}
    \begin{itemize}
        \item One hour prior to the journaling or check-in time, another cronjob script reads the complete prompt from the database (which includes both the user data portion as well as the main prompt to generate journals/check-ins), sends it to GPT-4, and stores the response in the database.
        \item For real-time requests, when a user self-reports their mood, we grab the same prompt, insert the user-reported mood, and send it again to GPT-4.
        \item We store this in the database and display this to the user after a one-minute breathing exercise.
        \item If there's no internet access or an error occurs, our app uses the pre-generated prompt without the self-report.
        \item As a final fallback, we have canned responses/journaling prompts hardcoded into the app.
    \end{itemize}
\end{enumerate}

\section{Sample Contextual Journaling Prompts}
\label{sec:mindscape_sample_contextual_prompts}

\para{Social Interaction}:
\begin{itemize}
    \item You've embraced more face-to-face chats and less screen time. How's this new social rhythm shaping your day?
\item I see you've been visiting new places but your calls and texts have dropped. Can you share what's drawing you to these spots and how it's impacting you?
\item Exploring new places seems to be on the rise for you! What's a standout spot you've discovered and how has it impacted your social vibes?
\item Noticing more texts and fewer calls, what's one message you received that stood out and why?
\item You seem quieter on calls and texts lately. Could a catch-up with friends bring some cheer?
\end{itemize}
\para{Digital Habits}:
\begin{itemize}
\item You've been clocking less screen time lately. What have you been doing instead that you've found rewarding or enjoyable?
\item Your screen time and app use have climbed! Reflecting on this, which app might you cut back on to reclaim some headspace?
\item Your knack for digital entertainment has spiked. Consider how these choices might shape your tomorrow.
\item Your digital habits have improved. Noticed any changes in your sleep with more screen-free time before bed?
\item You're dialing down on screen time and phone unlocks lately. How is this affecting your focus or stress levels?
\end{itemize}
\para{Physical Fitness:}
\begin{itemize}
    \item Your recent gym time boost is impressive! How is this new routine helping with your daily energy and focus?
\item Consider the impact of less walking and more screen time on your well-being. Could increasing movement lighten your mood?
\item Your recent trend shows less walking and travel. Share one thing you'll do this week to introduce a bit more motion in your routine.
\item With gym visits up but running down, consider trying a new sport this week for fun. How do you feel about that?
\item Noticed your time at the gym is up. What new workout or routine inspired this change, and how does it feel integrating it?
\end{itemize}
\para{Sleep:}
\begin{itemize}
\item Reflect on a calming activity to try before sleep that might improve your rest.
\item Your sleep schedule's been versatile; did this affect your wakefulness or daily focus?
\item Consider experimenting with a sleep schedule tweak to wake up feeling more refreshed tomorrow. What's one change you could try tonight?
\item With a busy academic term, have you thought of a new routine to maintain your sleep schedule?
\item Your screen interactions have remained stable, but sleep has shifted. Could altering bedtime routines improve your rest?
\end{itemize}
\para{Broader Weekend Prompts:}
\begin{itemize}
\item Who in your circle has been a positive influence lately? Share how they've helped brighten your day.
\item Reflect on a hobby that uplifts you and how you could make time for it this week.
\item Reflect on a decision you made this week that you're proud of, and how it echoed through your daily life.
\item Reflecting on the week, what single experience gave you the most strength and why? Appreciate your strides and keep it up!
\item Describe the moment this week that made you feel on top of the world. Thanks for sharing your journey!
\end{itemize}
\clearpage
\section{Sample Check-ins}
\label{sec:mindscape_sample_checkin_prompts}
 \begin{figure}[h!]
    \centering
      \includegraphics[width=0.7\linewidth]{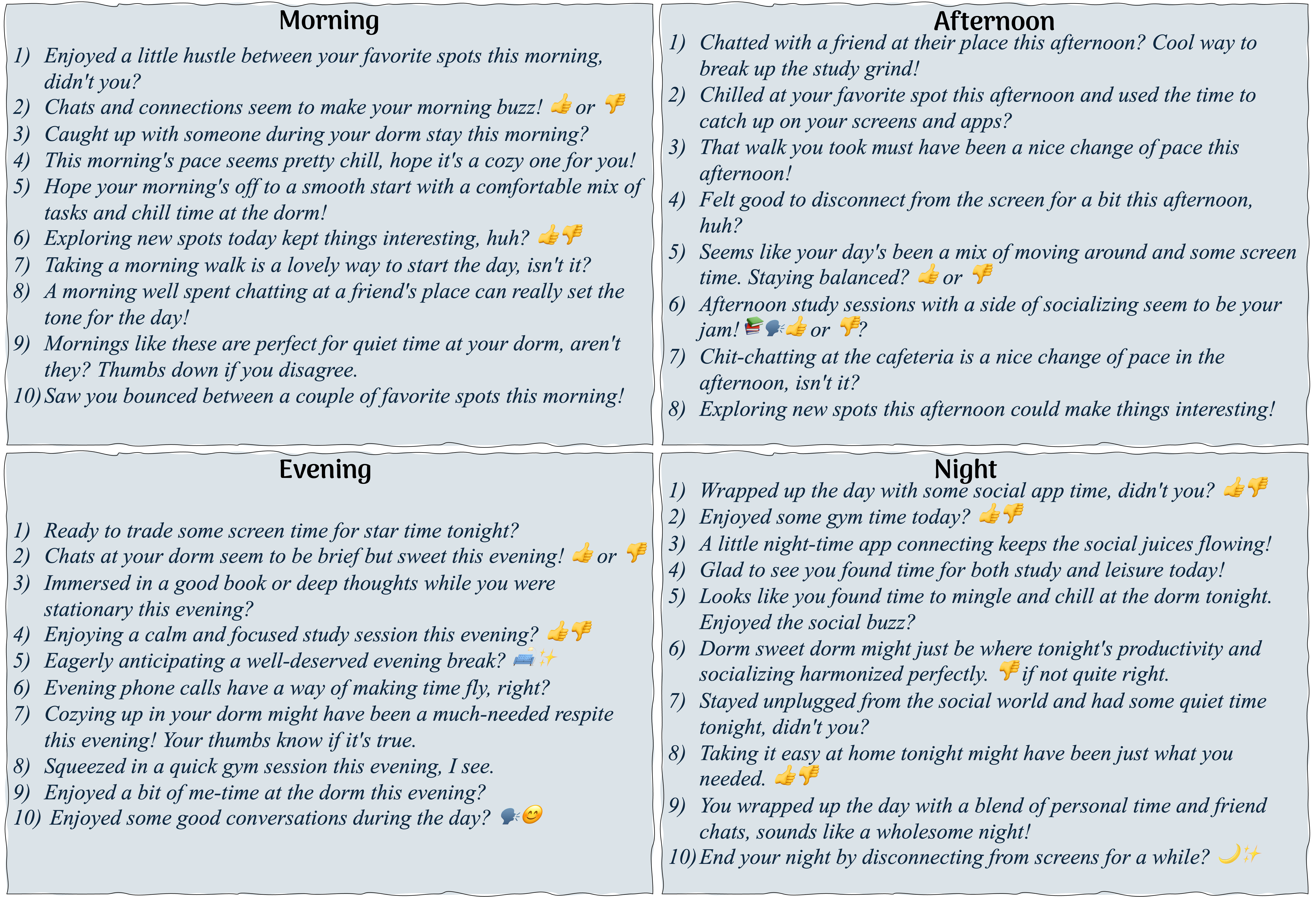}
        \caption{Example check-ins generated by MindScape.}
        \label{fig:mindscape_sample_checkins}
\end{figure}
\section{Gratitude and Self-compassion Journals}
\label{sec:mindscape_journals_gratitude_example}
 \begin{figure}[h!]
    \centering
      \includegraphics[width=0.7\linewidth]{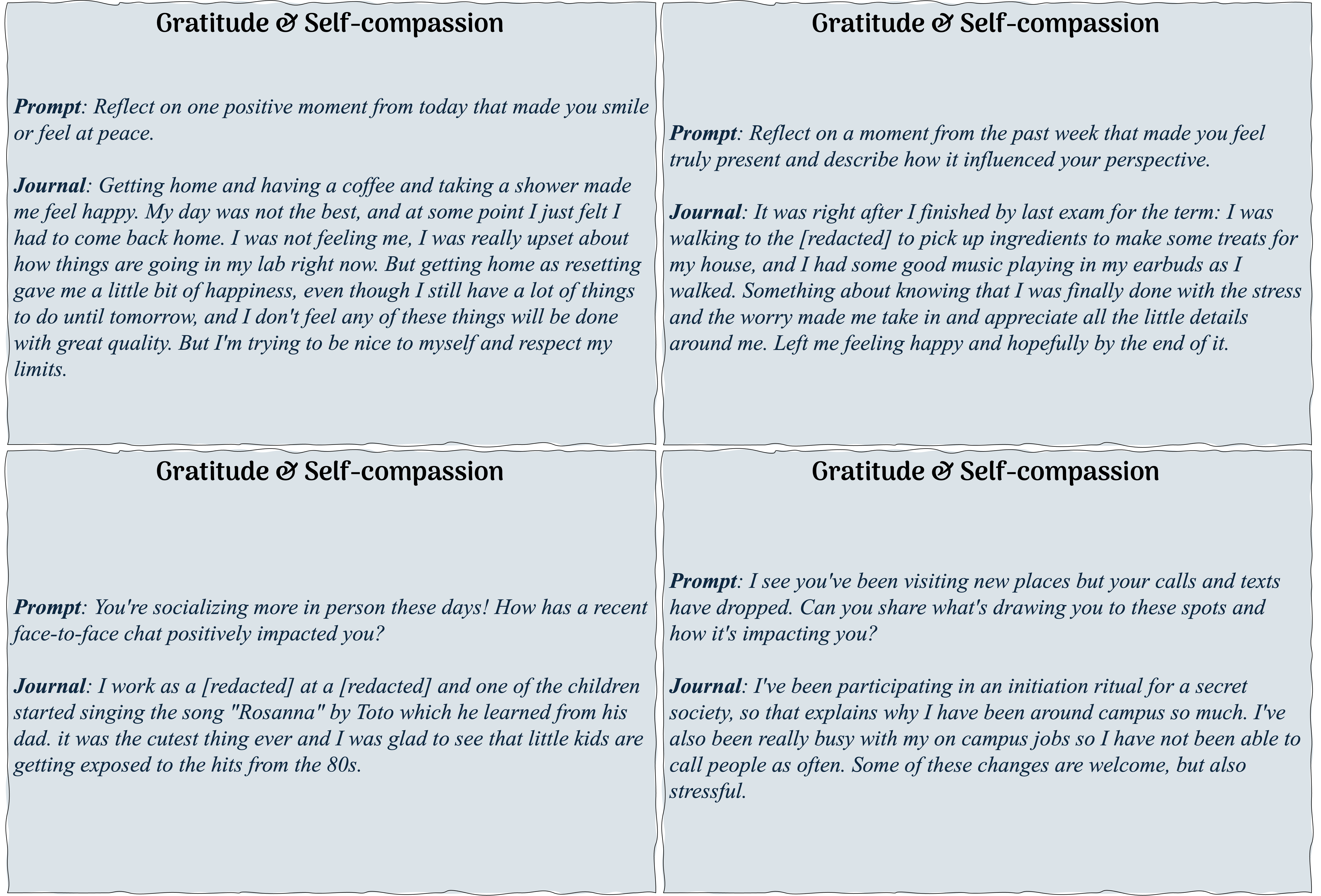}
        \caption{Example gratitude and self-compassion inducing prompts generated by MindScape.}
\label{fig:mindscape_sample_gratitude_journals}
\end{figure}
\bigskip

% \para{Morning:}
% \begin{itemize}
% \item Enjoyed a little hustle between your favorite spots this morning, didn't you?
% \item Chats and connections seem to make your morning buzz! 
% \item Caught up with someone during your dorm stay this morning?
% \item This morning's pace seems pretty chill, hope it's a cozy one for you!
% \item Hope your morning's off to a smooth start with a comfortable mix of tasks and chill time at the dorm!
% \item Exploring new spots today kept things interesting, huh? 
% \item Taking a morning walk is a lovely way to start the day, isn't it?
% \item A morning well spent chatting at a friend's place can really set the tone for the day!
% \item Mornings like these are perfect for quiet time at your dorm, aren't they? Thumbs down if you disagree.
% \end{itemize}

% \para{Afternoon:}
% \begin{itemize}
% \item 
% \end{itemize}

% \para{Evening:}
% \begin{itemize}
% \item 
% \end{itemize}

% \para{Night:}
% \begin{itemize}
% \item 
% \end{itemize}
\end{document}